\documentclass[a4paper,11pt]{iopart}
\pdfoutput=1 

\pdfoutput=1

\usepackage{graphicx}
\usepackage{bm}
\usepackage{braket}
\usepackage{amssymb}
\usepackage{cite}
\usepackage{mathrsfs}
\usepackage{amsmath}
\usepackage{xcolor}
\usepackage{hyperref}
\usepackage{enumerate}
\usepackage[toc,title]{appendix}
\usepackage{verbatim}
\usepackage[margin=30pt, bf, font=footnotesize, center, justification=justified]{caption}[2004/07/16]
\usepackage[T1]{fontenc} 
\usepackage{xcolor}
\usepackage{subfigure}
\usepackage{braket}

\hypersetup{colorlinks,bookmarksopen,bookmarksnumbered,
citecolor=red,
linkcolor=blue,
pdfstartview=green,
urlcolor=teal}

\catcode`@=11 
\renewcommand\tableofcontents{%
  \section*{\contentsname}%
  \@starttoc{toc}%
}
\catcode`@=12

\begin{document}

\title[Symmetry resolved entanglement in gapped integrable systems: a CTM approach
]
{Symmetry resolved entanglement in gapped integrable systems: a corner transfer matrix approach
}

\vspace{.5cm}

\author{Sara Murciano$^1$, Giuseppe Di Giulio$^1$, and Pasquale Calabrese$^{1,2}$}
\address{$^1$SISSA and INFN Sezione di Trieste, via Bonomea 265, 34136 Trieste, Italy.}
\address{$^{2}$International Centre for Theoretical Physics (ICTP), Strada Costiera 11, 34151 Trieste, Italy.}

\vspace{.5cm}

\begin{abstract}

We study the symmetry resolved entanglement entropies in gapped integrable lattice models.
We use the corner transfer matrix to investigate two prototypical gapped systems  with a $U(1)$ symmetry: the complex
harmonic chain and the XXZ spin-chain.
While the former is a free bosonic system, the latter is genuinely interacting. 
We focus on a subsystem being half of an infinitely long chain.  
In both models, we obtain exact expressions for the charged moments and for the symmetry resolved entropies. 
While for the spin chain we found exact equipartition of entanglement (i.e. all the symmetry resolved entropies are the same), 
this is not the case for the harmonic system where equipartition is effectively recovered only in some limits.
Exploiting the gaussianity of the harmonic chain, we also develop an exact correlation matrix approach to the symmetry resolved entanglement that allows us to 
test numerically our analytic results.

\end{abstract}

\maketitle

\newpage

\tableofcontents

\section{Introduction}
Entanglement is a characteristic treat of quantum mechanics since its early days. 
However, only in the last two decades it became clear that entanglement is an important concept also for many-body systems with ramifications 
to many different lines of research, ranging from high energy physics and gravity to quantum information and critical or topological  
extended quantum systems (see e.g. Refs. \cite{intro1, intro2, eisert-2010,intro3} as reviews). 
The most successful and used measures of the bipartite entanglement are surely the R\'enyi and von Neumann entropies, defined as follows.
 Let $\ket{\Psi}$ be a pure state of an extended quantum mechanical system and $\rho=\ket{\Psi}\bra{\Psi}$ its density matrix. 
Let us consider a bipartition of the system into $A$ and $ B$ and define the reduced density matrix (RDM) of the subsystem $A$ as the partial trace over the degrees 
of freedom of $B$, i.e. $\rho_A=\mathrm{Tr}_B\rho$. 
A measure of the entanglement between $A$ and $B$ is the R\'enyi entropy of order $n$
\begin{equation}
\label{eq:renydefinition}
S_n\equiv\dfrac{1}{1-n}\log \mathrm{Tr}\rho_A^n.
\end{equation}
The Von Neumann entanglement entropy $S_1\equiv-\mathrm{Tr}\rho_A\log \rho_A$
is the limit $n\to 1$ of the R\'enyi entropy. In a quantum field theory, $\mathrm{Tr}\rho_A^n$ for integer $n$ can be expressed in the path integral formalism 
as a partition function over suitable $n$-sheeted Riemann surfaces. 
For the ground state of critical one-dimensional systems with an underlying conformal field theory, 
this led to a remarkable universal scaling depending only on the central charge $c$ \cite{cc-04,cc-09,hlw-94, vidal,vidal1,vidal2}. 

Such a universal behaviour is not strictly a prerogative of the gapless models, but it also occurs
for gapped models in the vicinity of a quantum phase transition in the regime in which the correlation length $\xi$ is large but finite \cite{cc-04}.
Indeed, using ideas from the famous proof of the c-theorem by Zamolodchikov \cite{z-86}, it has been shown that for a 
bipartition of an infinite system into two semi-infinite halves, the leading behaviour of entanglement entropies is generically \cite{cc-04}
\begin{equation}
\label{eq:gappedcase}
S_n\simeq \dfrac{c}{12} \left( 1+\dfrac{1}{n} \right)\log \xi.
\end{equation}
This result can be elegantly recovered for integrable lattice models through the Baxter corner transfer matrix (CTM)  \cite{baxter},
as reported (and generalised) in many references \cite{cc-04,p-04,w-06,eer-10,cal2010,glen,eefr-12,df-13,br-16,Alba2018A}.
We will discuss explicitly this technique in the following sections. 
The CTM approach provided exact results not only close to the critical point, but gave generalisations also to the regime in which the 
correlation length is small. 
When the subsystem $A$ is a finite interval of length $\ell$, as long as $\ell\gg\xi$, the R\'enyi entropies are just twice the value in Eq. \eqref{eq:gappedcase}
as a consequence of cluster decomposition in the ground-state of these theories. 
However, as $\ell$ becomes of the order of $\xi$, a complicated crossover takes place that is not captured by CTM and requires more complicated 
techniques, see e.g. Refs. \cite{ccd-08,cd-09,lcd-13,c-17}. 

Only in very recent times, it became clear that it is also important to understand the relation between entanglement and symmetries and in particular how 
entanglement is shared between the various symmetry sectors of a theory \cite{lr-14,goldstein}. 
The physical motivations for shading light on the interplay between symmetry and entanglement are manifold. 
For example, one motivation comes from a recent experiment studying the time evolution of the symmetry resolved entanglement 
in systems with many body localisation \cite{fis}.
It has been shown that entanglement has two different contributions, called configurational and fluctuation entanglement (see below, cf. Eq. \eqref{eq:SvNEE}, for a precise definition). 
These two contributions account for the entanglement within symmetry sectors and fluctuations thereof, respectively. 
In the presence of both disorder and interaction, their dynamics occur over different time scales: 
the fluctuation entanglement quickly saturates to an asymptotic value while the configurational one exhibits a slow logarithmic growth \cite{fis}, 
providing a nice physical explanation of an older finding \cite{bpm-12,Vosk2014}. 
The possibility of measuring these quantities sparked the interest in further studying how the entanglement is related to the internal symmetries of a system,
leading to many results concerning critical ones  \cite{goldstein,goldstein1,xavier,goldstein2,riccarda,tr-19,fg-19,clss-19}. 
A surprising finding is that conformal invariance forces the entanglement entropy to be equally distributed among the different sectors of a $U(1)$ 
symmetric theory \cite{xavier}. 
It is an open issue to understand whether and when such {\it equipartition of entanglement} survives away from criticality.
However, to date there are no results concerning gapped systems (with the exception of Ref. \cite{fg-19} for a discrete symmetry, but here we are interested in continuous ones).
The goal of this work is to fill this gap and to study how the total entanglement splits into the contributions coming from disjoint symmetry sectors in gapped 
integrable models, using CTM techniques.
We carry out this analysis for two non-critical quantum lattice models with a $U(1)$ symmetry, namely the  double or complex harmonic chain (which is a free model) 
and the XXZ chain (which is genuinely interacting). 
To this aim, we first calculate the moments of the RDM in the presence of a charge flux, that we call charged moments, and 
then obtain the contributions of the sectors by Fourier transform.

 The manuscript is organised as follows. In Section \ref{sec:SR} we briefly review all the quantities of interest and we give an 
 overview of how the RDM of an off-critical quantum chain is related to Baxter's CTM. 
 For integrable models whose weights satisfy a Yang-Baxter relation, the eigenvalues of the RDM can be determined exactly. 
 In Sections \ref{sec:HC} and \ref{sec:XXZ} we exploit these exact results for the computation of the symmetry resolved entanglement entropy, 
 for the complex harmonic chain and XXZ spin-chain respectively. 
 We also benchmark the analytic results in Section \ref{sec:HC} against exact numerical computations. 
 We conclude in Section \ref{sec:concl} with some remarks and discussions. Many technical details of the calculations can be found in four appendices.

\section{Symmetry resolution, flux insertion, and corner transfer matrix}
\label{sec:SR}
We consider a system with a $U(1)$ symmetry, generated by the charge operator $Q$, which obeys $Q_A\oplus Q_B=Q$,  where $Q_i$ is the charge in the subsystem  $i$. If the system described by the density matrix $\rho$ is in an eigenstate of $Q$, then $[\rho,Q]=0$.  
We are interested in a bipartition of the total system into two semi-infinite halves, $A$ and $B$, and we denote by $\rho_A$ the reduced density matrix of $A$. 
Taking the trace over $B$ of $[\rho,Q] = 0$, we find that $[\rho_A,Q_A] = 0$. This means that $\rho_A$ is block-diagonal and each block corresponds to a different 
charge sector labelled by the eigenvalue $q$ of $Q_A$. Therefore we can write
\begin{equation}
\label{eq:sum}
\rho_A=\oplus_q p_{A}(q)\rho_{A}(q),
\end{equation}
where $p_{A}(q)$ is the probability of finding $q$ in a measurement of $Q_A$ in the RDM $\rho_A$, i.e. $p(q)= \mathrm{Tr} \Pi_q \rho_A$, where $\Pi_q$ is the projector 
on the sector of charge $q$. 
Within this convention, the density matrices $\rho_A (q)$ of different blocks are normalised as ${\rm tr}\rho_A (q)=1$.

Now, to understand how the total entanglement arranges into contributions coming from the disjoint charge sectors, 
we first define the {\it symmetry resolved entanglement entropy } as
\begin{equation}
\label{eq:SvNEE}
S_1(q) \equiv -\mathrm{Tr} \rho_A(q) \log \rho_A(q).
\end{equation}
The total von Neumann entanglement entropy associated to $\rho_A$ in Eq. (\ref{eq:sum}) can be then written as 
\begin{equation}
\label{eq:SvN}
S_{1}=\displaystyle \sum_q p(q) S_{1}(q)- \displaystyle \sum_q p(q) \log p(q).
\end{equation}
Let us describe the physical meaning of the two sums in Eq. ($\ref{eq:SvN}$) \cite{fis,wv-03,xavier}. The first contribution 
is known as  configurational entanglement entropy and it depends on the entropy of each charge sector, weighted with its probability. The second contribution is the fluctuation entanglement entropy which is due, as the name says, to the fluctuations of the charge within the subsystem. 
The configurational entropy is related to the operationally accessible entanglement entropy of Refs. \cite{wv-03,bhd-18,bcd-19}. 

For future use, we also define the symmetry resolved R\'enyi entropies as
\begin{equation}
\label{eq:RSREE}
S_{n}(q) \equiv \dfrac{1}{1-n}\log \mathrm{Tr} \rho^n_A(q) .
\end{equation}
In order to compute these quantities, following the approach of Ref. \cite{goldstein} we first define the normalised charged moments of $\rho_A$ as
\begin{equation}
\label{eq:firstdef}
Z_n(\alpha)\equiv\mathrm{Tr}\rho_A^ne^{iQ_A \alpha}.
\end{equation}
In a (1+1)-dimensional quantum field theory, this quantity is the partition function on a Riemann surface with the insertion of an Aharonov-Bohm flux $\alpha$, 
such that the field acquires a total phase $\alpha$ when moving on the entire worldsheet. 
Similar charged moments have been already considered in the context of free field theories \cite{CFH,d-16,ch-rev}, in holographic settings \cite{matsuura,cnn-16}, 
as well as in the study of entanglement in mixed states \cite{ssr-17,shapourian-19}.

The Fourier transforms of the charged moments are just the moments of the RDM restricted to the sector of fixed charge  \cite{goldstein}, i.e.
\begin{equation}
\label{eq:defF}
\mathcal{Z}_n(q)\equiv \mathrm{Tr} (\Pi_{q}\,\rho^n_A)=\displaystyle \int_{-\pi}^{\pi}\dfrac{d\alpha}{2\pi}e^{-iq\alpha}Z_n(\alpha).
\end{equation}
Hence the symmetry resolved entropies can be obtained as
\begin{equation}
\label{eq:SREE1}
S_n(q)=\dfrac{1}{1-n}\log \left[ \dfrac{\mathcal{Z}_n(q)}{\mathcal{Z}^n_1(q)}\right], \qquad S_{1}(q)=\lim_{n\rightarrow 1} S_n(q).
\end{equation}
Finally, also the probability $p(q)$ is simply related to the moments $\mathcal{Z}_n$ as
\begin{equation}
p(q)=\mathcal{Z}_1(q).
\end{equation}

\subsection{The corner transfer matrix and the entanglement entropy}

In dealing with the geometric bipartition considered in this paper (i.e. two semi-infinite half lines) the corner transfer matrix provides an exact form 
for the reduced density matrix \cite{peschel1} and hence it is a formidable tool for the derivation of the charged moments and symmetry resolved entropies. 
In order to understand how the CTM works, we give a brief review of the construction of the RDM. 

Generally, a direct computation of the density matrix of a system is tough. A trick to address this problem is to use the fact that the density matrix of the quantum chain 
is the partition function of a two-dimensional classical system on a strip \cite{it-87,nishino,nishino2}. 
The latter can be solved by means of the transfer matrix $T$ and we can identify the eigenstate $\ket{\Psi}$ of $T$ corresponding to its maximal eigenvalue. 
Given the Hamiltonian of the quantum chain $H$  and  its lattice spacing $a$, the transfer matrix is $T=e^{-aH}$ up to a prefactor; 
hence $\ket{\Psi}$ is the ground state of $H$. 
One then obtains the reduced density matrix of a subsystem $A$ of the chain by tracing over all the coordinates belonging to the complement of $A$.  
Therefore $\rho_A$ is the partition function of two half-infinite strips, one extending from $-\infty$ to $0$ and the other from $+\infty$ to $0$. 

The CTM plays a crucial role: it connects a horizontal row to a vertical one. 
Choosing the lattice in a clever way \cite{baxter}, when the model is isotropic, the four possible corner transfer matrices \cite{baxter} are all equivalent and the partition function
 is just $\mathrm{Tr}\hat{A}^4$, with $\hat A$ the CTM. Going back to our quantum problem, the reduced density matrix is \cite{peschel1}
\begin{equation}
\label{eq:red}
\rho_A=\dfrac{\hat{A}^4}{\mathrm{Tr}\hat{A}^4}.
\end{equation}
We will deal with integrable massive models satisfying the Yang-Baxter equations;
in this case, it is possible to show that Eq. (\ref{eq:red}) has an exponential form given by \cite{it-87,peschel1}
\begin{equation}
\label{eq:red1}
\rho_A=\dfrac{e^{-H_{\mathrm{CTM}}}}{\mathrm{Tr}e^{-H_{\mathrm{CTM}}}}.
\end{equation}
$H_{\mathrm{CTM}}$ is known as entanglement or modular Hamiltonian, that in the cases we are interested in can be diagonalised as \cite{peschel1}
\begin{equation}
\label{eq:diagonalform}
H_{\mathrm{CTM}}=\displaystyle \sum_{j=0}^{\infty} \epsilon_j n_j,
\end{equation} 
where $n_j$ are number operators and $\epsilon_j$ are the single-particle levels of the entanglement Hamiltonian. 
The result \eqref{eq:diagonalform} provides exact eigenvalues and degeneracies of the RDM (i.e. the entanglement spectrum of the system \cite{lh-08,cl-08}), 
from which one calculates straightforwardly the entanglement entropies \cite{cc-04}.

However, Eq. \eqref{eq:diagonalform} contains no information about the distributions of the eigenvalues $\epsilon_j$ into the various symmetry sectors
(indeed, it has exactly the same form for models with discrete and continuous symmetries).
In order to use it to compute the symmetry resolved entropies in gapped integrable models, we should complement Eq. \eqref{eq:diagonalform} with 
some other input providing the symmetry resolution, but this should be done on a case by case basis. 
The rest of the manuscript is devoted exactly to solve this problem for two specific 1D integrable lattice models: 
the complex harmonic chain and the non-critical XXZ chain in which we will exploit the results of Refs. \cite{Gaussian,Peschel} and \cite{albaES,ahl-12} respectively.

\section{The complex harmonic chain}\label{sec:HC}
In this section we use the CTM to derive the symmetry resolved entanglement entropies for a double or complex harmonic chain that is $U(1)$ 
symmetric and its continuum limit is a non-compact massive {\it complex} boson, i.e. a Klein-Gordon field theory. 
We will find an analytic expression for the charged moments as functions of $\alpha$ and we will discuss its limit close to the conformal invariant critical point, 
when the correlation length $\xi$ is finite but large. 
Then we will use this result to compute the symmetry resolved entropies.
All the analytical results will be compared against exact numerical computations based on correlation matrix techniques \cite{p-03,pe-09,p-12}.

\subsection{Brief recap of the free complex scalar field and its lattice discretisation}

The meaning of the symmetry of a double harmonic chain is clearer in the field theory language and so we first 
consider a free complex scalar field $\phi(x)$ described by the Euclidean action
\begin{equation}
S=\int d^2 x \left[\partial_\mu \phi^\dag(x)\partial_\mu \phi(x) +m^2\phi^\dag(x) \phi(x)\right].
\end{equation} 
This action is invariant under $U(1)$, i.e. the field $\phi$ can be rotated of an arbitrary phase $\phi(x)\to e^{i\theta}\phi(x)$ leaving the action unchanged. 
The Hamiltonian of this field theory is 
\begin{equation}
H=\int d x \left[\Pi^\dag(x)\Pi(x)+\partial_x \phi^\dag(x)\partial_x \phi(x) +m^2\phi^\dag(x) \phi(x)\right],
\label{hc}
\end{equation} 
with $\Pi(x)$ being the field conjugated to $\phi(x)$.

We can as well rewrite the model in terms of two scalar real fields $\phi_x(x)$ and $\phi_y(x)$
\begin{equation}
\phi(x)=\frac1{\sqrt2} (\phi_x(x)+ i\phi_y(x)),
\end{equation}
and the same for $\Pi(x)$.  In these variables the $U(1)$ symmetry is an $O(2)$ rotation in the plane $(\phi_x,\phi_y)$. 
The Hamiltonian \eqref{hc} in terms of these variables is 
\begin{multline}
H=\frac12\int d x \left[\Pi_x^2(x)+(\partial_x \phi_x(x))^2 +m^2\phi_x^2(x)\right]+\frac12\int d x \left[\Pi_y^2(x)+(\partial_x \phi_y(x))^2 +m^2\phi_y^2(x)\right] \\=
H_{\mathbb R}(\phi_x) +H_{\mathbb R}(\phi_y),
\label{h2r}
\end{multline} 
where in the second line we stressed that it is a sum of two identical Hamiltonians $H_{\mathbb R}$ for the real fields $\phi_x$ and $\phi_y$.
One introduces the modes $a^\dag_i(p)$ and $a_i(p)$ for each field $i=x,y$ and momentum $p$. 
The Hamiltonian and the conserved charge are instead better written in terms of particles and antiparticles modes operators
\begin{equation}
a(p)=\frac1{\sqrt2} (a_x(p) +i a_y(p)), \qquad b(p)=\frac1{\sqrt2} (a_x^\dag(p) +i a_y^\dag(p)).
\label{ab:op}
\end{equation}
The Hamiltonian is 
\begin{equation}
H=\int \frac{d p}{2\pi} \epsilon(p) (a^\dag(p) a(p) +b^\dag(p) b(p) ),
\label{Hpap}
\end{equation}
(with $\epsilon^2(p)=m^2+p^2$) 
while the conserved charge is
\begin{equation}
Q=\int \frac{d p}{2\pi}   (a^\dag(p) a(p) -b^\dag(p) b(p) ),
\end{equation}
i.e. the total number of particles {\it minus} the number of antiparticles. 
The conserved charge can be as well written in real space and its value in a given subsystem $A$ is the same integral restricted to $A$, i.e.
\begin{equation}
Q_A=\int_A d x  (a^\dag(x) a(x) -b^\dag(x) b(x) ).
\label{QAft}
\end{equation}

For the construction of the RDM for the lattice version of the complex Klein-Gordon field theory, we start from discretising each of the two real Hamiltonians in 
Eq. \eqref{h2r}. 
The lattice discretisation of each of them is the harmonic chain, i.e. a chain of $L$ harmonic oscillators of mass $M=1$ with equal frequency $\omega_0$, 
coupled together by springs with elastic constant $k$ (hereafter we set $\omega_0=1-k$), i.e. the lattice discretisation of the Hamiltonian $H_{\mathbb R}$ is
\begin{equation}
\label{eq:Haminiziale}
H_{\mathrm{HC}}(q)=\sum_{i=1}^L \Big(\dfrac{p_i^2}{2}+\dfrac{\omega_0^2 q_i^2}{2}\Big)+\sum_{i=1}^{L-1}\dfrac{1}{2}k(q_{i+1}-q_i)^2,
\end{equation}
where variables $p_i$ and $q_i$ satisfy standard bosonic commutation relations $[q_i,q_j] = [p_i,p_j] = 0$ and $[q_i,p_j] = i\delta_{ij}$.
Hence, the lattice version of the complex field theory is the sum of two of the above harmonic chains in the variables $q_x$ and $q_y$, i.e.
\begin{equation}
H_{CHC} (q_x+iq_y) = H_{\mathrm{HC}}(q_x)+ H_{\mathrm{HC}}(q_y).
\end{equation}
which we call complex or double harmonic chain. 

The  reduced density matrix, $\rho_A$, for half of the real harmonic chain was explicitly constructed by Peschel and Chung in \cite{Peschel} in the large $L$ limit. 
The trick is to relate $\rho_A$ to
the partition function of a two-dimensional massive Gaussian model in the geometry of an infinite strip of width $L$ with a cut perpendicular to it \cite{Gaussian}.
Due to the integrability of the Gaussian model, in the case where $L$
is much larger than the correlation length, the $H_{\mathrm{CTM}}$ for the harmonic chain may be written in a diagonal form as in Eq. \eqref{eq:diagonalform}, 
where now we explicitly have
\begin{equation}\label{eq:CTMsf}
H_{\mathrm{CTM}}=\displaystyle \sum_{j=0}^{\infty} (2j+1)\epsilon\, \beta_j^{\dagger}\beta_j, \qquad \epsilon=\dfrac{\pi I(\sqrt{1-k^2})}{I(k)}.
\end{equation}
Here $I(k)$ is the complete elliptic integral of the first kind, i.e.
\begin{equation}
\label{eq:elliptic}
I(k)=\displaystyle \int_0^{\pi/2} \dfrac{d\theta}{\sqrt{1-k^2\sin^2 \theta}},
\end{equation}
and $\beta_j, \beta_j^{\dagger}$ are bosonic annihilation and creation operators (satisfying $[\beta_i,\beta_j^\dag]=\delta_{i,j}$).
They are related to the ladder operators $a_i$ of the original chain by a generalised Bogoliubov transformation \cite{Peschel} as
\begin{equation}
\beta_j=\sum_{i\in A} g_{ji}a_i +h_{ji} a^\dag_i.  
\label{gen:bog}
\end{equation}
Notice that the transformation mixes $a$ and $a^\dag$ so it does not conserve the number operator. 

The RDM for the double chain is clearly factorised in $x$ and $y$ part, i.e. the entanglement Hamiltonian is the sum of two $H_{CTM}$ in 
Eq. \eqref{eq:CTMsf} one with $\beta_{x,i}$ and one with $\beta_{y,i}$ ladder operators. 
Now we proceed as follows. First we rewrite these two entanglement Hamiltonians in terms of the local ladder operators $a_{x,i}$ and $a_{y,i}$ using the inverse 
of the Bogoliubov transformation \eqref{gen:bog}. 
Then, using the lattice analogue of \eqref{ab:op}, i.e. 
 \begin{equation}
\label{eq:abi}
\begin{split}
a_{x,i}=\frac{1}{\sqrt{2}}(a_i+b_i), \qquad &a^{\dagger}_{x,i}=\frac{1}{\sqrt{2}}(a_i^{\dagger}+b_i^{\dagger}), \\
a_{y,i}=\frac{1}{\sqrt{2}i}(a_i-b_i), \qquad &a_{y,i}^{\dagger}=\frac{1}{\sqrt{2}i}(b_i^{\dagger}-a_i^{\dagger}) .
\end{split}
\end{equation}
we rewrite the entanglement hamiltonian in terms of local ladder operators for particles and antiparticles. 
This is clearly quadratic (it is the rewriting of a quadratic operator after two linear transformations and so it is quadratic) 
and commute with the charge operator. 
Hence, via another Bogoliubov transformation (see Appendix \ref{app:osc})
\begin{equation}
\label{eq:bogi}
\alpha_i=\sum_{j\in A} g_{ij} a_j+h_{ij} b_j^{\dagger},  \qquad 
\gamma_i^\dag=\sum_{j\in A} h^*_{ij} a_j^\dag+g^*_{ij} b_j,  
\end{equation}
which conserve the charge, the entire entanglement Hamiltonian of half-chain is brought into the form  
\begin{equation}
\label{eq:gen}
\mathcal{H}_A=\displaystyle \sum_{j=0}^{\infty} \epsilon (2j+1)  (\alpha_j^{\dagger} \alpha_j + \gamma_j^{\dagger}\gamma_j),
\end{equation}

The charge operator restricted to the semi-infinite line is just the discretisation of Eq. \eqref{QAft}, i.e.  
\begin{equation}
\label{eq:chargeanti}
Q_A=\displaystyle \sum_{j\in A} a^{\dagger}_ja_j-b^{\dagger}_jb_j.
\end{equation}
Once we apply the Bogoliubov transformation in Eq. (\ref{eq:bogi}), we have
\begin{equation}
\label{eq:chargeanti1}
Q_A=\displaystyle \sum_{j=0}^{\infty} \alpha^{\dagger}_j\alpha_j-\gamma^{\dagger}_j\gamma_j,
\end{equation}
up to an unimportant additive constant that we neglect. 

Since the $\alpha_i$ and $\gamma_i$ operators in Eq. \eqref{eq:gen} commute, the RDM factorises as 
\begin{equation}
\rho_A= \rho_A^{\alpha}\otimes \rho_A^{\gamma},
\end{equation}
where we denoted the RDM for $\alpha_i$ and $\gamma_i$ with $\rho_A^{\alpha}$ and $\rho_A^{\gamma}$ respectively.
For the charged moments, we need to compute ${\rm Tr}\rho_A^n e^{iQ_A\alpha}$, but using also that $Q_A$ is the difference of the number of 
$\alpha_i$'s and $\gamma_i$'s, see Eq. \eqref{eq:chargeanti1},
the trace factorises as 
\begin{equation}
Z_n(\alpha)={\rm Tr}\rho_A^n e^{iQ_A\alpha}= {\rm Tr}[(\rho_A^{\alpha})^n e^{iN_A^{\alpha}\alpha}]\times [{\rm Tr}(\rho_A^{\gamma})^n e^{-iN_A^{\gamma}\alpha}].
\end{equation}
where $N_A^{\alpha}=\sum_{j\in A} \alpha_j^{\dagger}\alpha_j$ and 
$N_A^{\gamma}=\sum_{j\in A} \gamma_j^{\dagger}\gamma_j$.  
The two factors are equal, except for the sign of $\alpha$. 
It is very instructive to see how this factorisation happens for a chain of two oscillators as we report in Appendix \ref{app:osc}.

If for a single harmonic chain, we introduce the quantity 
\begin{equation}
F_n(\alpha)=\log [{\rm Tr}\rho_A^n e^{iN_A\alpha}],
\end{equation}
then we have that the charged moments of the complex boson are given by 
\begin{equation}
\log Z_n(\alpha) =F_n(\alpha)+F_n(-\alpha).
\label{ZFF}
\end{equation}
We stress that $F_n(\alpha)$ is not the log of a local charged moment because in the single harmonic  chain there is no local $U(1)$ symmetry.

In the following we show how to compute $F_n(\alpha)$ by CTM methods for a single harmonic chain and after we use \eqref{ZFF} to get the charged moments.

\subsection{Charged moments from CTM}

\begin{figure}
\centering
\subfigure
  {\includegraphics[width=0.44\textwidth]{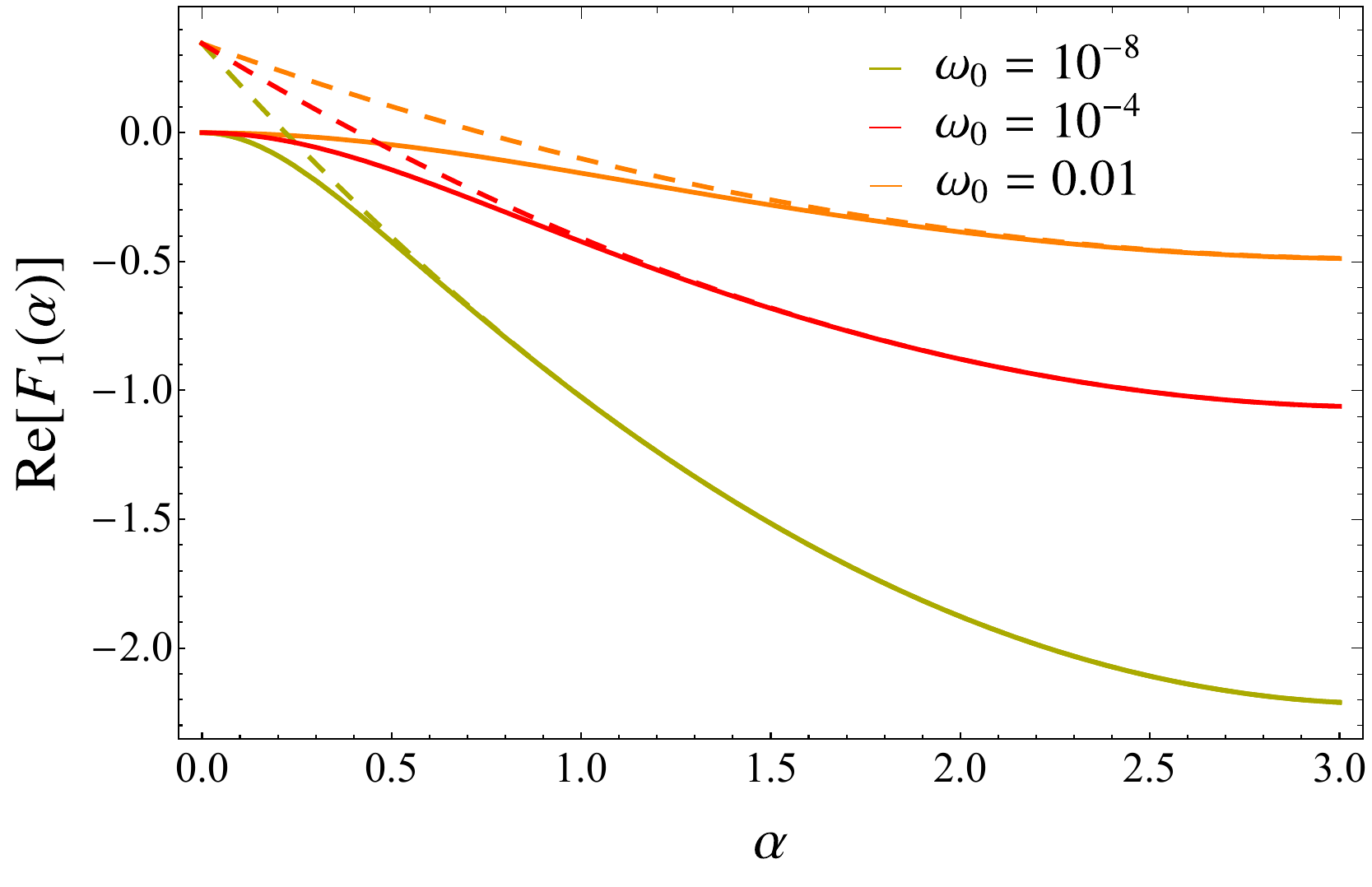}}
\subfigure
   {\includegraphics[width=0.44\textwidth]{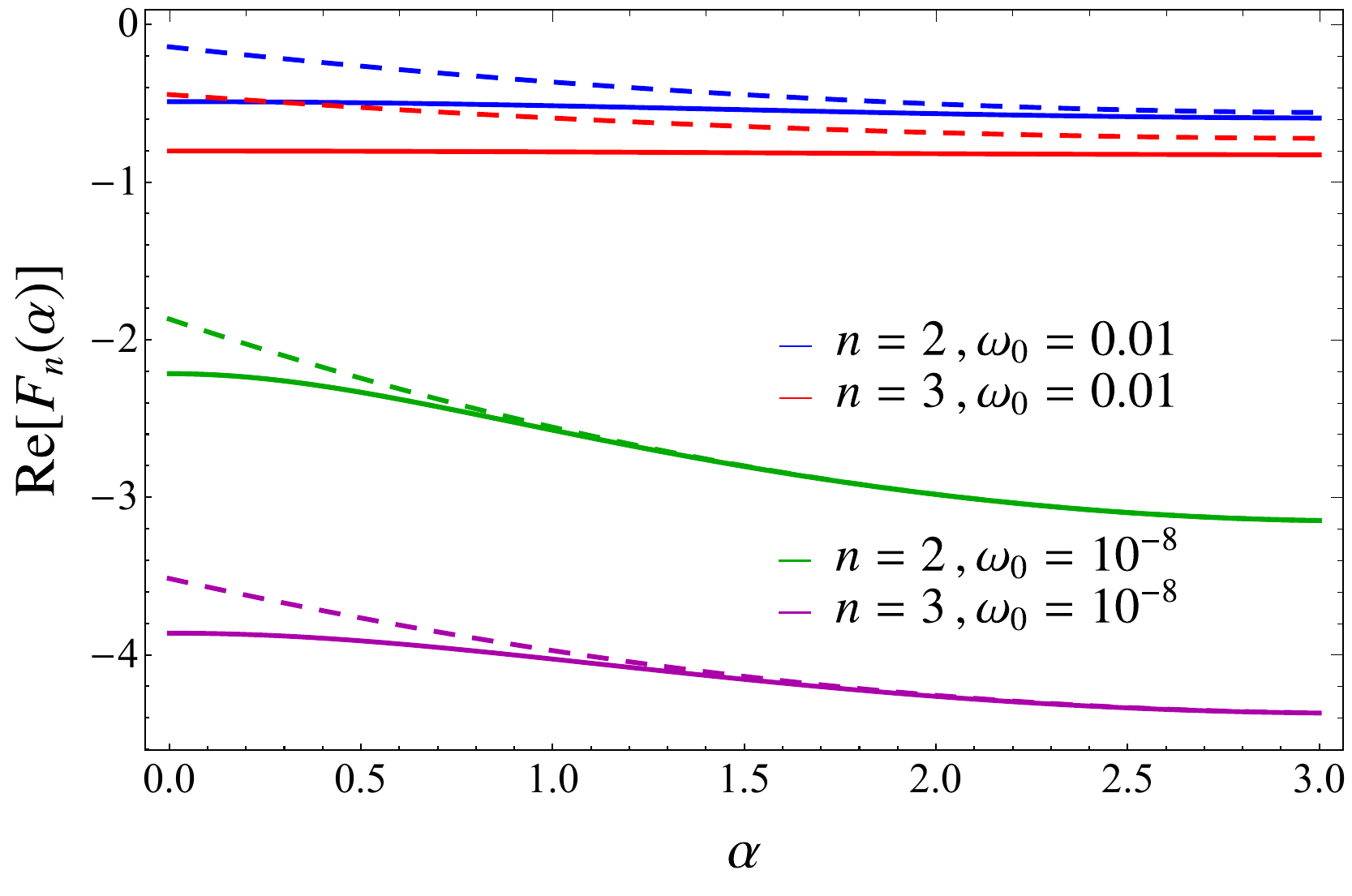}}
    \subfigure
  {\includegraphics[width=0.44\textwidth]{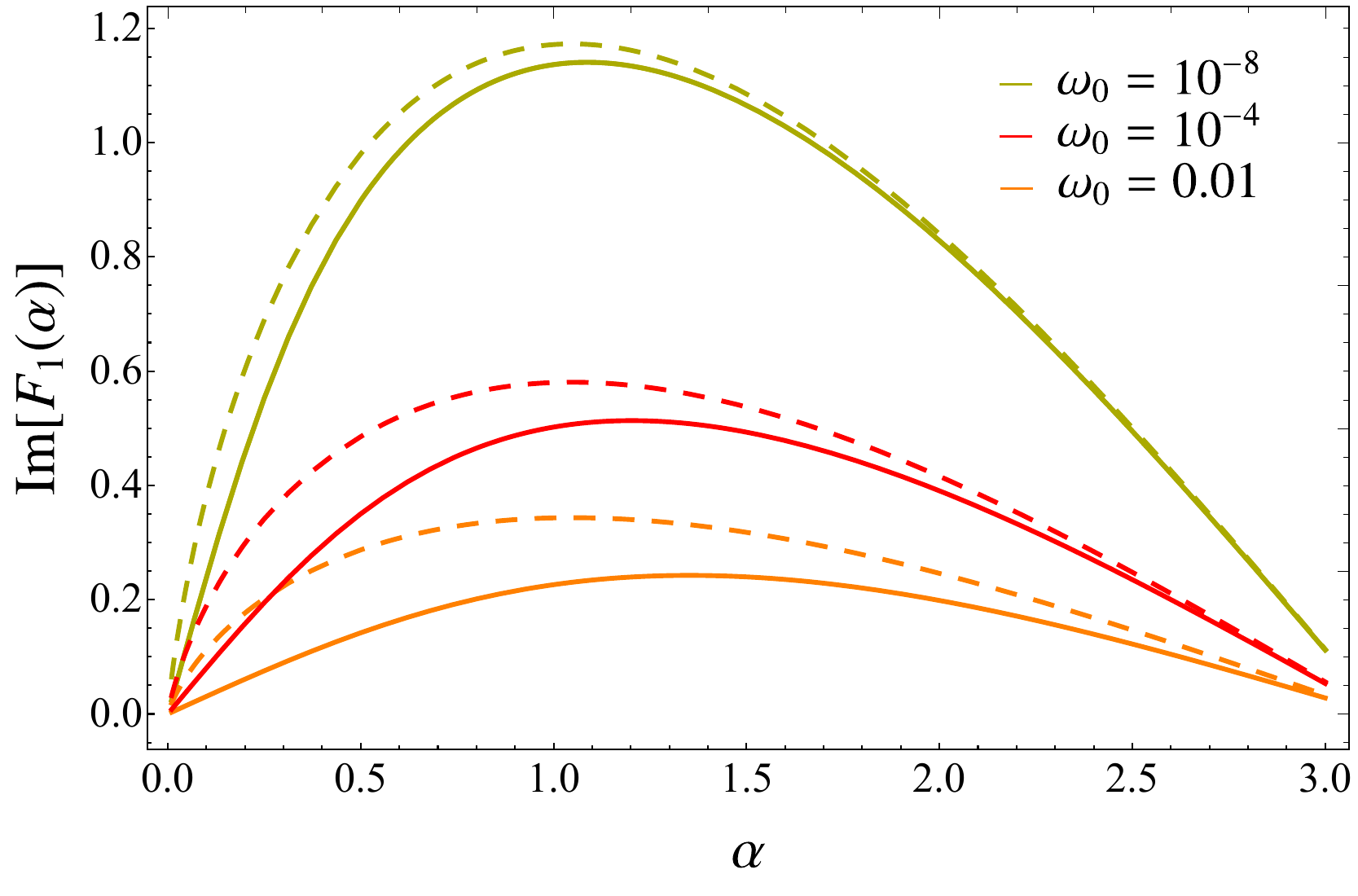}}
  \subfigure
  {\includegraphics[width=0.44\textwidth]{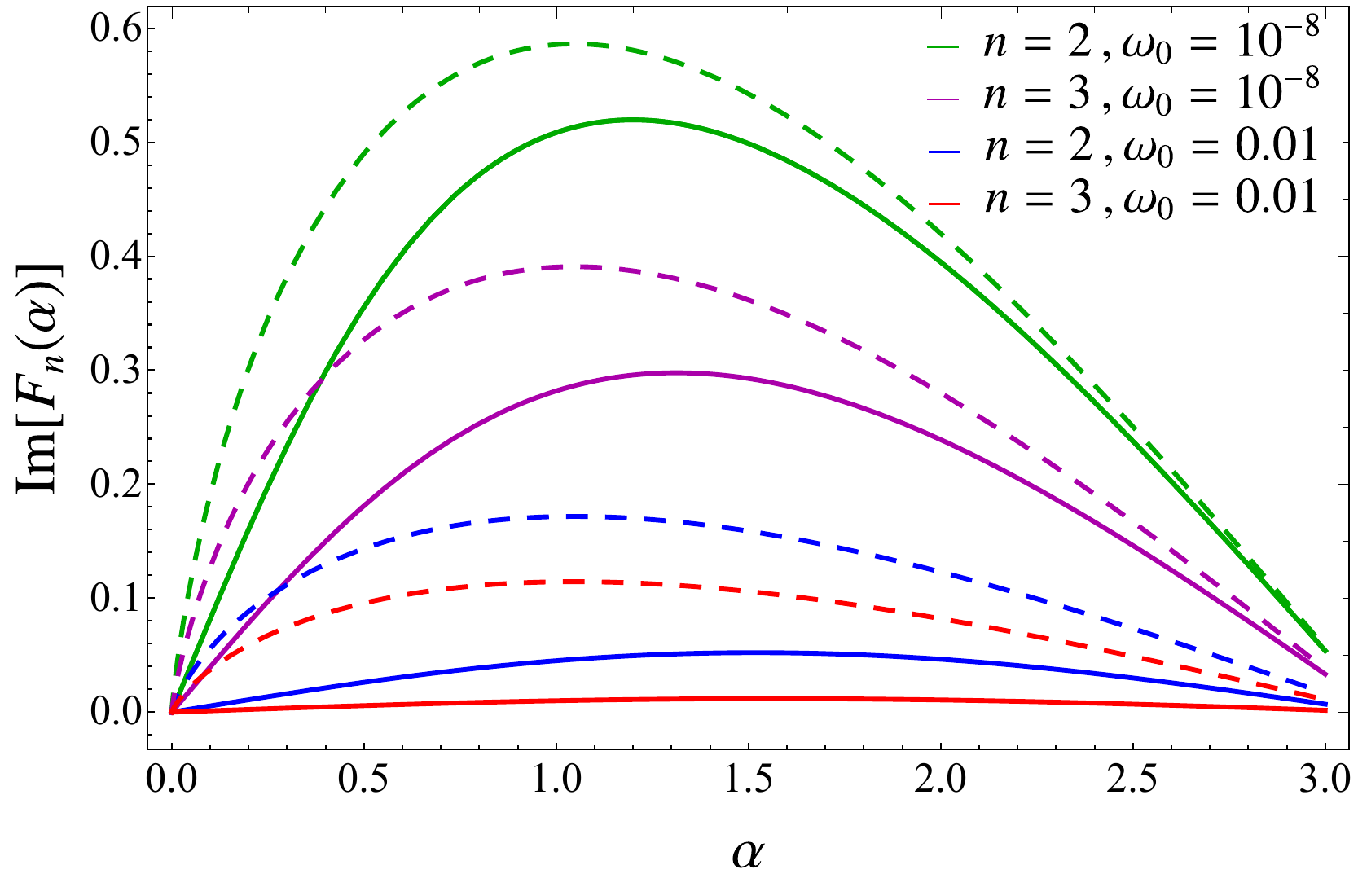}}
\caption{Charged moments for the harmonic chain: we report the real (top) and the imaginary (bottom) part of $F_n(\alpha)$, 
Eq. \eqref{eq:first}, as function of $\alpha$ for different values of $\omega_0$. 
Everywhere, the dashed lines are the asymptotic expansions for $\epsilon\to0$ and $\alpha\neq0$ up to ${\cal O}(\epsilon)$, cf. Eq. \eqref{eq:leadinghc1}. 
As discussed in the text, the convergence to the critical result is not uniform and it is slower for smaller $\alpha\neq0$.
The function $\log Z_n(\alpha)$ for the complex chain is twice the real part of $F_n(\alpha)$.
}\label{fig:pannel6}
\end{figure}

Here we first compute the quantity $F_n(\alpha)$ for a real harmonic chain and from this $Z_n(\alpha)$ is simply derived from Eq. \eqref{ZFF}. 
In the above subsection,  $N_A$ and $\rho_A$ for the single chain have been already written in the same basis and the derivation of $F_n(\alpha)$ amounts 
to compute the trace 
\begin{equation}
\label{eq:steps}
{e^{F_n(\alpha)}}
=\dfrac{\mathrm{Tr}e^{-\sum_{j=0}^{\infty}(\epsilon_j n  -i\alpha)n_j}}{\left( \mathrm{Tr} e^{-\sum_{j=0}^{\infty}\epsilon_j  n_j}\right)^n}=\dfrac{\displaystyle \prod_{j=0}^{\infty} \sum_{k=0}^{\infty}e^{-((2j+1)\epsilon n  -i\alpha)k}}{\left( \displaystyle \prod_{j=0}^{\infty} \sum_{k=0}^{\infty} e^{-(2j+1) \epsilon \,k}\right)^n}=
\dfrac{\displaystyle \prod_{j=0}^{\infty}(1-e^{-  (2j+1)\,\epsilon})^{n}}{\displaystyle \prod_{j=0}^{\infty}(1-e^{- (2j+1)\,\epsilon n+i\alpha})},
\end{equation}
whose logarithm is given by
\begin{equation}
\label{eq:first}
{F_n(\alpha)}=\sum_{j=0}^{\infty} n\log [1-e^{-(2j+1) \epsilon}]-\sum_{j=0}^{\infty}\log [1-e^{-(2j+1)\epsilon n+i\alpha}].
\end{equation}
This formula is exact and can be easily computed numerically, since it converges very quickly. 
It is plotted in Figure \ref{fig:pannel6} as a function of $\alpha $ for various values of $\omega_0$ and $n$, but we will discuss its properties later.

The charged moments for the complex harmonic chain, cf. Eq. \eqref{ZFF}, are
\begin{multline}
\label{eq:prodsv}
Z_n(\alpha)=e^{F_n(\alpha)}e^{F_n^*(\alpha)}  =\dfrac{\displaystyle  \prod_{j=0}^{\infty}(1-e^{-  (2j+1)\epsilon})^{2n}}{\displaystyle \prod_{j=0}^{\infty}(1-e^{- (2j+1)\epsilon n+i\alpha}) \displaystyle \prod_{j=0}^{\infty}(1-e^{- (2j+1)\epsilon n-i\alpha})}=\\
=Z_n \dfrac{\theta_4(0|e^{-\epsilon n})}{\theta_4(\frac{\alpha}{2}|e^{-\epsilon n})},
\end{multline}
where in the last equality we factor out the total partition sum 
\begin{equation} 
Z_n\equiv Z_n(\alpha=0)=  \prod_{j=0}^{\infty}\frac{(1-e^{-  (2j+1)\epsilon})^{2n}}{(1-e^{-  (2j+1)\epsilon n})^2},
\label{Zntot}
\end{equation}
and use the definition \eqref{Theta3Def} for $\theta_4(u|q)$.
Notice that the entire $\alpha$ dependence is encoded in the denominator of Eq. \eqref{eq:prodsv} and that $Z_1=1$, but $Z_1(\alpha)\neq1$.
Also the total R\'enyi entropies of the complex harmonic chains are
\begin{equation} 
S_n=\frac1{1-n}\log  Z_n=  \frac2{1-n} \sum_{j=0}^{\infty}[n \log{(1-e^{-  (2j+1)\epsilon})}-\log {(1-e^{-  (2j+1)\epsilon n})}],
\label{Sntot}
\end{equation}
i.e. the double of a real harmonic chain.

\subsubsection{Poisson resummation and critical regime.} 
A drawback of the form \eqref{eq:first} is that it does not directly allow a direct expansion in the critical regime, i. e. for small $\epsilon$. 
Moreover, we cannot perform an Euler Mac-Laurin summation (as for $\alpha=0$, see \cite{cc-04}) since the function $f(x)=\log (1-e^{-2x})$ diverges for $x\to 0$. 
However, following Ref. \cite{cal2010}, we can obtain the asymptotic expansion for small $\epsilon$ by using the (generalised) Poisson resummation formula:
\begin{equation}
\label{eq:poisson}
\displaystyle \sum_{j=-\infty}^{\infty} f(|\epsilon(b j+a) |)=\dfrac{2}{\epsilon b}\displaystyle \sum_{k=-\infty}^{\infty}\hat{f} \left( \dfrac{2\pi k}{\epsilon b}\right)e^{2\pi i k a/b},
\end{equation}
where  
\begin{equation}
\label{eq:cosineFourier}
\hat{f}(y)=\displaystyle \int_0^{\infty}f(x) \cos (yx) dx.
\end{equation}
In order to use this resummation formula for Eq. (\ref{eq:first}), we must choose $a=1/2$, $b=1$ and
\begin{equation}
f_{n,\alpha}(x)=-\log(1-e^{-2nx+i\alpha}),
\end{equation}
which allows us to rewrite the sum (\ref{eq:first}) as
\begin{equation}
\label{eq:second}
\begin{split}
 F_n(\alpha)&=\displaystyle \sum_{j=0}^{\infty} (n f_{1,\alpha=0}(\epsilon(j+1/2))-f_{n,\alpha}(\epsilon(j+1/2))) \\
&=\dfrac{1}{2} \sum_{j=-\infty}^{\infty} (n f_{1,\alpha=0}|(\epsilon(j+1/2)|)-f_{n,\alpha}(|\epsilon(j+1/2))|). 
\end{split}
\end{equation}
The cosine-Fourier transform of $f_{n,\alpha}(x)$ is 
\begin{equation}
\label{eq:FCT}
\hat{f}_{n,\alpha}(y)=\dfrac{ie^{i\alpha}}{2y}\left[\Phi(e^{i\alpha},1,1-\frac{iy}{2n})-\Phi(e^{i\alpha},1,1+\frac{iy}{2n})\right],
\end{equation}
where $\Phi$ is the Lerch transcendent function, defined as
\begin{equation}
\label{eq:lerch}
\Phi(z,s,a)=\displaystyle \sum_{j=0}^{\infty}\dfrac{z^j}{(j+a)^s}.
\end{equation}
If $\alpha=0$ and $n=1$, Eq. (\ref{eq:FCT}) simplifies to the known value \cite{cal2010}
\begin{equation}
\hat{f}_{1,0}(y)=\dfrac{1}{y^2}-\dfrac{\pi}{2y} \coth \left( \dfrac{\pi y}{2}\right).
\end{equation}
Plugging Eq. \eqref{eq:FCT} into the Poisson resummation formula, we rewrite $\log Z_{n}(\alpha)$ in such a way to 
isolate the contribution of the term $k=0$ which gives the leading divergence in the limit $\epsilon \rightarrow 0$, i.e.
\begin{equation}
\label{eq:leadinghc}
\begin{split}
 F_n(\alpha)=&\dfrac{\mathrm{Li}_2(e^{i\alpha})}{2\epsilon n}-\dfrac{n\pi^2}{12 \epsilon}+\displaystyle \sum_{k=1}^{\infty} \left[(-1)^k\dfrac{n \epsilon}{2\pi^2 k^2}+(-1)^{k+1}\dfrac{n}{2k} \coth \frac{\pi^2 k}{\epsilon}   \right]+ \\
&\dfrac{ie^{i\alpha}}{2\pi}\displaystyle \sum_{k=1}^{\infty} \dfrac{(-1)^k}{k}\left[\Phi(e^{i\alpha},1,1-\frac{i\pi k}{\epsilon n})-\Phi(e^{i\alpha},1,1+\frac{i \pi k}{\epsilon n}) \right].
\end{split}
\end{equation}
Here we have introduced the polylogarithm of order $2$
\begin{equation}
\label{eq:polylog}
\mathrm{Li}_2(z)=\displaystyle \sum_{m=1}^{\infty}\dfrac{z^m}{m^2}.
\end{equation}

We are now interested in the critical region of the parameter space in which the correlation length $\xi$ (inverse gap) is large but finite.  
In the critical regime $\xi \gg 1$ (or equivalently $\epsilon\ll1$), the correlation length of the model behaves like 
\begin{equation}
\label{eq:unscaling}
\log \xi \simeq\dfrac{\pi^2}{\epsilon}+\mathcal{O}(\epsilon^0).
\end{equation}
Using the results of Ref. \cite{ferreira}, the last sum over $k$ in Eq. (\ref{eq:leadinghc}) in the limit $\epsilon\to0$ behaves like 
\begin{equation}\label{eq:divergent}
\dfrac{ie^{i\alpha}}{2\pi}\displaystyle \sum_{k=1}^{\infty} \dfrac{(-1)^k}{k}\left[\Phi(e^{i\alpha},1,1-\frac{i\pi k}{\epsilon n})-\Phi(e^{i\alpha},1,1+\frac{i \pi k}{\epsilon n}) \right] \to \dfrac{n\epsilon}{12}\dfrac{e^{i\alpha}}{1-e^{i\alpha}}.
\end{equation}
and hence the only non-vanishing terms in the asymptotic expansion close to $\epsilon=0$ are 
\begin{equation}
\label{eq:leadinghc1}
{ F_n(\alpha)}=\dfrac{\mathrm{Li}_2(e^{i\alpha})}{2\epsilon n}-\dfrac{n\pi^2}{12 \epsilon}+\dfrac{n}{2}\log2+	\mathcal{O}(\epsilon),
\end{equation}
whose real part is 
\begin{equation}
\label{eq:leadinghc3}
\mathrm{Re}[{F_n(\alpha)}]=\left[ \dfrac{1}{2n}\left( \dfrac{\alpha}{2\pi}\right)^2- \dfrac{|\alpha|}{4\pi n}+\dfrac{1}{12 n}-\dfrac{n}{12 } \right] \log \xi+\dfrac{n}{2}\log2+	\mathcal{O}(\epsilon),
\end{equation}
because
\begin{equation}
\dfrac{\mathrm{Re}[\mathrm{Li}_2(e^{i\alpha})]}{n}=\dfrac{1}{n}\left( \dfrac{\alpha}{2}\right)^2- \dfrac{\pi |\alpha|}{2 n}+\dfrac{\pi^2}{6n}. 
\end{equation}

The charged moments for the complex harmonic chain are now given by Eq. \eqref{ZFF}, i.e.
$\log Z_n(\alpha)=F_n(\alpha)+ F_n(-\alpha)$
and, in the limit $\epsilon\to 0$,
\begin{equation}
\log Z_n(\alpha)= \left[ \dfrac{1}{n}\left( \dfrac{\alpha}{2\pi}\right)^2- \dfrac{|\alpha|}{2\pi n}+\dfrac{1}{6 n}-\dfrac{n}{6} \right] \log \xi+n\log 2+\mathcal{O}(\epsilon).
\label{Znvero}
\end{equation}
Notice that while $F_n(\alpha)$ is generically complex, $\log Z_n(\alpha)$ for the complex chain is real and even in $\alpha$. 

\subsubsection{Discussions.}
We concluded our exact computation of the charged moments and we are now ready to critically discuss our findings. 
Eq. \eqref{Znvero} is very suggestive. 
It tells us that the leading term in the ``charged entropies'' diverges logarithmically with $\xi$ but with a non-standard prefactor. 
Indeed, in the conformal field theory of the compactified boson, it has been found that when $\alpha\neq 0$, the additional term in the 
logarithm is proportional to $\alpha^2$ \cite{goldstein} , while here we also have a linear contribution in $\alpha$.
Obviously the two results are not in contradiction, because the continuous limit of the harmonic chain is non-compact and the prefactor of $\alpha^2$ in 
Ref. \cite{goldstein} diverges when the compactification radius is sent to infinity.   
These results are very intriguing and it would be interesting to recover them directly in a field theory approach; work in this direction is in progress \cite{mdc-20}.   

Another interesting fact is that the limit $\alpha \to 0$ and the expansion for $\epsilon $ around $0$ do not commute, as a difference with other known cases  
(we believe that the origin of the non-commutativity is the non compact nature of the continuum limit).  
Indeed, if we consider first the limit $\alpha \rightarrow 0$ in Eq. (\ref{eq:leadinghc}), the last sum gives
\begin{equation}
\label{eq:leadingnoncommute}
\displaystyle \sum_{k=1}^{\infty}(-1)^k\left[ -\frac{\epsilon n }{2k^2\pi^2}+\dfrac{1}{2k} \coth \left( \frac{k\pi^2}{\epsilon n}\right) \right],
\end{equation}
leading to the known formula of the R\'enyi entropies of a  real harmonic chain, that in the critical regime $\epsilon \rightarrow 0$ is \cite{cc-04,cal2010} 
 (see Eq. \eqref{Sntot})
\begin{equation}
\label{eq:leadingsF}
S_n=\dfrac{\pi^2}{12\epsilon }\dfrac{1+n}{n}-\dfrac{\log 2}{2}+	\mathcal{O}(\epsilon).
\end{equation}
On the other hand, if we invert the order of these two operations, we obtain the divergent term in Eq. (\ref{eq:divergent}).
Considering now the charged moments of the complex chain, $\ln Z_n(\alpha)= 2{\rm Re} F_n(\alpha)$,  the divergent term (\ref{eq:divergent}) 
cancels, but the finite part is not the total moment $\ln Z_n$ in Eq. \eqref{Zntot}.
This fact implies that the approach of $\ln Z_n(\alpha)$ to the critical limit $\epsilon\to0$ is non-uniform in $\alpha$: 
exactly at $\alpha=0$ the charged entropy approaches \eqref{Sntot},
but for any non-zero $\alpha$ the limit is \eqref{eq:leadinghc1} that as a consequence is reached for smaller and smaller $\epsilon$ (i.e. $\omega_0$) as
$\alpha$ gets closer to $0$.

All these aspects are evident in Figure \ref{fig:pannel6} where we show (for $\alpha\geq 0$ since $F_n(-\alpha)=F_n^*(\alpha)$)
the exact result Eq. (\ref{eq:leadinghc}) (or equivalently \eqref{eq:first}) together with its critical limit, Eq. (\ref{eq:leadinghc1}). 
As we discussed above, the former converges to the latter as $\omega_0$, therefore $\epsilon$, decreases, but in a non-uniform way. 
Indeed, while for large $\alpha$ (i.e. close to $\pi$) the two curves are very close also when $\omega_0$ is not so small, 
for smaller and non-zero values of $\alpha$, we need much smaller $\omega_0$ to approach the critical limit. 
For $\alpha=0$ the limit is different.  
It is also clear that for higher values of $n$, the convergence is slower and starts at smaller values of $\omega_0$. 
The last observation is a well known fact for $\alpha=0$, cf. Ref. \cite{cal2010}, and it is not surprising that the effect is amplified in the presence of a flux.

\subsection{Symmetry resolved moments and entropies via Fourier trasform}
\label{Res moments FT HC}

The symmetry resolved moments $\mathcal{Z}_n(q)$ are obtained as Fourier transform of $Z_n(\alpha)$ in Eq. \eqref{eq:prodsv}, i.e. 
\begin{equation}
\mathcal{Z}_n(q)= \int_{-\pi}^{\pi}\dfrac{d\alpha}{2\pi}e^{-iq\alpha}Z_n(\alpha)=
Z_n \theta_4(0|e^{-\epsilon n})  \int_{-\pi}^{\pi}\dfrac{d\alpha}{2\pi}e^{-iq\alpha} \dfrac{1}{\theta_4(\frac{\alpha}{2}|e^{-\epsilon n})}.
\label{eq:firstFv}
\end{equation}
The integral in the rhs of the above equation can be found in Ref. \cite{book} (exercise 14 at page 489), obtaining 
\begin{equation}
\frac{\mathcal{Z}_n(q)}{Z_n}=
\prod_{k=1}^\infty \left(\frac{1-e^{-n \epsilon(2 k-1)}}{1-e^{-2 n \epsilon k}}\right)^2 e^{-n \epsilon |q|} \sum_{k=0}^\infty (-1)^k e^{-n \epsilon k^2} e^{-n \epsilon (2 |q| + 1) k}\,,
\label{Znq:fi}
\end{equation}
which is our final result for the symmetry resolved moments. 
It is likely that the sum in Eq. \eqref{Znq:fi} can be rewritten in terms of some special functions, but we did not find any particularly useful expression. 
We define the sum as 
\begin{equation}
\Phi_q(u)= u^{ |q|} \sum_{k=0}^\infty (-1)^k u^{k^2} u^{ (2 |q| + 1) k},
\label{Phi}
\end{equation}
which can be written in few different equivalent ways that are useful for investigating diverse properties:
\begin{equation}
\Phi_q(u) =
\sum_{k=0}^\infty (-1)^k u^{k^2 + k+|q|(2 k + 1)}=
u^{|q| - \frac14}  \sum_{k=0}^\infty (-1)^k  u^{(k + \frac12)^2}   u^{2 |q|  k}\,.
\label{Phi2}
\end{equation}
Clearly in terms of this function we have 
\begin{equation}
\mathcal{Z}_n(q)=
\prod_{k=1}^\infty \left(\frac{(1-e^{- \epsilon(2 k-1)})^n}{1-e^{-2 n \epsilon k}}\right)^2 \Phi_q(e^{-n \epsilon } ),
\label{Znq:fi2}
\end{equation}
where we used the explicit form of $Z_n$ in Eq. \eqref{Zntot}.


%

The symmetry resolved R\'enyi entropies are now easily deduced from Eq. (\ref{eq:SREE1}), obtaining
\begin{multline}\label{eq:resv}
S_n(q)=\frac{1}{1-n}\log \left[\dfrac{\mathcal{Z}_n(q)}{\mathcal{Z}_1(q)^n} \right]=\\ =
 \frac{2}{1-n}\sum_{k=1}^{\infty}\Big[n\log (1-e^{-2\epsilon  k})-\log (1-e^{-2n\epsilon  k}) \Big] 
+ \dfrac{1}{1-n}\log \dfrac{\Phi_q(e^{-n \epsilon } )}{(\Phi_q(e^{- \epsilon } ))^n} .
\end{multline}
Taking the limit $n\rightarrow 1$, we get the von Neumann entropy
\begin{equation}
\label{eq:res1v}
S_1(q)=-2 \sum_{j=1}^{\infty}\left[ \log (1-e^{-2\epsilon j})-\frac{2\epsilon j e^{-2\epsilon j}}{1-e^{-2\epsilon j}}\right]+
\log\Phi_q(e^{-n \epsilon })+ \epsilon e^{-\epsilon} \frac{\Phi'_q(e^{-n \epsilon })}{\Phi_q(e^{-n \epsilon })}.
\end{equation}

The critical limit $\epsilon\to0$ is easily understood if one focuses on the variation in $q$ of moments and entropies, rather than on their absolute values.
Indeed from Eq. \eqref{Znq:fi2}, it is easy to see that 
\begin{equation}
\label{eq:FcritV}
\frac{\mathcal{Z}_n(q)}{\mathcal{Z}_n(q=0)}= \frac{\Phi_q(e^{-n\epsilon})}{\Phi_0(e^{-n\epsilon})}\xrightarrow{\epsilon\to0} e^{-n^2 q^2 \epsilon^2/2}\,,
\end{equation}
where the last limit is performed by expanding to the second order in $\epsilon$ each term in the sum \eqref{Phi}, making carefully the sum in terms of 
$\zeta$ functions, and finally re-exponentiating the result. 
We stress that this critical limit is {\it not} the Fourier transform of the critical limit for $Z_n(\alpha)$ in Eq. \eqref{Znvero} because the two limiting procedures do not commute.  
The critical behaviour of the resolved entropies is then easily worked out as 
\begin{equation}
S_n(q)=\frac{1}{1-n}\log \frac{\mathcal{Z}_n(q)}{\mathcal{Z}_1^n(q)}=S_n(q=0)+ \frac{n \epsilon^2 q^2}2 +O(\epsilon^3),\\
\label{Sncr}
\end{equation}
which is valid also for $n=1$ without any particular limit. 
Also in the critical limit, it is worth to mention the behaviour 
\begin{equation}
S_n(q=0)=S_n -\log \frac{8\pi}{\epsilon}+\frac{\log n}{1-n}  +o(1)\,,
\end{equation}
which signals the presence of a subleading term proportional to $\log\epsilon\sim \log(\log \xi)$.
Such a term has not a unique interpretation and origin for the (complex) harmonic chain. 
Indeed, we know that the total entropy of a massive free non-compact boson has such subleading terms in $\log(\log \xi)$ \cite{casini} in the small mass limit, 
but even that double logarithmic terms appear generically in the symmetry resolution, also for the critical compact boson \cite{goldstein,riccarda}.

\begin{figure}
\centering
\subfigure
  {\includegraphics[width=0.45\textwidth]{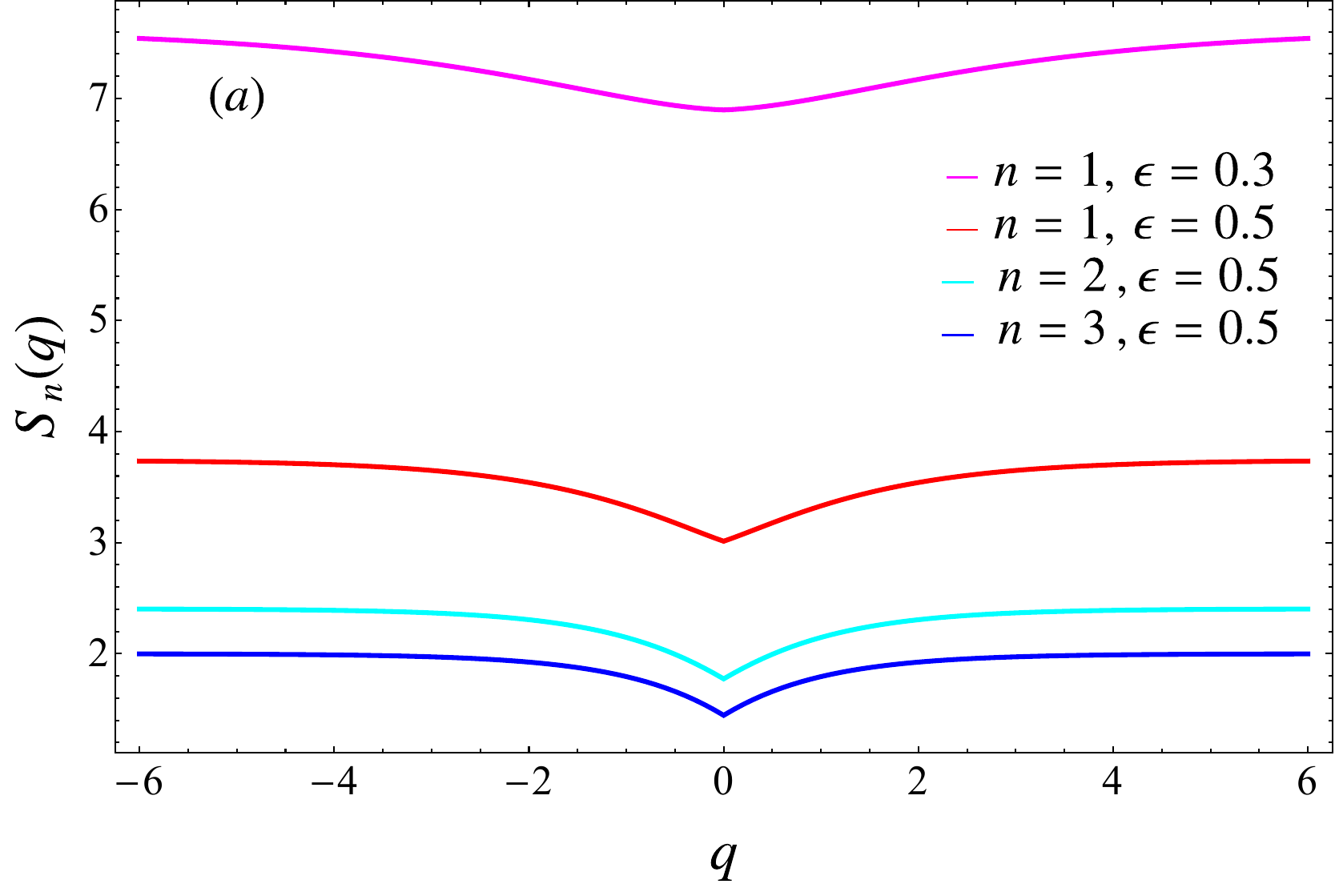}}
\subfigure
   {\includegraphics[width=0.45\textwidth]{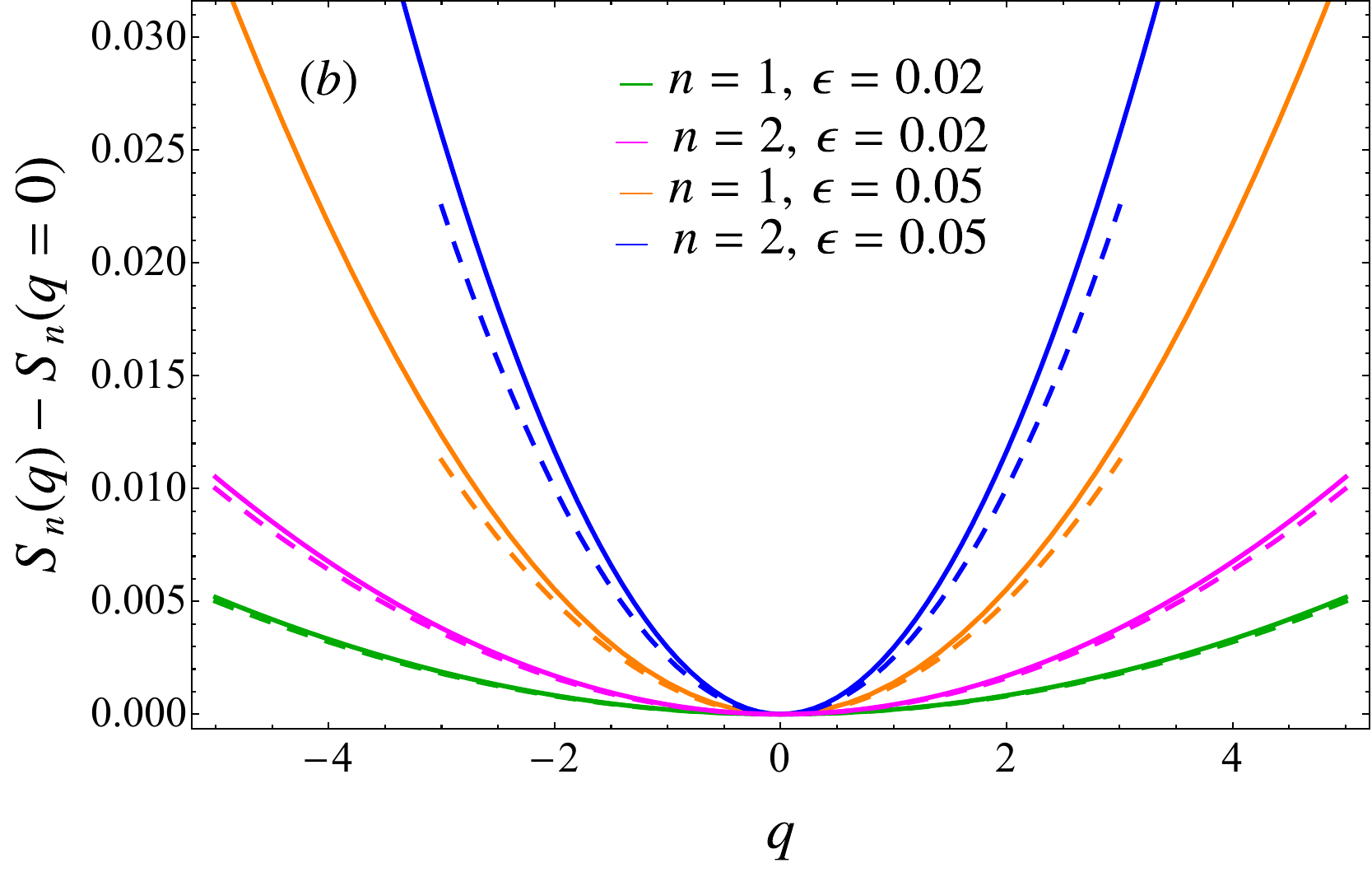}}
    \subfigure
   {\includegraphics[width=0.45\textwidth]{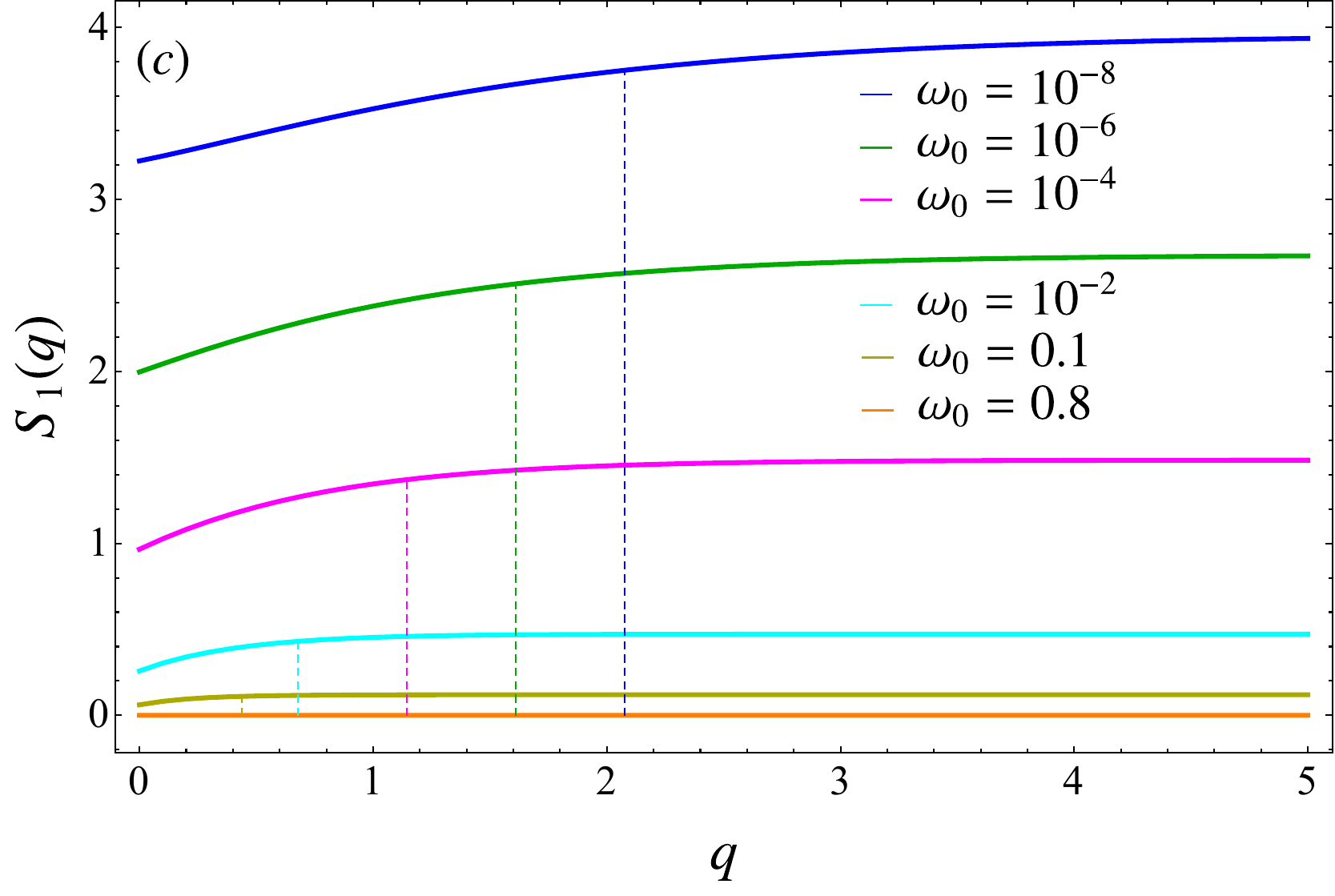}} 
   \subfigure
   {\includegraphics[width=0.45\textwidth]{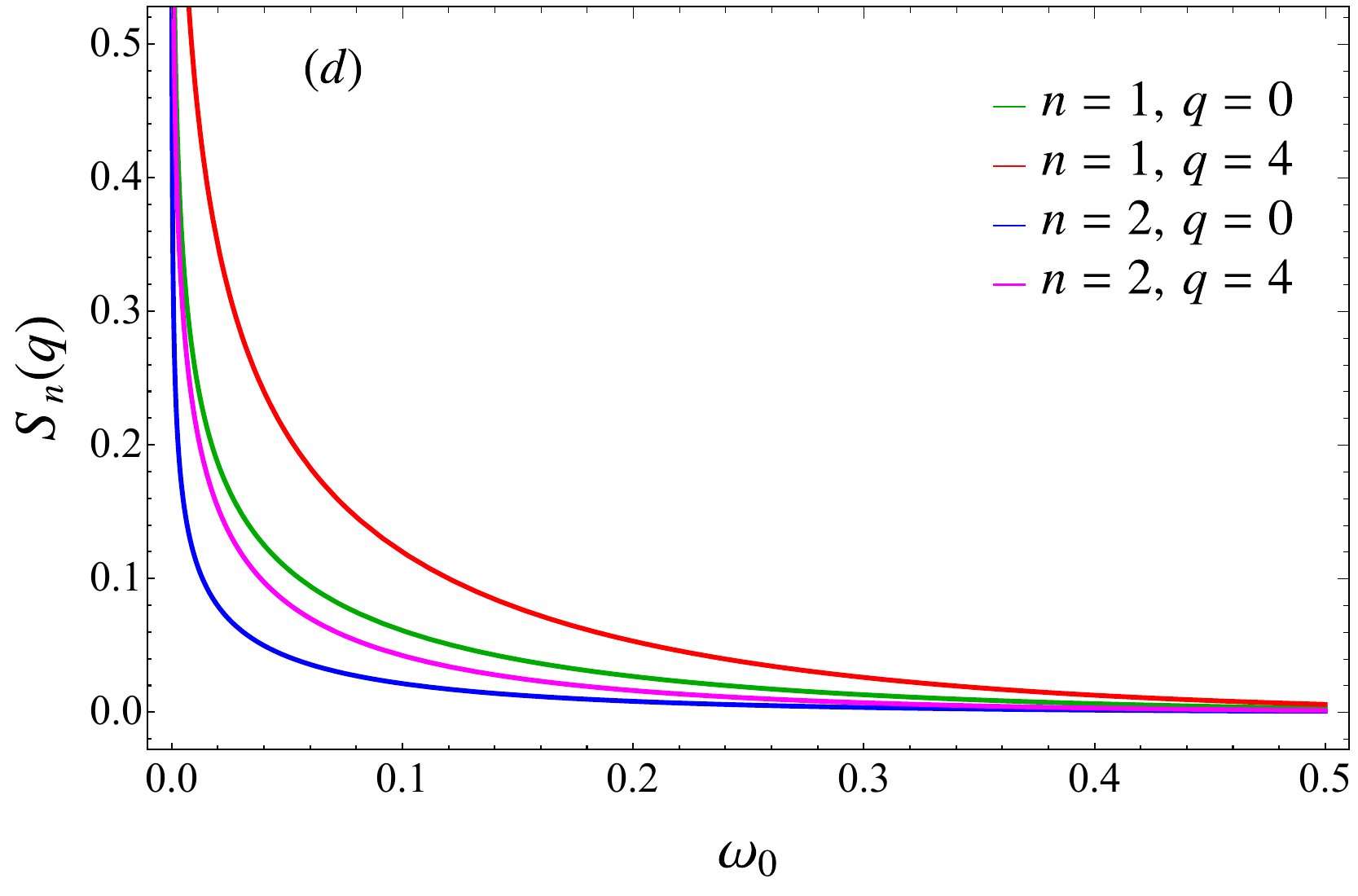}}
\caption{Symmetry resolved entanglement entropies for the  complex harmonic chain. 
Panel (a) shows $S_n(q)$ as functions of $q$ for different values of $\epsilon$ and $n$. 
Panel (b) reports $S_n(q)-S_n(q=0)$ for small values of $\epsilon$, showing the validity of the expansion in the critical regime \eqref{Sncr}. 
The critical limits in Eq. (\ref{Sncr}) are also reported as dashed lines  showing its accuracy for 
small $n\epsilon$. 
The panel (c)  shows the effective equipartition of entanglement for $q\gtrsim 1/\epsilon$
(these crossover values are reported as dashed vertical lines). 
The panel (d) shows $S_n(q)$ as function of $\omega_0$ for different values of $q$ and $n$. 
}\label{fig:pannel2v}
\end{figure}

Let us now critically discuss our results. 
First of all, there is a very important difference compared to the conformal gapless case \cite{goldstein}, i.e. the absence of equipartition of entanglement \cite{xavier}:
the R\'enyi entropies \eqref{eq:resv} depend explicitly on $q$. 
This dependence is explicitly reported in Figure \ref{fig:pannel2v} (a) where, in order to show its variation, we plot it as a continuous function of $q$, 
although only integer values are physical. 
The lack of exact equipartition is not surprising; also in critical models the leading terms for large $\ell$ show equipartition \cite{xavier}, 
while some subleading terms depend explicitly on $q$ \cite{goldstein,riccarda}.
In panel (b) of Figure \ref{fig:pannel2v} we focus on the critical limit of R\'enyi entropies  \eqref{Sncr} plotting $S_n(q)-S_n(q=0)$. 
As $\epsilon\to 0$, the result approaches the critical form \eqref{Sncr}, but clearly the convergence is not uniform: 
it is faster for smaller $q$ and $n$. 
Indeed, since this dependence is all encoded in the function $\Phi_q(e^{-n\epsilon})$, the  parameter that must be small is not $\epsilon$, but $n\epsilon$.
On the other hand, the higher order terms in $\epsilon$, that have been neglected in \eqref{Sncr}, become important for large $q$.
%

Another interesting feature of the symmetry resolved entropies for this complex harmonic chain is an {\it effective equipartition} in two limits. 
The first one is the limit of large $q$. 
Indeed, in Eq. \eqref{eq:resv} the entire $q$-dependence is encoded in the function $\Phi_q(e^{-n\epsilon})$.
Looking at Eq. \eqref{Phi2}, it should be clear that all the terms with $q n \epsilon \gg 1$ are exponentially suppressed.  
Practically, the total sum is more or less the same for all $q$ such that $n \epsilon q \gtrsim 1$ (from Eq. \eqref{eq:unscaling} this is equivalent to 
$nq \pi^2 \gtrsim  \log \xi$ in the critical region).
Hence, there is an {\it effective equipartition} among all $q\gtrsim 1/(n\epsilon)$. 
Actually, since the only physical values of $q$ are the integers, this fact implies that there is an almost exact equipartition (with the exception of $S_n(0)$)
of the entropy if $n\epsilon \gtrsim 1$, which corresponds to $\omega_0\gtrsim 10^{-4}$ (for $n=1$). 
In panel (c) we report the von Neumann entropies $S_1(q)$ for several values of $\omega_0$, showing that, as $q$ becomes large enough, the entropies $S_n(q)$ 
do not depend on $q$ anymore. We also explicitly report the (approximate) crossover values for $q\sim 1/\epsilon$ (as function of $\omega_0$ is given by Eq. \eqref{eq:CTMsf}), 
showing that it correctly captures the change of behaviour. 
Finally, we have effective equipartition also in the critical regime, but in this case also for small $q$. 
In fact, Eq. \eqref{Sncr} shows that the $q$-dependent term is proportional to $\epsilon^2$, while the leading term of $S_n(q)$ (say for $q=0$)
diverges as $\epsilon^{-1}$. Thus the $q$-dependence is suppressed as $\epsilon^3$ and there is an effective equipartition.
Even if for large $q$, the expansion \eqref{Sncr} breaks down, we do not expect that $S_n(q)-S_n(0)$ becomes of the order $S_n(0)$ and so there is 
an effective equipartition for all $q$: the numerical analysis of Eq. \eqref{eq:resv} seems to confirm this expectation.  
The functional form of the leading $q$-dependent term in Eq. \eqref{Sncr} is reminiscent of the one found for free fermions \cite{riccarda}.

\subsubsection{The total entanglement entropy as a consistency check.}
As a non-trivial consistency check of our results, we compute the total von Neumann entanglement entropy starting from the symmetry resolved ones 
using Eq. \eqref{eq:SvN}. 
The probability $p(q)$ is given by Eq. (\ref{Znq:fi2}) with $n=1$  and Eq. (\ref{eq:res1v}) provides the symmetry resolved entropies.
Plugging these two results  into Eq. \eqref{eq:SvN} leads to
\begin{multline}
S_1
=
-2\sum_{j=1}^{\infty} 
\log\left(
1-e^{-\epsilon 2j}
\right)+\sum_{j=1}^{\infty} \frac{4\epsilon j}{e^{2\epsilon j}-1}
+\\ -
2\log \prod_{j=1}^{\infty}\frac{(1-e^{-(2j-1)\epsilon})}{(1-e^{-2\epsilon j})} +\sum_{q=-\infty}^{\infty} \epsilon e^{-\epsilon}\prod_{j=1}^{\infty}\frac{(1-e^{-(2j-1)\epsilon})^2}{(1-e^{-2\epsilon j})^2} \Phi'_q(e^{-\epsilon}) .
\end{multline}
The last sum over $q$ above can be written as the following derivative
\begin{equation}
-2\epsilon\frac{d}{d\epsilon}\Big[ \displaystyle \sum_{j=1}^{\infty}  \log \frac{(1-e^{-2j\epsilon})}{(1-e^{-\epsilon(2 j-1)})} \Big],
\end{equation}
where we have used that 
\begin{equation}
\sum_{q=-\infty}^{\infty} \Phi_q(e^{-\epsilon})
=
\prod_{j=1}^\infty\frac{(1-e^{-2j\epsilon})^2}{(1-e^{-\epsilon (2j-1)})^2},
\end{equation}
reflecting that $\mathcal{Z}_1(q)$ is normalised to 1.
Taking now the derivative with respect to $\epsilon$, we finally obtain 
\begin{equation}
S_1
=
2\sum_{j=1}^{\infty} 
\left[
\frac{ \epsilon (2j-1)}{e^{\epsilon (2j-1)}-1}
-\log\left(
1-e^{-\epsilon(2j-1)}
\right)
\right],
\end{equation}
which is the entanglement entropy of a complex harmonic chain (i.e. the double of a real one).

\subsection{Numerical checks}
In this subsection we test the validity of the results in the previous ones against exact numerical computations. 
We work only with an infinite real harmonic chain \eqref{eq:Haminiziale} with finite $\omega_0$. 
For the complex case, we just combine the results for two real chains.
Let us consider a bipartition where the subsystem $A$ is given by $\ell$ contiguous lattice sites.
Let us call $X_A$ and $P_A$ the $\ell\times\ell$ matrices of the correlators restricted to the subsystem $A$, where $X_{ij}=\braket{q_iq_j}$ and $P_{ij}=\braket{p_ip_j}$.
The explicit forms of these correlators in the ground state of the gapped harmonic chain have been already reported many times in the literature (see e.g. 
Refs. \cite{br-04,p-12, eisert-2010}) 
and we are not going to rewrite them here.  
Let us  denote by $\sigma_k$, with $k=1,\dots,\ell$, the eigenvalues of the matrix $\sqrt{X_A P_A}$. 
We introduce the vectors $|\boldsymbol{n}\rangle\equiv\bigotimes_{k=1}^\ell| n_k \rangle$, products of Fock states of the number operator in the subsystem $A$, namely $N_A$.

\begin{figure}[t]
\centering
\subfigure
  {\includegraphics[width=0.45\textwidth]{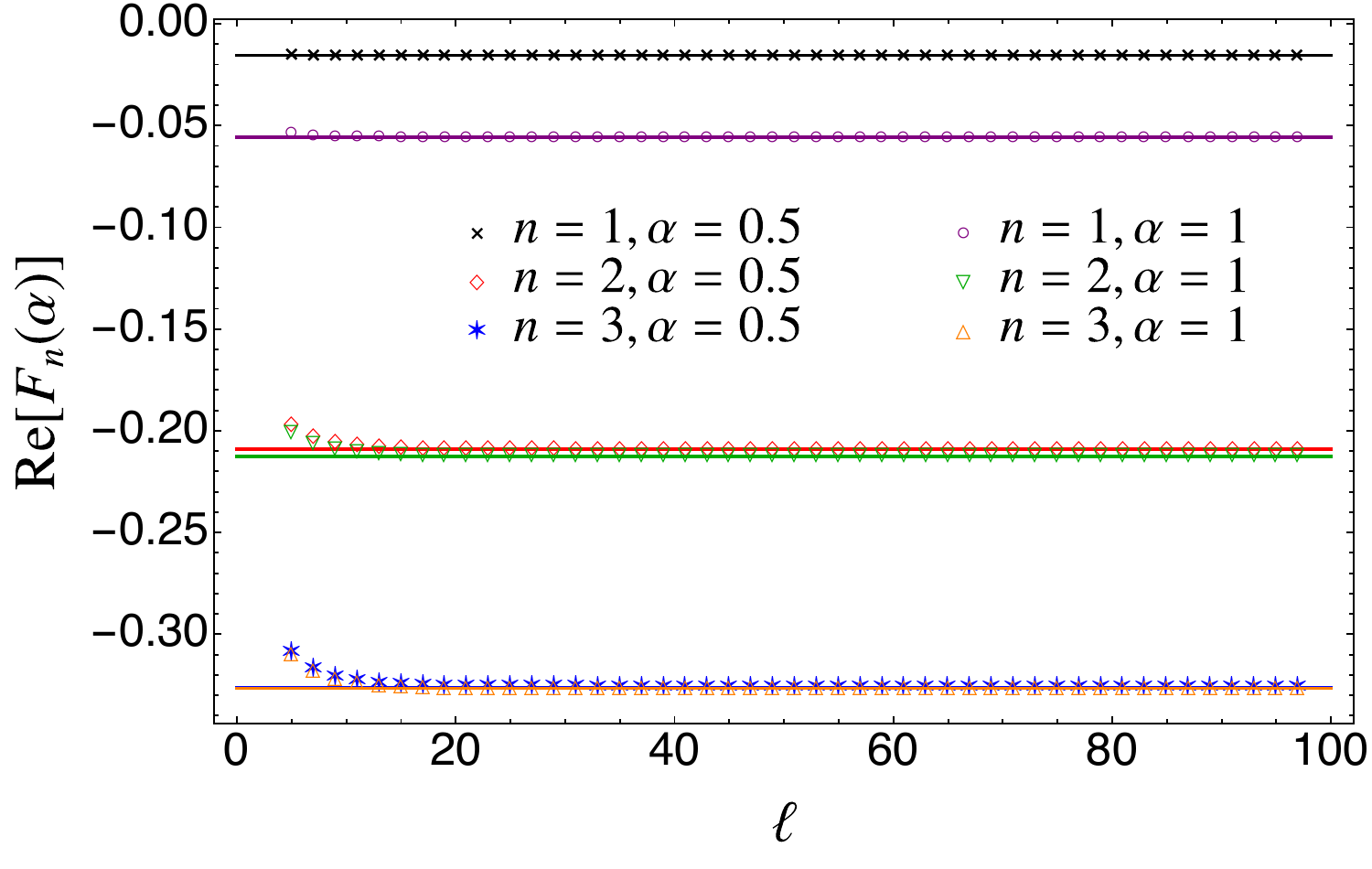}}
\subfigure
   {\includegraphics[width=0.45\textwidth]{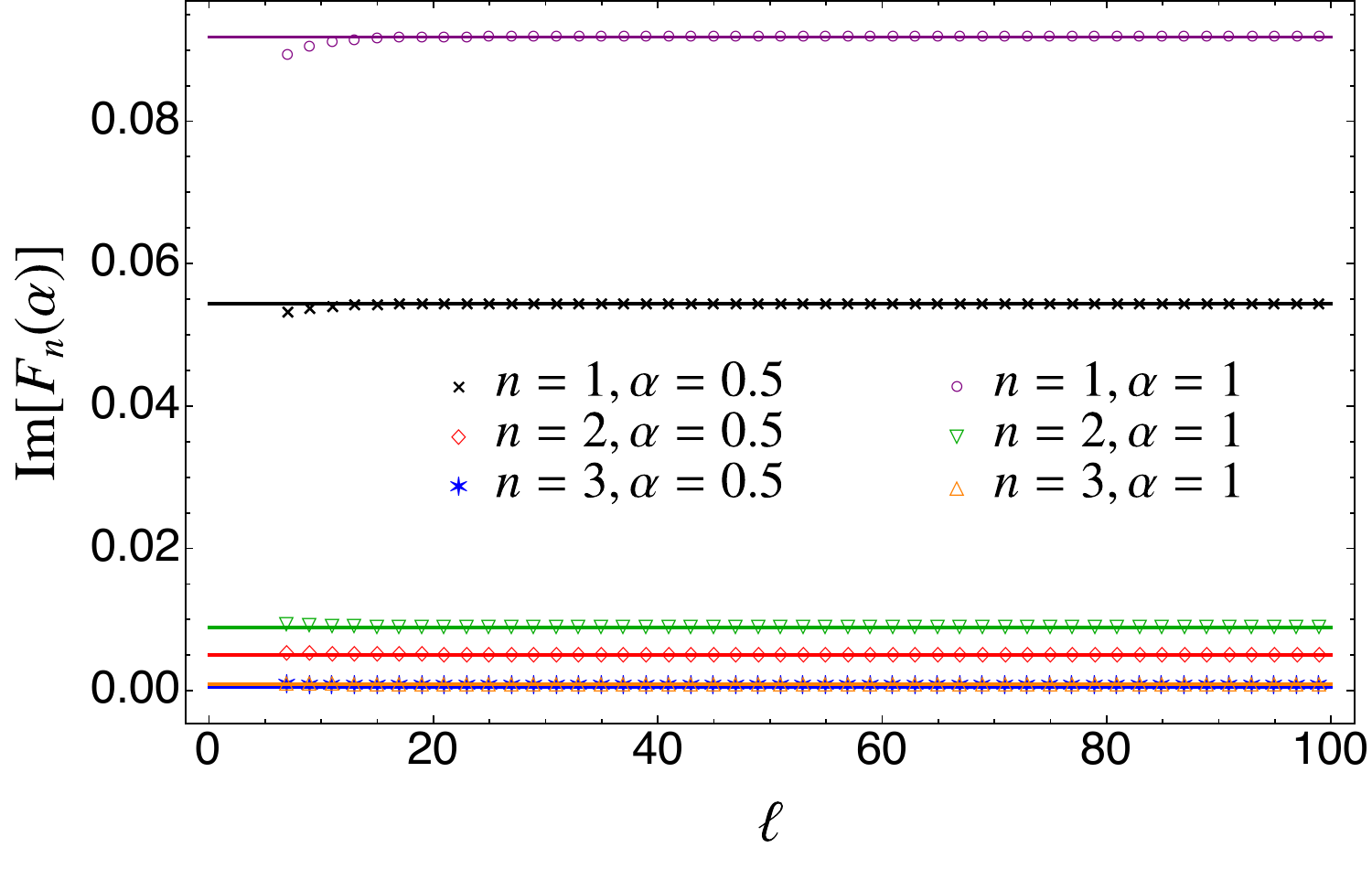}}
\caption{Numerical results for the charged moments for an interval of length $\ell$ embedded in the infinite harmonic chain.  
We report the real (left)  and the imaginary (right) parts of  $F_n(\alpha)$ as a function of the subsystem length $\ell$, 
for different values of $n=1,2,3$ and fixed $\omega_0=0.1$. 
The numerical data for an interval of length $\ell$ (divided by 2) are compared to the analytic CTM prediction \eqref{eq:first}:
as $\ell$ is moderately large, the agreement is perfect.
 The charged moments are just $\log Z_{n}(\alpha)=2 {\rm Re}[ F_n(\alpha)]$.
}\label{fig:plotHC1}
\end{figure}

The reduced density matrix of $A$ can be written as \cite{holevo1,holevo2}
\begin{equation}
\label{eq:rho_A lambda_n}
\rho_A
=
\sum_{\boldsymbol{n}} 
\prod_{k=1}^\ell
\frac{1}{\sigma_k+1/2}
\left( \frac{\sigma_k-1/2}{\sigma_k+ 1/2}\right)^{n_k}
\;
|\boldsymbol{n}\rangle \langle \boldsymbol{n}|,
\end{equation}
where the non-negative integer $n_k$ is the $k$-th element of the $\ell$-dimensional vector $\boldsymbol{n}$. 
Since $N_A=\sum_{j\in A}n_j$ is the number operator in the  orthonormal basis made of the states $|\boldsymbol{n}\rangle$, we can write 
\begin{equation}
 {\rm Tr} [\rho_A^n e^{iN_A \alpha}]
=
\sum_{\boldsymbol{n}} 
\prod_{k=1}^\ell
\left[
\frac{1}{\sigma_k+1/2}
\left( \frac{\sigma_k-1/2}{\sigma_k+ 1/2}\right)^{n_k}
\right]^n e^{i n_k \alpha}.
\end{equation} 
Summing over the possible occupation numbers $n_k$ from 0 to $\infty$, we get
\begin{equation}
\label{eq:logZ n alfa lattice} 
 {\rm Tr} [\rho_A^n e^{iN_A \alpha}]=
\prod_{k=1}^\ell
\frac{1}{
\left(
\sigma_k+\frac{1}{2}
\right)^n
-
e^{i\alpha}
\left(
\sigma_k-\frac{1}{2}
\right)^n
}.
\end{equation}
%
This relation holds also in higher dimensions and for a generic shape of the subsystem $A$ provided that $\ell$ is the number of sites in $A$. 
Notice the similarity of Eq. (\ref{eq:logZ n alfa lattice}) with the analogous result for fermions (cf. Refs. \cite{goldstein,riccarda}): there 
are only some different signs, reflecting the different statistics. 
The formula (\ref{eq:logZ n alfa lattice}) allows us to check numerically the results obtained via the CTM approach. 
Finally, the charged moments for an arbitrary subsystem $A$ for a complex harmonic lattice model are
\begin{equation}
Z_n(\alpha)=  \left|{\rm Tr} [\rho_A^n e^{iN_A \alpha}]\right|^2=
\prod_{k=1}^\ell
\frac{1}{
\left(
\sigma_k+\frac{1}{2}
\right)^n
-
e^{i\alpha}
\left(
\sigma_k-\frac{1}{2}
\right)^n
}
\frac{1}{
\left(
\sigma_k+\frac{1}{2}
\right)^n
-
e^{-i\alpha}
\left(
\sigma_k-\frac{1}{2}
\right)^n
}.
\label{cm2}
\end{equation}

\begin{figure}[t]
\centering
\subfigure
  {\includegraphics[width=0.45\textwidth]{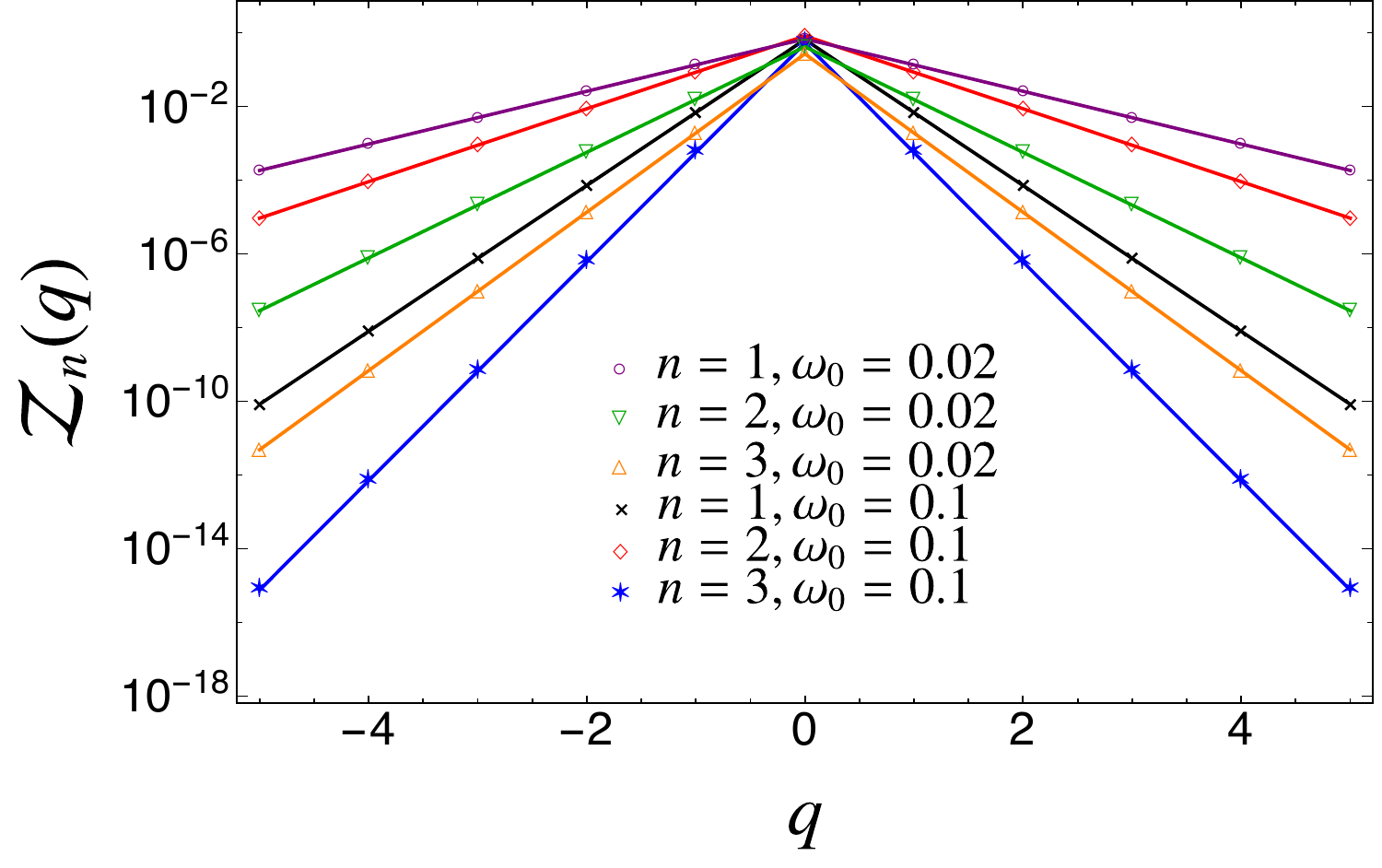}}
\subfigure
   {\includegraphics[width=0.44\textwidth]{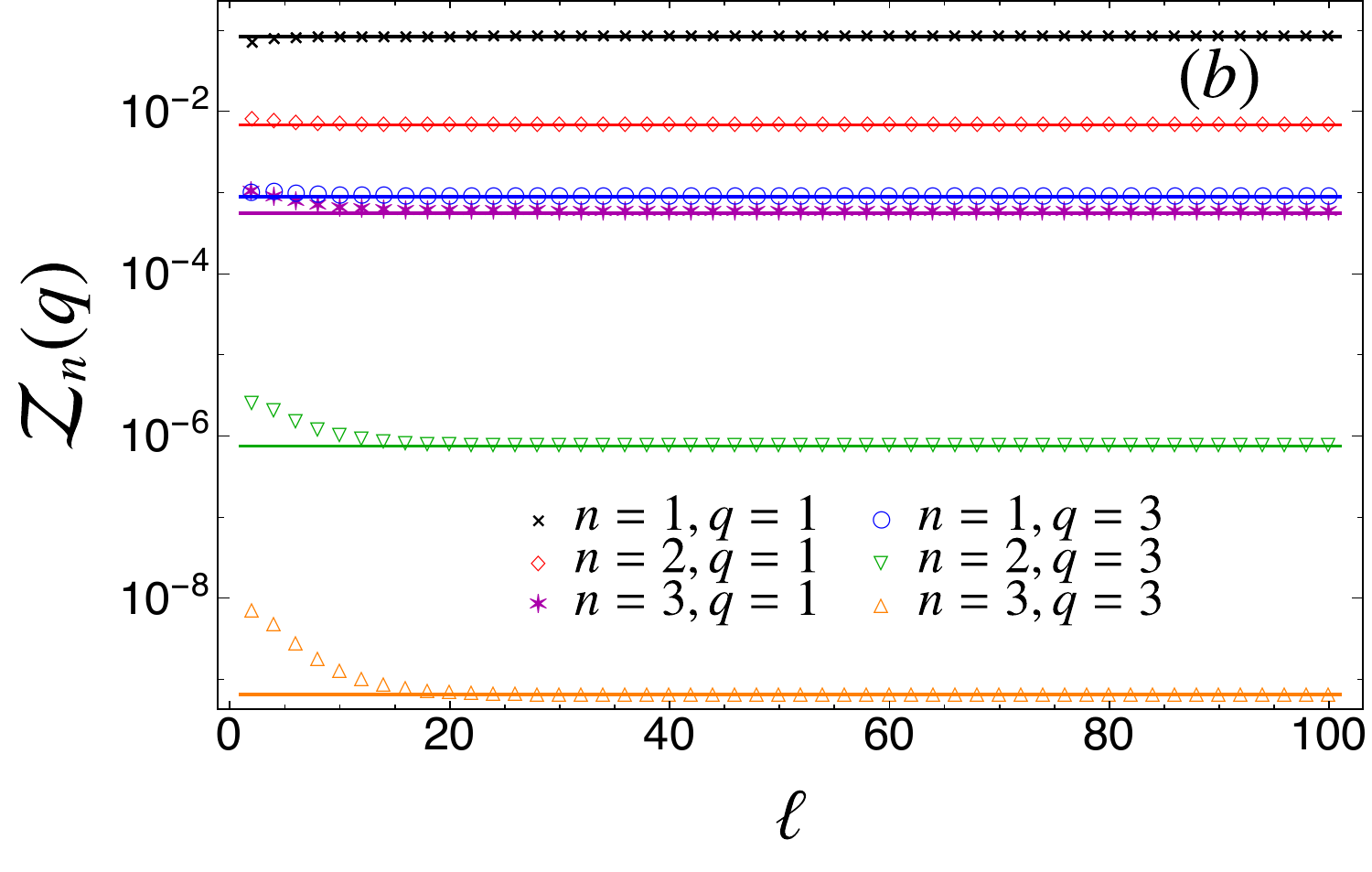}}
\caption{Numerical results for the symmetry resolved moments  for the  complex harmonic chain. 
(a): (Square root of the) symmetry resolved partition sums $\mathcal{Z}_n(q)$ as function of $q$. 
The numerical data for $n = 1, 2, 3$ are compared with the CTM prediction (\ref{Znq:fi}) for two values of $\omega_0$. 
(b): The same quantity is plotted against the subsystem size $\ell$ for different values of $q=20, 40$ and fixed $\omega_0=0.1$, showing the convergence towards the CTM 
prediction (\ref{Znq:fi}) for $n = 1, 2, 3$. 
}\label{fig:plotHC2}
\end{figure}

We now consider $F_n(\alpha)=\log {\rm Tr} [\rho_A^n e^{iN_A \alpha}]$ for a real harmonic chan.
The numerical data for $F_n(\alpha)$ for an interval of length $\ell$ should converge to the double 
(because of the two end-points) of the CTM prediction for the semi-infinite line (with one-endpoint) 
as soon as $\ell$ becomes larger than the correlation length $\xi$.
In Figure \ref{fig:plotHC1} we report the numerical data for (half of) the real and the imaginary parts of $F_n(\alpha)$ for different values of $n$ and $\alpha$. 
We have set $\omega_0=0.1$, so that after a short crossover in $\ell$, the data saturate. 
The CTM prediction (\ref{eq:first}) is also reported for comparison, showing that the analytical result perfectly describes the saturation values. 
The charged moments for the complex harmonic chain are just $\log Z_{n}(\alpha)=2 {\rm Re}[ F_n(\alpha)]$ both for numerics and analytics and 
so Figure \ref{fig:plotHC1} is a direct test also for them.

\begin{figure}[t]
\centering
   {\includegraphics[width=0.65\textwidth]{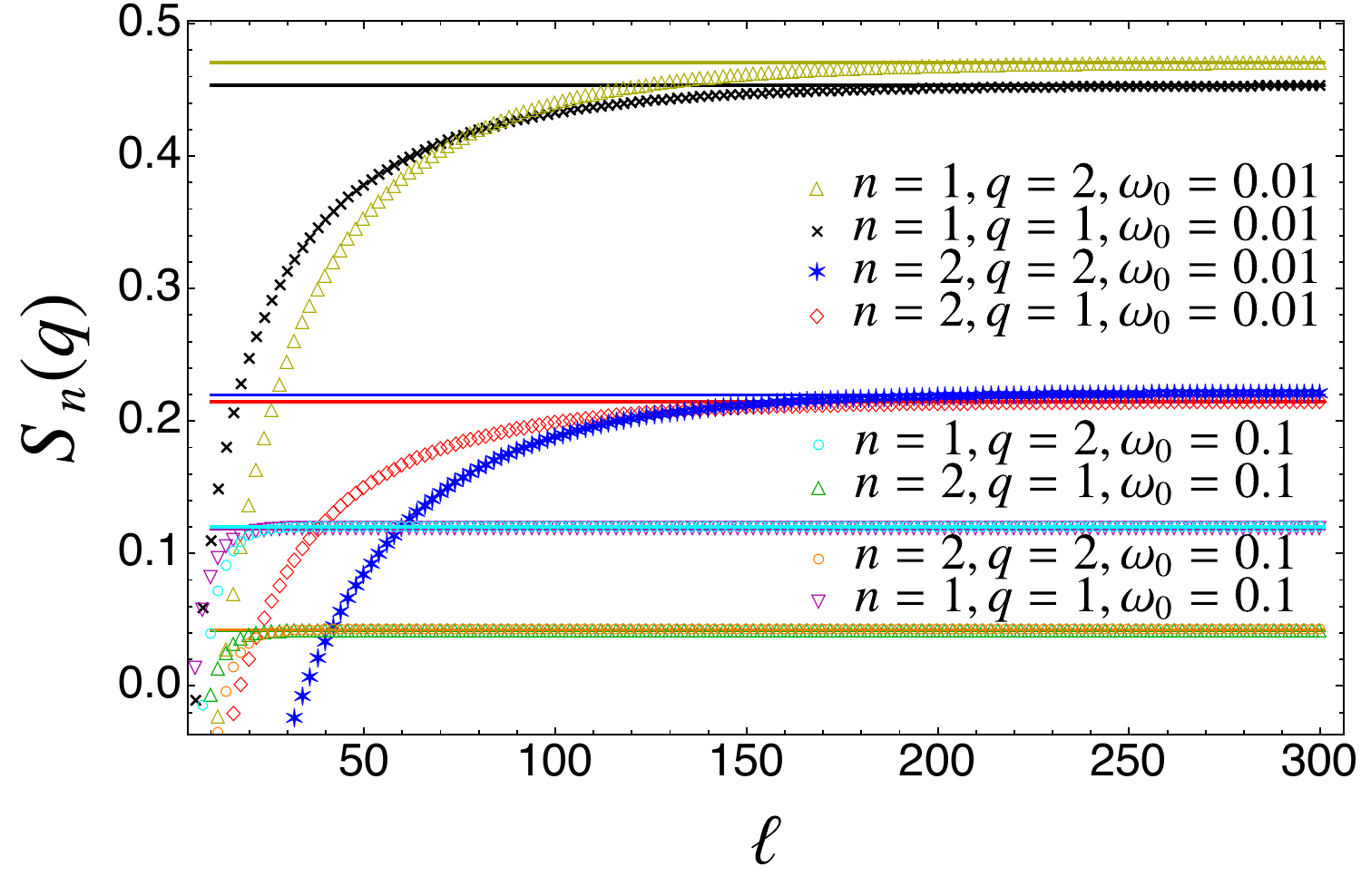}}
\caption{Numerical results for the symmetry resolved  entropies for the complex harmonic chain. 
The numerical data for $q=1, 2$, $n=1,2$ and $\omega_0=0.1$ and $0.01$ are compared with the CTM 
predictions (\ref{eq:resv}) and (\ref{eq:res1v}), to which they clearly approach. Notice that the convergence is slower for smaller $\omega_0$.
For $\omega_0=0.1$ we have an approximate equipartition, but this is not the case for $\omega_0=0.01$.   
}\label{fig:plotHC2b}
\end{figure}

We now take the Fourier transform of the numerical data for $Z_n(\alpha)$ to test the validity and the accuracy of the CTM predictions for the symmetry resolved 
moments and entropies. 
In Figure \ref{fig:plotHC2} we report the (square roots of the) numerically calculated symmetry resolved partition sums $\mathcal{Z}_n(q)$.
We compare the data for $n=1,2,3$ with the CTM prediction (\ref{Znq:fi}). 
The latter perfectly captures the $q$-dependence, as shown in the panel (a), and gives the value at which the data saturate 
when studied as functions of $\ell$, panel (b). 
Finally, in Figure \ref{fig:plotHC2b} we report the symmetry resolved entropies for several values of $q,n,\omega_0$. 
For large $\ell$, the numerical data converge to (twice) the CTM predictions in Eqs. (\ref{eq:resv}) and (\ref{eq:res1v}).
Notice that for the larger values of $\omega_0$ the saturation values do not depend on $q$ because of the effective equipartition, 
but for smaller $\omega_0$ they clearly do.  
As $\omega_0$ becomes much smaller (such that $\epsilon\sim0.1$), we expect again effective equipartition, although we do not report such data here because 
they require very large $\ell$.

\section{Gapped XXZ spin-chain}\label{sec:XXZ}
In this section we study the symmetry resolved entanglement in the anisotropic Heisenberg model in the gapped antiferromagnetic regime using the CTM approach. 
The resolved moments are computed starting from the explicit expressions for the eigenvalues of the RDM and their degeneracies. 
Then the symmetry resolved entropies are deduced and their critical regime is investigated. 
The discrete Fourier transform of the resolved moments allows us to compute the charged moments and to discuss their behaviour in the critical regime. 

\subsection{Symmetry resolved moments and entropies}
\label{sec:RMXXZ}
The Hamiltonian of the anisotropic Heisenberg model (also known as XXZ chain) is
\begin{equation}
\label{eq:XXZ Hamiltonian}
H_{\mathrm{XXZ}}
=
\sum_{j}
\left[
\sigma_j^x \sigma_{j+1}^x
+
\sigma_j^y \sigma_{j+1}^y
+
\Delta
\sigma_j^z \sigma_{j+1}^z
\right],
\end{equation}
where $\sigma^i$, $i=x,y,z$ are the Pauli matrices. The model has a conformal quantum critical point for $\Delta=1$, 
it is gapless when $|\Delta|\leq1$ and gapped when $|\Delta|>1$. 
We consider this model in the antiferromagnetic gapped regime with $\Delta>1$.

The XXZ chain is solvable by Bethe Ansatz techniques; unfortunately this framework is not very effective to study the entanglement properties both in the 
coordinate \cite{vidal1} and in the algebraic \cite{bjms-06,ss-08,sst-06,ncc-08,amt-09,km-10,ms-19,s-19,g-19} approach. 
On the other hand, the CTM solution for the XXZ chain is a powerful tool  to compute the entanglement entropies; in this approach, the reduced density matrix is related 
to the partition function of the six-vertex model on a strip with a cut. 
In Ref. \cite{peschel1} $H_{\mathrm{CTM}}$ has been found to be of the form (\ref{eq:diagonalform}) with
\begin{equation}
\label{eq:singleparticlelevelXXZ}
\epsilon_j=2 \epsilon j, \qquad \epsilon = \mathrm{arccosh}\Delta,
\end{equation}
and $n_j$ being some fermionic number operators.
Since in the thermodynamic limit, the ground-state of the gapped XXZ spin-chain is doubly degenerate we should
clarify  which state we are going to deal with in this section.  
The entanglement Hamiltonian \eqref{eq:diagonalform} together with \eqref{eq:singleparticlelevelXXZ} selects 
by construction the ground state that does not break the inversion symmetry, i.e. the one that in the limit 
of large $\Delta$ is $(|N_1\rangle+|N_2\rangle)/\sqrt2$  where $|N_i\rangle$ are the two possible N\'eel states.
However, we prefer to work with the more physical symmetry breaking state $|N_i\rangle$.
In CTM approach this can be constructed with an entanglement Hamiltonian of the form \eqref{eq:diagonalform} 
where the sum over $j$ starts from $1$ rather than $0$, i.e. 
\begin{equation}
\label{eq:diagonalform2}
H_{\mathrm{CTM}}=\displaystyle \sum_{j=1}^{\infty} \epsilon_j n_j,\qquad  \epsilon_j=2 \epsilon j, \qquad \epsilon = \mathrm{arccosh}\Delta.
\end{equation} 
In the remaining part of this section we always focus on the symmetry breaking ground state with the above $H_{\mathrm{CTM}}$.
If one is interested into the other state, analogous results may easily be derived.

The entanglement spectrum is obtained by filling in all the possible ways the single particle levels in (\ref{eq:diagonalform2}) (i.e. setting all $n_j$ equal either to 0 or 1). 
The resulting levels are equally spaced with spacing $2 \epsilon$ and highly degenerate. 
The degeneracy of the level $2\epsilon s$, with $s=\sum_j j$ (see (\ref{eq:singleparticlelevelXXZ})) is  $\mathcal{Q}(s)$, the number of partitions of $s$ 
into smaller non-repeated integers (including zero). 
(Notice we use the non-standard symbol $\mathcal{Q}(s)$ instead of $q(s)$ to avoid confusion with $q$, the charge sector.) 

We want to characterise how the entanglement of the semi-infinite line $A$ with respect to its complement splits into the different sectors with fixed magnetisation $S_z\equiv \sum_j \sigma^z_j/2$. 
We indicate with $q$ the possible values, in the subsystem $A$, of the {\it difference} of the magnetisation with respect to the antiferromagnetic N\'eel state chosen as a 
reference configuration.
Such variable $q$ is quantised in terms of integer numbers (each spin flip leads to a change of magnetisation of $\pm1$), i.e. $q\in {\mathbb Z}$. 
With a slight abuse of language, we will refer to $q$ as the magnetisation, although it is a magnetisation difference. 
To derive the symmetry resolved entanglement, we first write $\mathcal{Z}_n(q)$, defined in (\ref{eq:defF}), as 
\begin{equation}
\label{eq:ZnqXXZcountingESdef}
\mathcal{Z}_n(q)=\sum_{s\,\in\, \mathcal{S}_q} \lambda_s^{\,n}\,,
\end{equation}
where $\lambda_s$ are the eigenvalues of the RDM and the sum is restricted to the levels with fixed value of $q$.
Using Eq. (\ref{eq:red1}) and the explicit expression of the entanglement spectrum from Eq. (\ref{eq:diagonalform2}), we can write

\begin{equation}
\label{eq:ZnqXXZcountingES}
\mathcal{Z}_n(q)=\dfrac{\displaystyle\sum_{s}   \mathcal{F}(q,s) e^{- 2 n \epsilon s}}{\Big(\displaystyle\sum_{s}   \mathcal{Q}(s)  e^{- 2  \epsilon s}\Big)^n}\,,
\end{equation}
where $\mathcal{F}(q,s)$ is the number of eigenvalues at level $s$ with magnetisation $q$.
The degeneracies $\mathcal{F}(q,s)$ have been studied in Ref. \cite{albaES} with a combination of perturbation theory and integrability arguments. 
The final result for the bipartition of our interest is $\mathcal{F}(q,s)= \mathcal{P}(\frac{s-m(q)}{2})$  \cite{albaES}, with $\mathcal{P}(n)$ the number of integer partitions of 
$n$ and $m(q)=q(2q-1)$.
Using this result and changing variable in the sum of the numerator in Eq. (\ref{eq:ZnqXXZcountingES}) as $(s-m(q))/2 \to s$, we obtain
\begin{equation}
\label{eq:ZnqXXZcountingES2}
 \mathcal{Z}_n(q)=e^{-2n\epsilon q(2q-1)} \frac{\displaystyle\sum_{s}  \mathcal{P}(s)e^{- 4 n\epsilon s}}{\Big(\displaystyle\sum_{s}   \mathcal{Q}(s) e^{- 2  \epsilon s}\Big)^n},
\end{equation}
where we have also exploited that $\mathcal{P}(n)$ is non vanishing only if $n$ is a positive integer.

\begin{figure}
\centering
\subfigure
  {\includegraphics[width=0.325\textwidth]{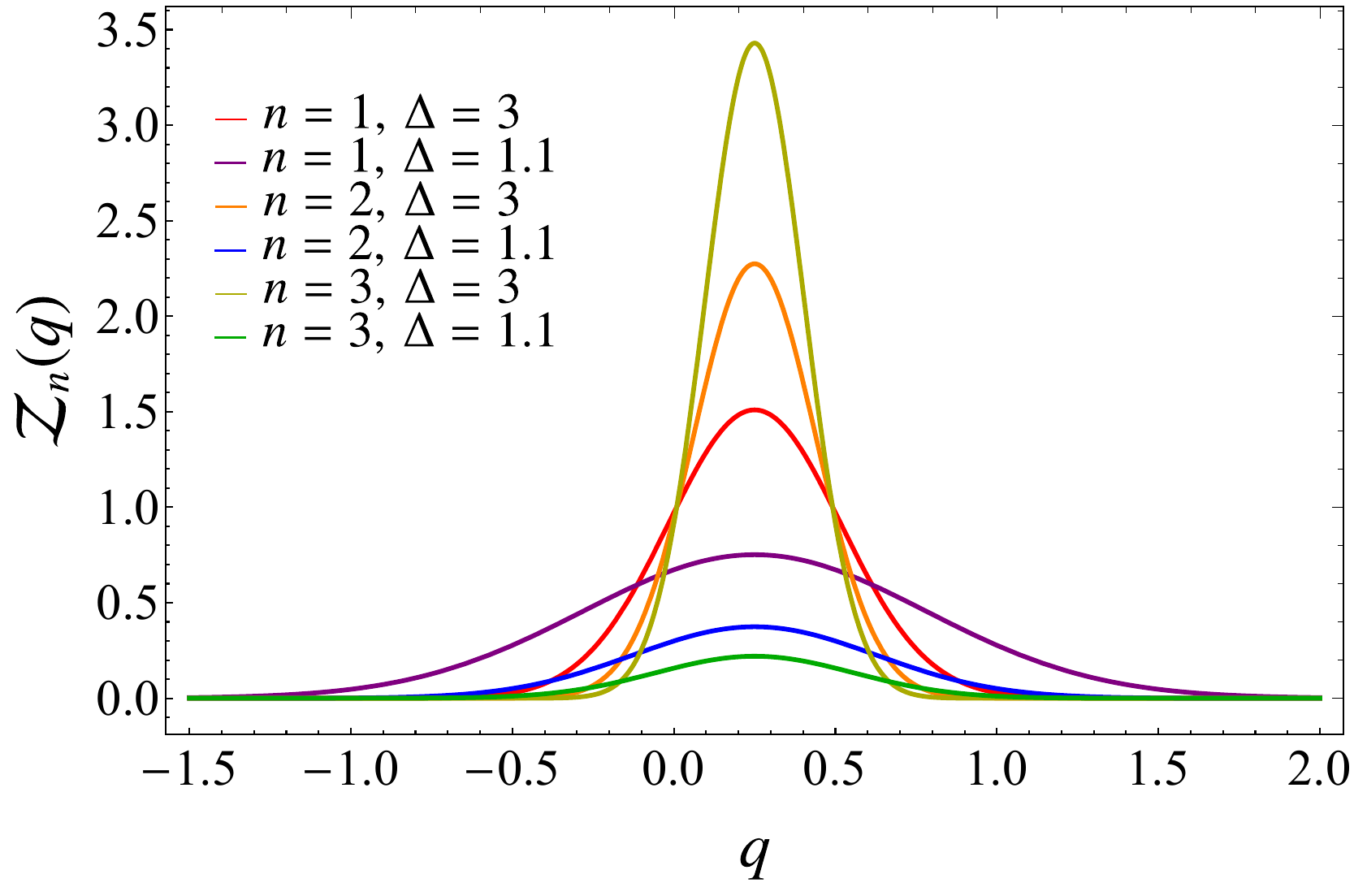}}
\subfigure
  {\includegraphics[width=0.325\textwidth]{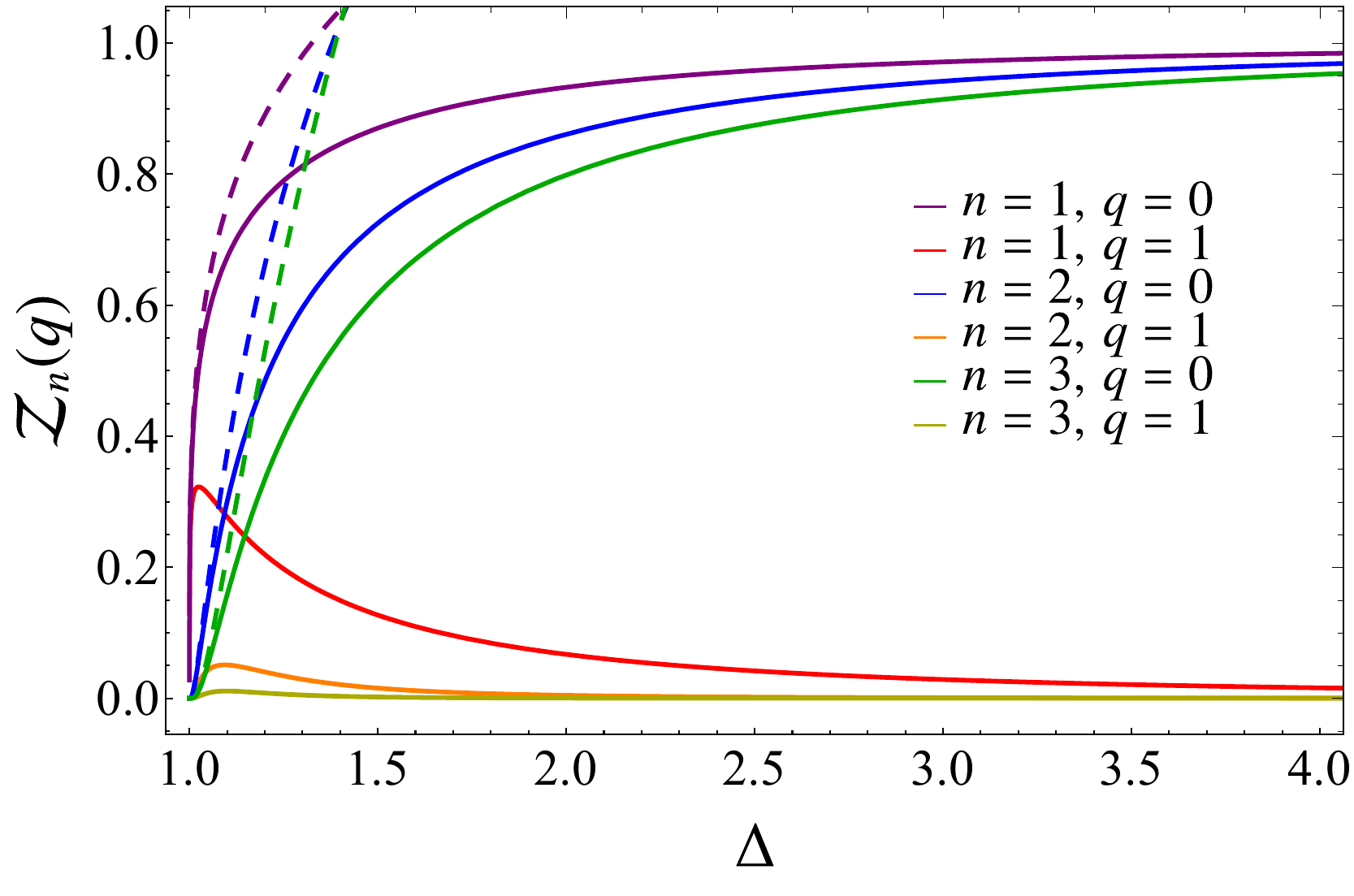}}
\subfigure
   {\includegraphics[width=0.325\textwidth]{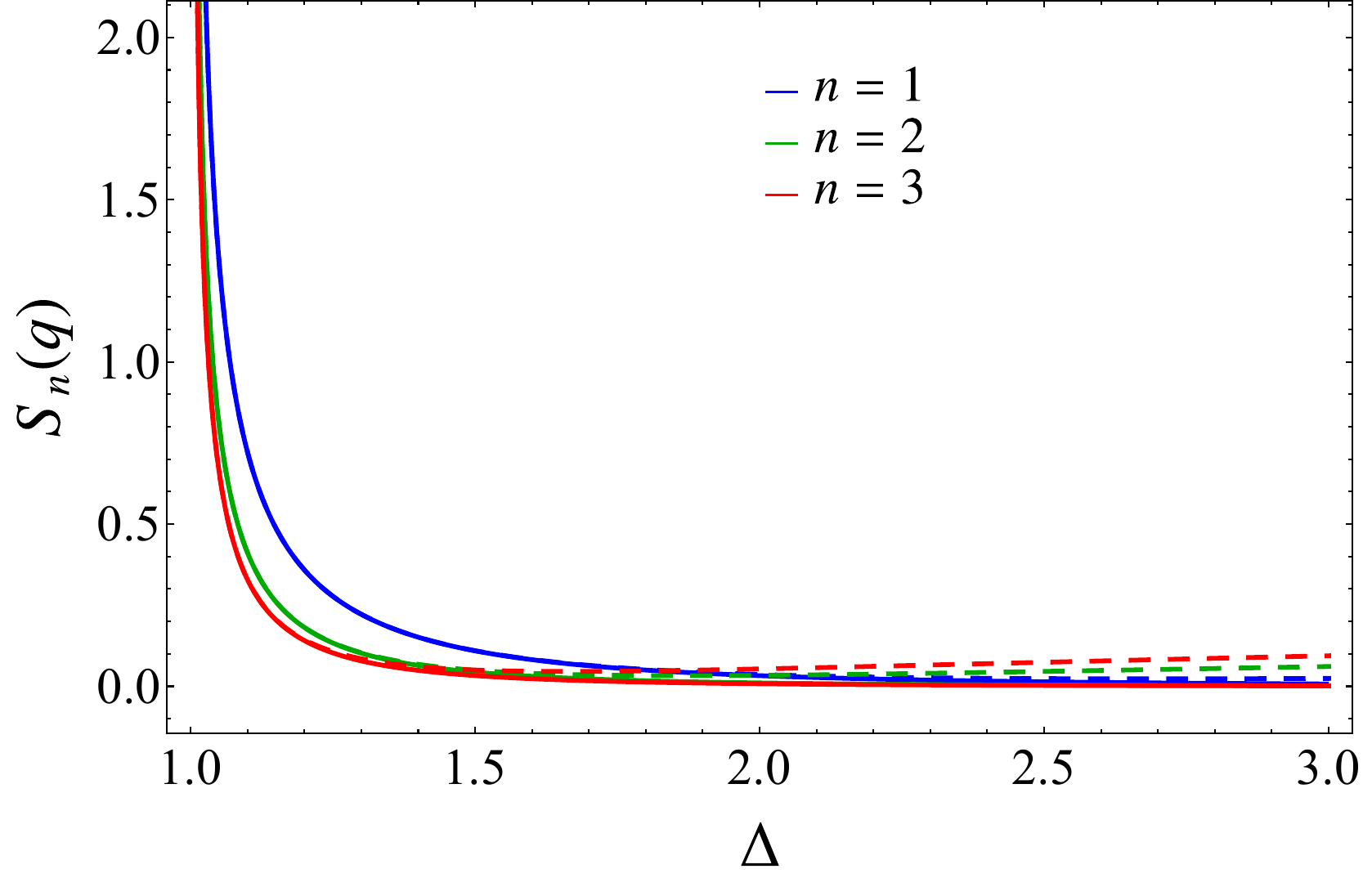}}
\caption{Magnetisation resolved moments and entropies for the XXZ spin-chain.
The left panel shows the results for $\mathcal{Z}_n(q)$, Eq. (\ref{eq:ZnqXXZcountingES3}), against $q$ for different values of $n = 1, 2, 3$ and $\Delta=1.1, 3$. 
In the middle panel, we report again $\mathcal{Z}_n(q)$ at fixed $q$ and as function of $\Delta$ (full lines).
As a comparison, we also report the asymptotic expansion (\ref{eq:ZnqXXZcountingEScrit}) for $\Delta$ close to $1$ (dashed lines).
In the right panel, we report  $S_n(q)$ and its critical limit, respectively Eq. (\ref{eq:SREEXXZn}) and Eq. (\ref{eq:SREEXXZcrit}), 
as function of $\Delta$ for $n=1,2,3$. We recall that $S_n(q)$ does not depend on $q$ because of entanglement equipartition. 
}\label{fig:2}
\end{figure}

The two sums in (\ref{eq:ZnqXXZcountingES2}) can be conveniently rewritten in terms of generating functions
\begin{equation}
\label{eq:generatingfcts}
\sum_{s=0}   \mathcal{P}(s) x^s=
\prod_{k=1}^\infty \frac{1}{1-x^k},
\,\qquad
\sum_{s=0}   \mathcal{Q}(s) y^s =
\prod_{k=1}^\infty (1+y^k).
\end{equation}
Setting $x= e^{- 4 n \epsilon}$ and $y= e^{- 2 \epsilon }$ in (\ref{eq:generatingfcts}) and plugging them into (\ref{eq:ZnqXXZcountingES2}) we obtain
\begin{equation}
\label{eq:ZnqXXZcountingES3}
\mathcal{Z}_n(q)= \dfrac{  e^{-2 n\epsilon q(2q-1)}}
{\displaystyle \prod_{k=1}^\infty \left(1-e^{- 4 n\epsilon k }\right) \prod_{k=1}^\infty \left(1+e^{- 2 \epsilon k }\right)^n}.
\end{equation}
We remark that $\mathcal{Z}_1(q)$ is normalised to one, i.e. $\sum_{q\in {\mathbb Z}} \mathcal{Z}_1(q) =1$, as it should be from the definition (\ref{eq:ZnqXXZcountingESdef}). 
This is consistent with the interpretation of $\mathcal{Z}_1(q)$ as a probability, see Section $\ref{sec:SR}$.
The denominator of Eq. (\ref{eq:ZnqXXZcountingES3}) can be expressed in terms of elliptic theta functions (see Appendix \ref{app:appendixB}) and then $\mathcal{Z}_n(q)$ reads
\begin{equation}
\label{eq:ZnqXXZcountingES4}
\mathcal{Z}_n(q)=\dfrac{2^{\frac{1+n}{3}}\left[\kappa(e^{-\epsilon })\right]^{\frac{n}{12}} e^{-4n\epsilon (q-\frac{1}{4})^2} }{ \left[\kappa(e^{-2\epsilon n})\kappa'(e^{-2\epsilon n})  \right]^{\frac{1}{6}}\,\left\lbrace  \left[\kappa'(e^{-\epsilon })\right]^{-\frac{2}{3}}-\left[\kappa'(e^{-\epsilon })\right]^{\frac{4}{3}} \right\rbrace^{\frac{n}{8}} \theta_3\left( e^{-2\epsilon n}\right)},
\end{equation}
where $\kappa$ and $\kappa'$ are defined in (\ref{eq:kappa}).
Notice that $q=1/4$ is exactly the mean magnetisation of the subsystem in the critical limit $\epsilon \to 0$, as we can check by computing $\bar{q}= \int dq q\mathcal{Z}_1(q)$,
since we are dealing with the symmetry breaking ground state.  
Notice that the dependence on $q$ in Eq. \eqref{eq:ZnqXXZcountingES4} is entirely encoded in the Gaussian factor and it is symmetric for $q\to 1/2- q$.
Moreover, exploiting the asymptotic behaviours in (\ref{eq:asymptotictheta}) and (\ref{eq:asymptotickappa}) in appendix  \ref{app:appendixB}, 
we have that in the critical regime  $\mathcal{Z}_n(q)$ becomes
\begin{equation}
\label{eq:ZnqXXZcountingEScrit}
\mathcal{Z}_n(q)
\simeq
\sqrt{\frac{2^{1+ n}\epsilon n}{\pi}} e^{-\frac{\pi^2}{24 \epsilon}\left(n-\frac{1}{n}\right)} e^{-4n\epsilon (q-\frac{1}{4})^2},
\end{equation}
where we keep the Gaussian factor in order to have a meaningful result. 
Once the resolved moments $\mathcal{Z}_n(q)$ have been worked out,  the symmetry resolved entropies follow straightforwardly 
\begin{equation}
\label{eq:SREEXXZn}
S_n(q) =
\frac{1}{1-n}\sum_{k=1}^{\infty}\left[ n\log\left(1-e^{-4\epsilon k}\right)
- \log\left(1-e^{-4n\epsilon k}
 \right) \right],
\end{equation}
and, taking the limit $n\to 1$,
\begin{equation}
\label{eq:SREEXXZ1}
S_1(q) =
\sum_{k=1}^{\infty}
\left[
\frac{4 \epsilon k}{e^{4\epsilon k}-1}
-\log\left(1-e^{-4\epsilon k}\right)
\right].
\end{equation} 
Notice that as  $\Delta \gg 1$, $S_n(q)\to 0$ (see also Figure \ref{fig:2}), 
since in this limit the selected antiferromagnetic ground state is a product state. 
If we would have considered the non-symmetry breaking ground state $(|N_1\rangle+|N_2\rangle)/\sqrt2$, $\Delta \gg 1$ we would have found $S_n(q)\to\log 2$, 
as for the total entropy \cite{cc-04,cal2010,Alba2018A}. 
We stress that although there is entanglement equipartition, the functions $S_n(q)$ are not equal to the total entropies $S_n$ because 
there is a non-vanishing fluctuation term like in Eq. \eqref{eq:SvN} for $n=1$.

Remarkably, the expressions \eqref{eq:SREEXXZn} and \eqref{eq:SREEXXZ1} for the symmetry resolved R\'enyi and von Neumann entanglement entropies do not depend 
on $q$ for any value of $n$, i.e. they {\it exactly satisfy the equipartition of entanglement} for any value of $\Delta$.
In the critical case, only the leading terms satisfy such equipartition \cite{xavier,riccarda}.
 
The relation between the correlation length of the model and $\epsilon$, in the critical regime $\xi\gg 1$, is \cite{baxter}
\begin{equation}
\label{eq:xifunctioneps}
\log \xi
\simeq
\frac{\pi^2}{2 \epsilon} + O(\epsilon^0)\,,
\end{equation}
which combined with  Eqs. (\ref{eq:ZnqXXZcountingEScrit}) provides
the expansions of the symmetry resolved entropies in the critical regime
\begin{equation}
\label{eq:SREEXXZcrit}
\begin{split}
S_n(q)
&=\frac{1}{12}\left(1+\frac{1}{n} \right)\log\xi -\frac{1}{2}\log\left(\frac{\log\xi}{\pi}\right) +
\frac{1}{2}\log 2+\frac{\log n}{2 (1-n)}, \\
S_1(q)
&=\frac{1}{6} \log\xi -\frac{1}{2}\log\left(\frac{\log\xi}{\pi}\right) + \frac{\log 2 -1}{2} .
\end{split}
\end{equation}
We notice that the term $-\frac{1}{2}-\frac{1}{2}\log(\log \xi/\pi)$ appearing in $S_1(q)$ in Eq. (\ref{eq:SREEXXZcrit}) is canceled exactly by the fluctuation entanglement entropy once we consider the total von Neumann entanglement entropy. Indeed, using that the probability is $p(q)=\mathcal{Z}_1(q)$, we write the fluctuation entropy as 
$-\int dq \mathcal{Z}_1(q)\log\mathcal{Z}_1(q) $. Using (\ref{eq:ZnqXXZcountingES4}), computing the gaussian integral in $q$ and then taking the critical limit, we find   
\begin{equation}
-\displaystyle \int_{-\infty}^{\infty} dq  \mathcal{Z}_1(q) \log  \mathcal{Z}_1(q) = \frac{1}{2}+\frac{1}{2}\log(\log \xi/\pi),
\end{equation} 
which exactly cancels the contribution from the configurational entropy.
This is in complete analogy with what has been found for critical systems for the $\log \log\ell$ term \cite{riccarda}.

As for the harmonic chain, another useful check  is to recover the total von Neumann entanglement entropy from  $S_1(q)$ in Eq.\,(\ref{eq:SREEXXZ1}).
Using the expression of $\mathcal{Z}_1(q)=p(q)$ in Eq.\,(\ref{eq:ZnqXXZcountingES3}) once we set $n=1$, the total von Neumann entropy is 
\begin{equation}\label{eq:prima}
S_1=\sum_q\mathcal{Z}_1(q)S_1(q)-\sum_q \mathcal{Z}_1(q)\log \mathcal{Z}_1(q).
\end{equation}
Let us introduce the constants $\prod_{k=1}^\infty \left(1-e^{- 4 \epsilon k }\right) =\mathcal{N}_1$
 and $ \prod_{k=1}^\infty \left(1+e^{- 2 \epsilon k }\right)=\mathcal{N}_2$. Because of normalisation of $\mathcal{Z}_1(q)$, the first term in Eq.\,(\ref{eq:prima}) just gives $S_1(q)$ (since, as already stressed, it does not depend on $q$), while the second one leads to
\begin{equation}
\sum_{q=-\infty}^{\infty} \mathcal{Z}_1(q)\log \mathcal{Z}_1(q)=\dfrac{1}{\mathcal{N}_1\mathcal{N}_2} \epsilon\partial_{\epsilon} {(\mathcal{N}_1 \mathcal{N}_2) 
}-\sum_{j=1}^{\infty}\log (1-e^{-4\epsilon j})-\sum_{ j=1}^{\infty}\log (1+e^{-2\epsilon j}).
\end{equation}
Performing explicitly the derivative with respect to $\epsilon$ and summing all contributions in Eq.\,(\ref{eq:prima}), we obtain
\begin{equation}
S_1=\sum_{ j=1}^{\infty}\log (1+e^{-2\epsilon j})+\sum_{ j=1}^{\infty}\dfrac{2\epsilon j}{e^{2\epsilon j}+1},
\end{equation}
which is the known entanglement entropy found in Refs. \cite{cc-04,Alba2018A} for the symmetry breaking ground state.

In Figure \ref{fig:2} we report symmetry resolved moments and entropies. 
The possible values of $q$ are just integers, but since $\mathcal{Z}_n(q)$ becomes quickly 
small as $q$ increases, we consider arbitrary real values. 
As anticipated, $\mathcal{Z}_n(q)$ has a peak at $q=1/4$ and shows a clear Gaussian shape for all $\Delta$. 
The exact result (\ref{eq:ZnqXXZcountingES3}) is well approximated by its critical limit (\ref{eq:ZnqXXZcountingEScrit}) for $\Delta$ close to $1$, 
but the approach is not uniform and it is worse for larger $q$ (as well as larger $n$).
Clearly, the maximum of $\mathcal{Z}_n(q)$ is a decreasing function of $n$.
In the last panel of Figure \ref{fig:2}, we report the symmetry resolved entropies as functions of $\Delta$ (as we stressed because of equipartition, they do not depend on $q$). 
Notice that the window of $\Delta$ for which the critical limit of $S_n(q)$ in Eq.  (\ref{eq:SREEXXZcrit}) is a good approximation of 
the exact expression (\ref{eq:SREEXXZn}) is wider for smaller values of $q$.

\subsection{Charged moments via Fourier series}
The charged moments  are obtained from the resolved ones $\mathcal{Z}_n(q)$ by inverting the formula (\ref{eq:defF}), i.e.
\begin{equation}
\label{eq:FourierSeriesDef}
Z_n(\alpha)
=
\sum_{q=-\infty}^{\infty}
\mathcal{Z}_n(q) e^{i q \alpha}.
\end{equation}
Plugging in the above equation the result for $\mathcal{Z}_n(q)$ in Eq. (\ref{eq:ZnqXXZcountingES3}) and using the definition  
of the elliptic function $\theta_3(z|u)$ (see Eq. (\ref{Theta3Def}) in appendix \ref{app:appendixB}),  we obtain
\begin{equation}
\label{eq:ChargedMomentsXXZ}
Z_n(\alpha)
=
\dfrac{\theta_3(\,\frac{\alpha}{2}-i n\epsilon | e^{-4\epsilon n})}
{\displaystyle \prod_{k=1}^\infty \left(1-e^{- 4 n\epsilon k }\right) \prod_{k=1}^\infty \left(1+e^{- 2 \epsilon k }\right)^n}.
\end{equation}
Setting $\alpha=0$ and exploiting the infinite product representation (\ref{eq:Theta3Prodznonzero}) of $\theta_3(z|u)$, we get
\begin{equation}
\label{eq:retrievingCC}
Z_n(0)
=
\dfrac{\displaystyle\prod_{k=1}^\infty \left(1+e^{- 2 \epsilon n k }\right)}
{\displaystyle \prod_{k=1}^\infty \left(1+e^{- 2 \epsilon k }\right)^n},
\end{equation}
as found in \cite{cc-04}.
As for $\mathcal{Z}_n(q)$ in Section \ref{sec:RMXXZ}, we can express $Z_n(\alpha)$ in terms of elliptic functions obtaining
\begin{equation}
\label{eq:ChargedMomentsXXZ2}
Z_n(\alpha)
=
\dfrac{2^{\frac{1+n}{3}} e^{- \frac{n}{4}\, \epsilon}\,\left[\kappa(e^{-\epsilon })\right]^{\frac{n}{12}}}{ \left[\kappa(e^{-2 \epsilon n})\kappa'(e^{-2 \epsilon n})  \right]^{\frac{1}{6}}\,\left\lbrace  \left[\kappa'(e^{-\epsilon })\right]^{-\frac{2}{3}}-\left[\kappa'(e^{-\epsilon })\right]^{\frac{4}{3}} \right\rbrace^{\frac{n}{8}} }\,
\dfrac{\theta_3(\frac{\alpha}{2}-i n\epsilon\,|\,e^{-4\epsilon n}) }{\theta_3\left( e^{-2\epsilon n}\right)}.
\end{equation}
$Z_n(\alpha)$ in the critical regime is obtained using the asymptotic expansions reported in appendix \ref{app:appendixB} (i.e. Eqs. (\ref{eq:asymptTheta3znonzero}),  (\ref{eq:asymptotictheta}) 
and (\ref{eq:asymptotickappa})), finding
\begin{equation}
\label{eq:logZnalfaXXZcountingEScrit0}
Z_n(\alpha)
\simeq
2^{ - \frac{1-n}{2}}\, e^{-\frac{\pi^2}{24 \epsilon}\left(n-\frac{1}{n}\right)}e^{-\frac{\alpha^2}{16 n\epsilon}+ i\frac{\alpha}{4}}.
\end{equation}
Taking the logarithm of $Z_n(\alpha)$ and using (\ref{eq:xifunctioneps}) we have
\begin{equation}
\label{eq:logZnalfaXXZcountingEScrit}
\log Z_n(\alpha)
\simeq
\left[\frac{1}{12}\left(\frac{1}{n}-n\right) -\frac{\alpha^2}{8\pi^2 n} \right]\log\xi
+ i\frac{\alpha}{4}- (1-n)\frac{\log 2}{2}.
\end{equation}
Here, the linear term in $\alpha$ is just the mean magnetisation in $A$,  $\bar{q}={1}/{4}$.

The leading term in Eq. (\ref{eq:logZnalfaXXZcountingEScrit}) is very suggestive.
Indeed,  for the critical compact boson (aka, Luttinger liquid), in the case of $A$ being an interval of length $\ell$ embedded in an infinite 1D system, 
$\log Z_n(\alpha)$ diverges logarithmically with $\ell$ as \cite{goldstein} 
\begin{equation}
\label{eq:CFT}
\log Z_n(\alpha)\simeq
\left[\frac{1}{6}\left(\frac{1}{n}-n\right) -\frac{\alpha^2}{2\pi^2 n} K \right]\log\ell +\dots, 
\end{equation}
where $K$ is the Luttinger liquid parameter (related to compactification radius). 
The prefactor of Eq. (\ref{eq:logZnalfaXXZcountingEScrit}) is exactly half of the conformal result \eqref{eq:CFT} for $K=\frac{1}{2}$, which is the 
Luttinger parameter at $\Delta=1$. 
The multiplicative factor $1/2$ is simply understood because in our geometry there is a single endpoint instead of two as in the conformal case. 
It is natural to wonder under what hypotheses this can happen since we have seen that it is not true for the harmonic chain. 
Moreover, for the symmetry resolved entropies, the CFT result is $S_n(q)-S_n= -\frac12 \log ((2K/\pi) \log \ell)+O(\ell^0)$ \cite{goldstein,xavier}, 
which is the same as in Eq. \eqref{eq:SREEXXZcrit} with the replacement $\ell\to\xi$ and with $K=1/2$.

\begin{figure}
\centering
   \subfigure
  {\includegraphics[width=0.49\textwidth]{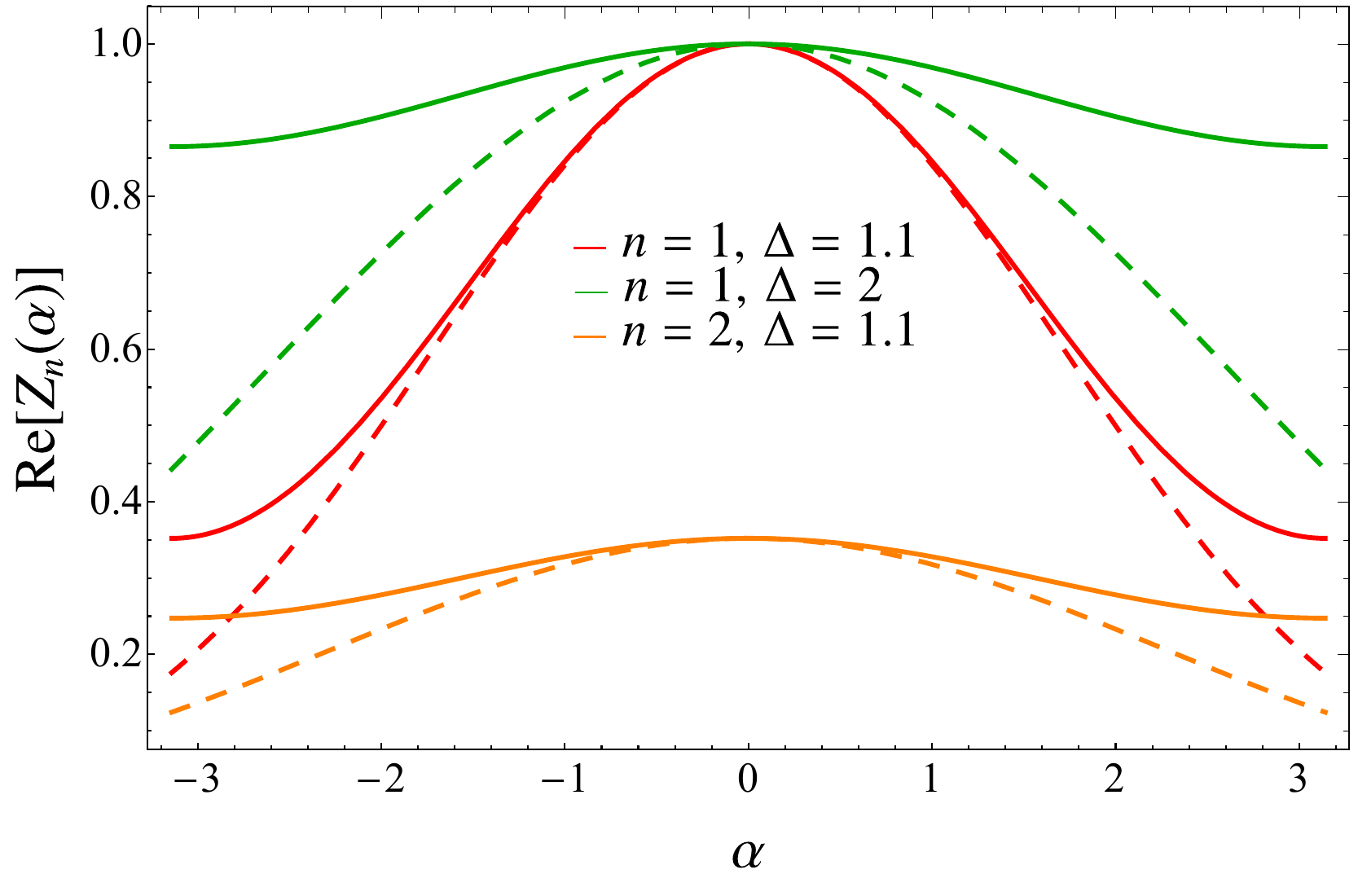}}
  \subfigure
  {\includegraphics[width=0.49\textwidth]{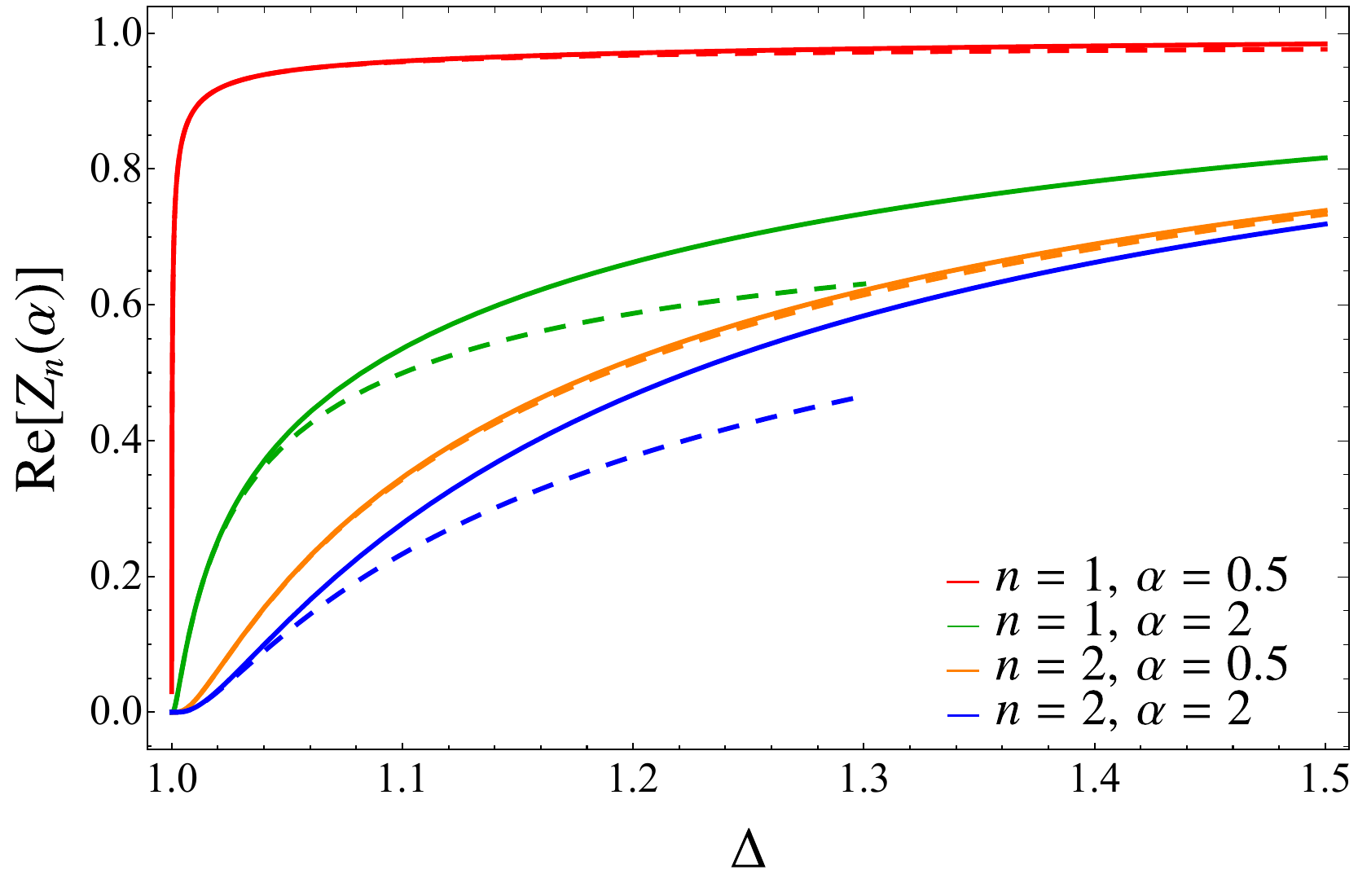}}
\subfigure
  {\includegraphics[width=0.49\textwidth]{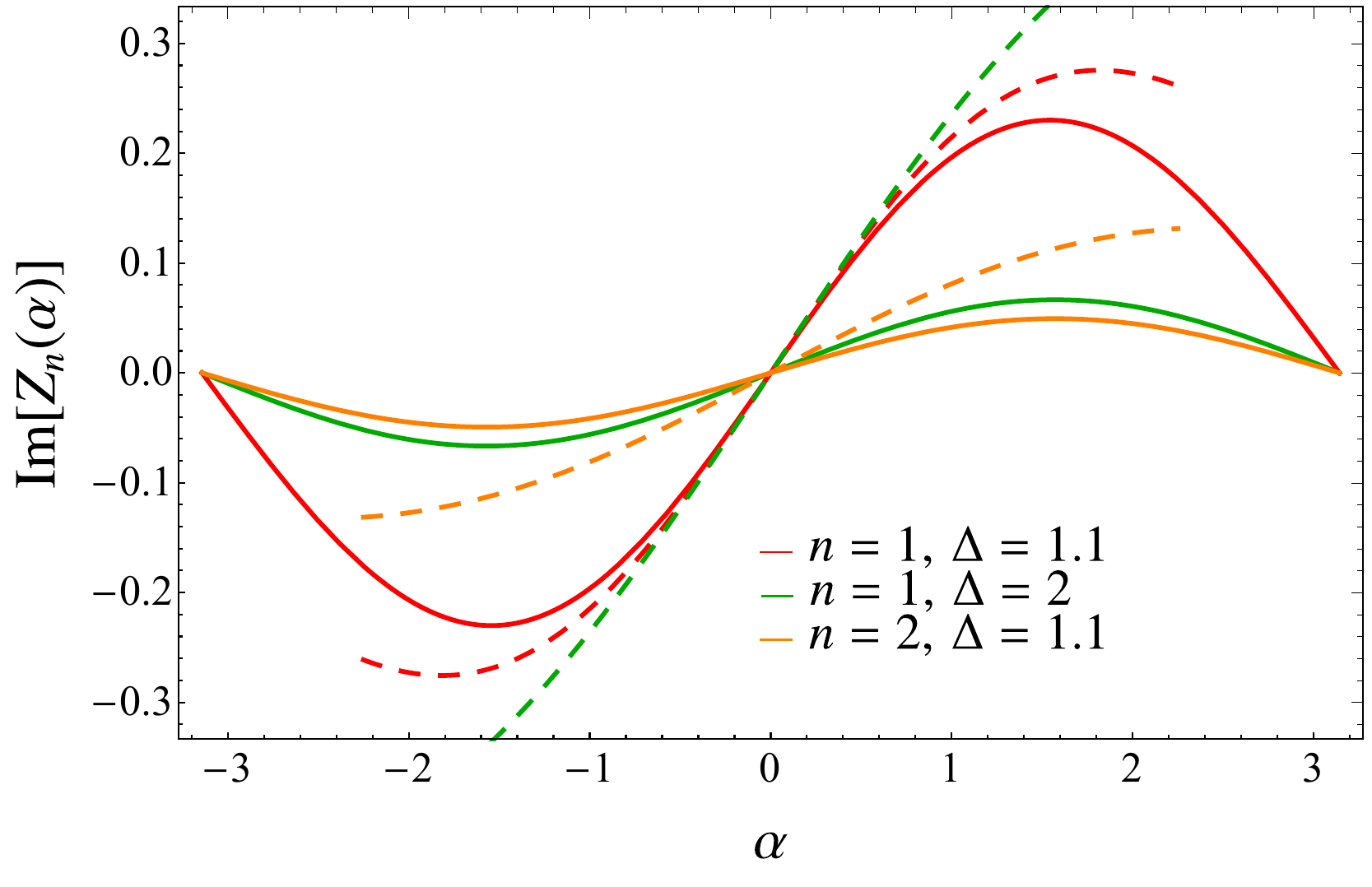}}
  \subfigure
  {\includegraphics[width=0.49\textwidth]{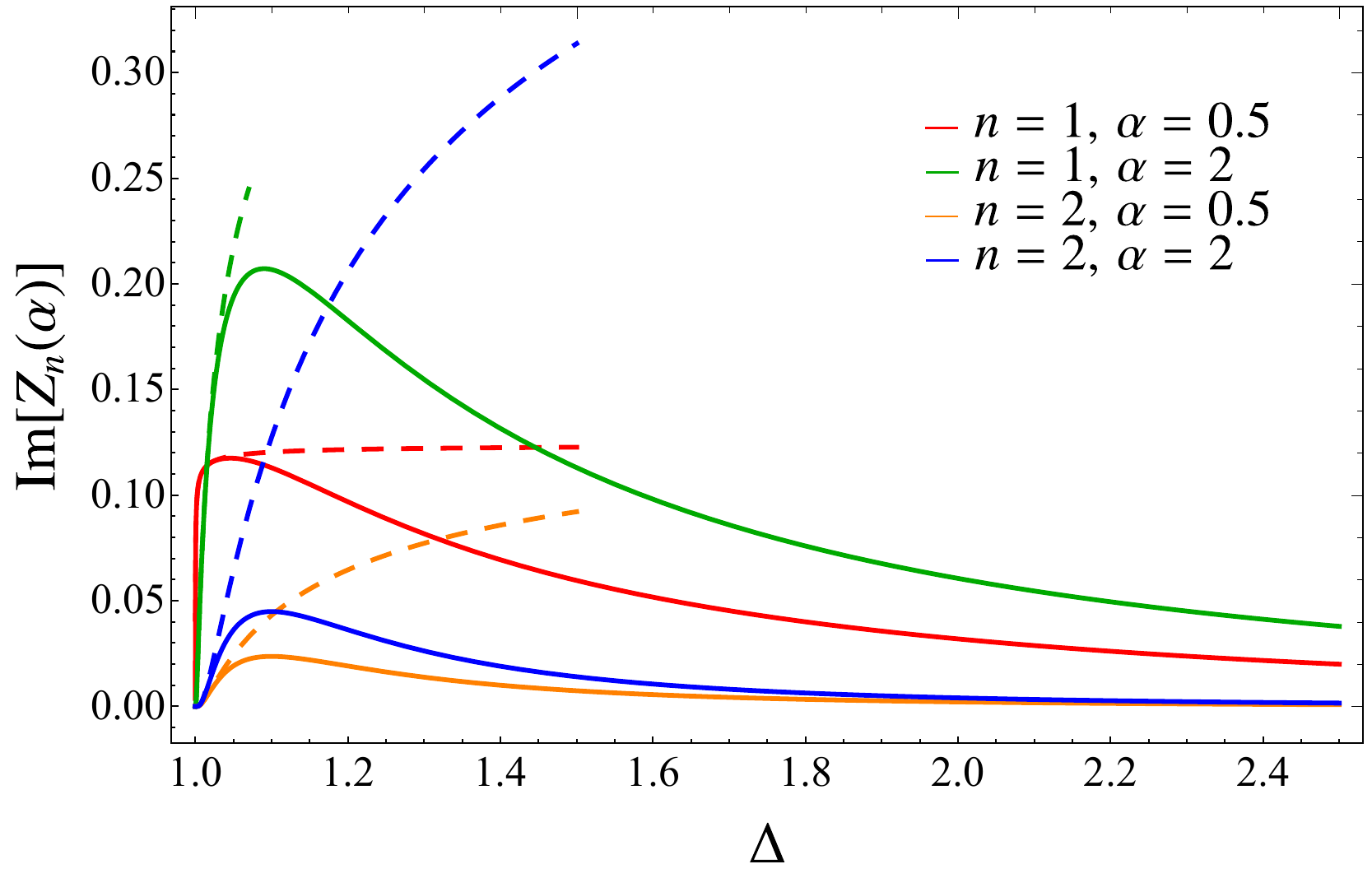}}
\caption{Charged moments for the XXZ spin chain. Top and Bottom plots correspond to real and imaginary parts respectively. 
In the left panels, the plots are against $\alpha$ for different values of $\Delta$, while in the right panels are against $\Delta$ at fixed $\alpha$. 
The dashed lines are the expansions close to the critical points (Eq. (\ref{eq:logZnalfaXXZcountingEScrit0})) that are approached only for $\Delta$ very close to $1$. 
 }\label{fig:1}
\end{figure}

In Figure \ref{fig:1} we report the plots of the charged moments as functions of $\alpha$ and $\Delta$. 
Also in this case, the approach to the critical regime is not uniform and it is faster for $\alpha$ closer to $0$ (and $n$ close to $1$).
This is very different compared to what we have seen in the previous section for the harmonic chain for which the limit $\alpha\to 0$ is singular. 
This is a further confirmation that the anomalous behaviour of the harmonic chain is due to its non-compact nature of the continuum limit.

\section{Conclusions}\label{sec:concl}
In this manuscript we found exact results for the symmetry resolved entanglement entropies of half line in infinite
 integrable systems in the gapped regime. 
We considered two models for which the RDM (and therefore the entanglement spectrum) of the subsystem can be obtained through the Baxter CTM.

In Section \ref{sec:HC} we considered the massive regime of the complex harmonic chain that has a $U(1)$ symmetry corresponding 
to the conservation of the electric charge. 
In order to obtain the symmetry resolved entanglement entropies, we first computed the charged moments of the RDM
in Eqs. \eqref{ZFF}, \eqref{eq:first}, and (\ref{eq:leadinghc}). 
Their critical behaviour is also discussed and an interesting discontinuity for $\alpha\to 0$ has been pointed out.  
Then we computed the Fourier transform of the charged moments and the symmetry resolved entanglement entropies
(see Eqs. \eqref{eq:resv} and \eqref{eq:res1v}); we also discussed their leading behaviour in the critical regime, see  \eqref{Sncr}. 
Interestingly we found that there is no entanglement equipartition, i.e. the symmetry resolved entropies $S_n(q)$ explicitly depend on $q$.
However, entanglement equipartition is effectively recovered in two limits: i) 
for large $q$, i.e. as soon as $q$ becomes larger than the logarithm of the correlation length and ii) in the critical region for $n\epsilon\ll1$. 

We also derived an exact expression for the charged moments valid for a generic harmonic system
in the correlation matrix approach \cite{p-12}.
The final results are the formulas (\ref{eq:logZ n alfa lattice}) and \eqref{cm2}) which hold in any dimension and for any shape of the subsystem.
Here we limit ourselves to use these relations to check numerically the results derived in the CTM approach.
We considered a finite interval of length $\ell$ in an infinite chain and we found that for large $\ell$ the results converge to the CTM predictions.

In Section \ref{sec:XXZ} the symmetry resolved entanglement entropies have been computed for the XXZ chain in the antiferromagnetic gapped regime 
(Eqs. (\ref{eq:SREEXXZn}) and (\ref{eq:SREEXXZ1})). 
Here,  the conserved $U(1)$ symmetry corresponds to the rotations in the plane perpendicular to the anisotropy.
Somehow surprisingly, for this model, the symmetry resolved entropies {\it exactly satisfy the equipartition of entanglement} for any anisotropy $\Delta\geq1$. 
We found this result very remarkable, although its physical origin is not clear: it would be very interesting to establish a priori which properties 
guarantee an exact equipartition of entanglement and how they are related to integrability.
The computation has been performed exploiting the explicit expressions of the elements of the entanglement spectrum and the degeneracies of each level in a 
given magnetisation sector \cite{albaES}. 
We also computed the charged moments (Eq. (\ref{eq:ChargedMomentsXXZ})) checking that for $\alpha=0$ the result of \cite{cc-04,Alba2018A} was retrieved.
We found that $Z_n(\alpha)$ have no discontinuities, as a difference with the complex harmonic chain.

Let us conclude this manuscript with some possible directions for future investigations motivated by the results we have found.
A first and natural question is whether some of the results we found here may be also recovered in massive two-dimensional field theories both free and integrable.
Work in this direction is in progress \cite{mdc-20}. It is also interesting to understand what happens when integrability is absent: while a general treatment 
seems impossible, the results for the entanglement spectrum in Refs. \cite{albaES,ahl-13} suggests that in some non-trivial regimes general results may be derived.
Another natural extension is to study symmetry resolved entanglement in higher dimension for which there are only few works for free fermions \cite{fg-19,tr-19}.
Our Eq. (\ref{eq:logZ n alfa lattice}) paves the way for general numerical studies in arbitrary dimension for bosonic systems as well, 
also in the presence of a spherical constraint \cite{waa-19}. In some cases, also analytical results can be explicitly worked out \cite{mrc-20}.
Finally, one expects that the symmetry resolved entanglement should help in reconstructing the entanglement (or modular) Hamiltonian, but 
it is still unclear how. This issue is very timely given the large current effort devoted to understand  the structure of the entanglement Hamiltonians
both in field theories \cite{chm-11,ct-16,achp-18,fr-19} and lattice models \cite{dvz-17,gmcd-18,ep-17,pa-18, etp-19}.

\section*{Acknowledgments}
We are very grateful to Moshe Goldstein and Ingo Peschel for pointing out a major flaw in the calculations for the harmonic chain
in the first version of this manuscript.
We thank Mario Collura and Paola Ruggiero for useful discussions and collaborations on related topics. 
PC and SM acknowledge support from ERC under Consolidator grant number 771536 (NEMO).

\begin{appendices}

\section*{Appendices}

\section{Details for the complex harmonic chain}
\label{app:osc}

\subsection{A two-site chain with complex oscillators}

For a single harmonic chain with two sites, the RDM has been worked out e.g. in \cite{Peschel}.
The entanglement Hamiltonian of one site is  
\begin{equation}
\label{eq:1}
\mathcal{H}_A=\varepsilon \beta^{\dagger} \beta.
\end{equation}
For this site, the $\beta$'s are related to the $a$'s as
\begin{equation}
\beta= a \cosh \theta  + a^\dag\sinh \theta \,,
\end{equation} 
which is the specialisation of Eq. \eqref{gen:bog} to the case of A being one site. 
Here $e^\theta= (1+\omega_0^2/4)^{1/4}$, but its explicit value is unimportant.
Hence, in terms of the ladder operators $a, a^{\dagger}$, ${\cal H}_A$ can be rewritten as
\begin{equation}
\label{eq:2}
\mathcal{H}_A=\varepsilon \left( \frac{1}{2}( a^{\dagger 2} +a^2)\sinh 2\theta  + a^{\dagger} a \cosh^2\theta  +a a^{\dagger} \sinh^2 \theta   \right).
\end{equation}
Rather then one real harmonic oscillator, we consider a complex one, which is the same as two real harmonic oscillators described by the ladder operators 
$a_i^{\dagger},a_i, \, i=x,y$ such that the only non-vanishing commutators are $[a_i,a_i^{\dagger}]=1, \, i=x,y$. 
Therefore, the entanglement Hamiltonian of these two real harmonic oscillators is simply the sum of two single ones:
\begin{equation}
\label{eq:3}
\mathcal{H}_A=\sum_{i=x,y}\varepsilon \left( \frac{1}{2}( a^{\dagger 2}_i +a_i^2)\sinh 2\theta  + a_i^{\dagger} a_i \cosh^2\theta  +a_i a^{\dagger}_i  \sinh^2 \theta \right).
\end{equation}
Let us rewrite Eq. (\ref{eq:3}) in terms of the particle and antiparticle ladder operators $a$ and $b$  in Eq. \eqref{eq:abi}, i.e. 
 \begin{equation}
\label{eq:ab}
\begin{split}
a_x=\frac{1}{\sqrt{2}}(a+b) \qquad &a^{\dagger}_x=\frac{1}{\sqrt{2}}(a^{\dagger}+b^{\dagger}) \\
a_y=\frac{1}{\sqrt{2}i}(a-b) \qquad &a_y^{\dagger}=\frac{1}{\sqrt{2}i}(b^{\dagger}-a^{\dagger}) .
\end{split}
\end{equation}
One can check that $[a,a^{\dagger}]=[b,b^{\dagger}]=1$, while all other commutators vanish. Plugging Eqs.\,(\ref{eq:ab}) into Eq.\,(\ref{eq:3}), we obtain 
\begin{equation}
\label{eq:4}
\mathcal{H}_A=\varepsilon \left(( a^{\dagger } b^{\dagger}+a b) \sinh 2\theta  + (a^{\dagger} a +b^{\dagger}b) \cosh^2\theta +( a a^{\dagger}+b b^{\dagger}) \sinh^2 \theta\right),
\end{equation}
or, equivalently (up to an additive constant we can absorb in the normalisation factor of the RDM)
\begin{equation}
\label{eq:5}
\mathcal{H}_A=\varepsilon \left(( a^{\dagger } b^{\dagger}+a b) \sinh 2\theta   + (a^{\dagger} a +bb^{\dagger}) \cosh 2\theta  \right).
\end{equation}
One can bring Eq.\,(\ref{eq:5}) into a diagonal form through Bogoliubov transformations, i.e.
\begin{equation}
\label{eq:bog}
\begin{split}
\alpha=\cosh \theta \, a+\sinh \theta \, b^{\dagger},  \qquad &\alpha^{\dagger}=\cosh \theta\, a ^{\dagger}+\sinh \theta \, b,  \\
\gamma=\sinh \theta \, a+\cosh \theta \, b^{\dagger},  \qquad &\gamma^{\dagger}=\sinh \theta \, a^{\dagger}+\cosh \theta\, b,
\end{split}
\end{equation}
where $[\alpha,\alpha^{\dagger}]=[\gamma,\gamma^{\dagger}]=1$, while $[\alpha,\gamma]=0$.
As a result, one finds that the RDM for one single complex harmonic oscillator $\rho_1$  has the form
\begin{equation}
\label{eq:rho1c3}
\rho_1=K e^{-\mathcal{H}_A}, \qquad \mathcal{H}_A=\varepsilon \left( \alpha^{\dagger} \alpha + \gamma^{\dagger}\gamma \right).
\end{equation}
Since the operators $\gamma$ and $\alpha$ commute, we can rewrite Eq.\,(\ref{eq:rho1c3}) 
\begin{equation}
\label{eq:rhopp}
\rho_1=K e^{-\mathcal{H}_A^{(\alpha)}}\otimes e^{-\mathcal{H}_A^{(\gamma)}}.
\end{equation}

\subsection{The Bogoliubov transformation for a chain of arbitrary length}

For a real harmonic chain of arbitrary length $2L$, the entanglement Hamiltonian for half system is \cite{Peschel}
\begin{equation}
{\cal H}_A= \sum_{j=0}^{L-1} \varepsilon_j \beta_j^\dag \beta_j,
\end{equation}
where the eigenvalues $\varepsilon_j$ depend on $L$ and in the thermodynamic limit are given by Eq. \eqref{eq:CTMsf}
while for $L=1$ by Eq. \eqref{eq:1}. 

The ladder operator $\beta_j$ as function of the local ladder operators are given by Eq. \eqref{gen:bog}, i.e. 
\begin{equation}
\beta_j=\sum_{i\in A} g_{ji}a_i +h_{ji} a^\dag_i.  
\label{gen:bog:app}
\end{equation}
Hence, the entanglement Hamiltonian in terms of local operators is
\begin{equation}
\label{eq:2:2}
\mathcal{H}_A=\sum_j \varepsilon_j  \sum_{i_1,i_2} 
\left( g^*_{i_1 j} g_{j i_2}  a^{\dagger}_{i_1} a_{i_2}+ g^*_{i_1 j} h_{j i_2}  a^{\dagger}_{i_1} a^\dag_{i_2}+h^*_{i_1 j} g_{j i_2}  a_{i_1} a_{i_2}
+h^*_{i_1 j} h_{j i_2}  a_{i_1} a^\dag_{i_2}   \right).
\end{equation}
Therefore, the entanglement Hamiltonian of a complex chain is just the sum of two real ones with local ladder operators $a_{a,j}$ with $a=x,y$ as in the case of two 
oscillators in the previous subsection.  
Such ${\cal H}_A$ can be rewritten in terms of the particle and antiparticle ladder operators  in Eq. \eqref{eq:abi},  obtaining (up to constants) 
\begin{equation}
\label{eq:4:2}
\mathcal{H}_A=\sum_j \varepsilon_j  \sum_{i_1,i_2} 
\left( (g^*_{i_1 j} g_{j i_2} + h^*_{i_1 j} h_{j i_2} )  (a^{\dagger}_{i_1} a_{i_2} +b^{\dagger}_{i_1} b_{i_2}) + g^*_{i_1 j} h_{j i_2}  a^{\dagger}_{i_1} b^\dag_{i_2}+ 
h^*_{i_1 j} g_{j i_2}  a_{i_1} b_{i_2}   \right),
\end{equation}
which we can  put in the diagonal form
\begin{equation}
\mathcal{H}_A=\displaystyle \sum_{j=0}^{\infty}  \varepsilon_j (\alpha_j^{\dagger} \alpha_j + \gamma_j^{\dagger}\gamma_j),
\end{equation}
by the transformation \eqref{eq:bogi}.

\section{A generalisation of the binomial theorem}\label{app:A}
In this Appendix we report a proof (based on Refs. \cite{comb,sm-12}) of a generalisation of the binomial theorem that has been used in Eq. \eqref{eq:prods}.
We also discuss some corollaries of the theorem used in the main text. 

The generalisation of the binomial theorem is:
\begin{equation}
\label{eq:qanalog111}
\displaystyle \prod_{j=0}^{n-1} (1+xt^j)=\displaystyle \sum_{k=0}^n t^{k(k-1)/2}\binom{n}{k}_{t} x^k,
\end{equation}
where $\binom{n}{k}_{t}$ is the generating function (in the variable $t$) for the number of integer partitions with at most $k$ parts, whose largest part is at most $n-k$, i.e.
\begin{equation}
\label{eq:qnumber}
\binom{n}{k}_{t}=\displaystyle \prod_{\ell=0}^{k-1}\dfrac{(1-t^{n-\ell})}{(1-t^{\ell +1})}.
\end{equation}
We give a combinatorial proof of Eq. (\ref{eq:qanalog111}). 
Take the left hand side of Eq. (\ref{eq:qanalog111}) and think of it as a polynomial in $x$ (of degree $n$) with coefficients being polynomials in $t$, i.e.
rewrite it as $\sum_{k=0}^n  a_k(t) x^k$.
Clearly, $a_k(t)$ is the generating function for partitions with exactly $k$ parts not exceeding $n$. 
In fact, expanding the product on the left hand side, the term $x^k$  comes from taking $xt^j$ exactly in $k$ factors.  
In each of them, $x$ comes together with some power of $t$, which is different for each factor and does not exceed $n$; 
hence they are parts of our partition. 
These partitions can be thought as Young tableaux with $k$ rows and at most $n$ columns. 
Choosing a given partition, denote as $\lambda_j$ the length of the row $j$ (starting from the bottom). 
We then have $1 \leq \lambda_1 < \lambda_{2} \dots \lambda_{k-1}< \lambda_{k} \leq n$. 
From this partition, we can produce another one with $k$ rows and  at most $n-k$ columns. 
Just proceed as follows: remove zero boxes from the first row, one box from the second row and, in general,  $i-1$ boxes from the $i$-th row. 
So we obtain a partition of $\mu$'s, $1 \leq \mu_1 < \mu_{2} \dots \mu_{k-1}  < \mu_{k}  \leq n-k$, where $\mu_1=\lambda_1$, 
$\mu_2=\lambda_2-1, \dots \mu_k=\lambda_k-(k-1)$. 
The generating function for $\mu$'s is exactly $\binom{n}{k}_{t}$. On the other hand, the generating function for $\lambda$'s is obtained from the generating function on $\mu$'s by multiplying it by $t^{k(k-1)/2}$, which takes into account the total number of  removed boxes. Therefore we have 
\begin{equation}
a_k(t)=t^{k(k-1)/2}\binom{n}{k}_{t},
\end{equation}
which proves Eq. (\ref{eq:qanalog111}). 

When $n \rightarrow \infty$, the limit of Eq. (\ref{eq:qanalog111}) is \cite{sm-12}
\begin{equation}
\label{eq:qanalog}
\displaystyle \prod_{j=0}^{\infty} (1+xt^j)=\displaystyle \sum_{k=0}^{\infty} \dfrac{t^{k(k-1)/2}}{\prod_{\ell=0}^{k-1}(1-t^{\ell})} x^k.
\end{equation}
Another useful property derived from this theorem is the identity \cite{comb}
\begin{equation}
\label{eq:qanalog1}
\displaystyle \prod_{j=0}^{n-1} (1+xt^j)^{-1}=\displaystyle  \sum_{k=0}^{\infty}\binom{n+k-1}{k}_tx^k,\quad \xrightarrow[]{n\rightarrow \infty} \quad  \displaystyle \prod_{j=0}^{\infty} (1+xt^j)^{-1}=\displaystyle  \sum_{k=0}^{\infty}\dfrac{x^k}{ \prod_{j=1}^{k}(1-t^{j})},
\end{equation}
 that we used to derive Eq. (\ref{eq:prods}). 
 
A final observation is that, through this binomial theorem, one can prove that $e^{-\epsilon n q (q-1)}/ \prod_{j=0}^{q-1}(1-e^{-2\epsilon n(j+1)})$ is 
the generating function of the partitions of an integer into $q$ distinct parts. This result will be useful in  appendix \ref{app:C}. 

\section{Some properties of the Jacobi theta functions} \label{app:appendixB}
In this Appendix we report and discuss some properties of the Jacobi theta functions that we exploited  to get some results in the main text.

The Jacobi theta functions $\theta_r(z|u)$, $r=2,3,4$ are defined as \cite{book}
\begin{equation}
\begin{split}\label{Theta3Def}
\theta_2(z|u)
&=
\sum_{k=-\infty}^{\infty} 
u^{\left(k+\frac{1}{2}\right)^2}\, 
 e^{ i(2k+1) z}, \\
\theta_3(z|u)
&=
\sum_{k=-\infty}^{\infty}
 u^{k^2}\,
 e^{2 i k z}, \\
\theta_4(z|u)
&=
\sum_{k=-\infty}^{\infty}
(-1)^k\, u^{k^2}\,
 e^{2 i k z},
 \end{split}
\end{equation}
and we use the standard shorthand $\theta_r(u)\equiv\theta_r(0|u)$, $r=2,3,4$.
The functions $\theta_r(u)$, $r=2,3,4$ can be expressed in terms of infinite products \cite{book}
\begin{equation}
\label{eq:Theta2Prod}
\begin{split}
\theta_2(u)
&=
2 u^{\frac{1}{4}}\prod_{k=1}^\infty
\left[
(1-u^{2k})
(1+u^{2k})^2
\right]
,\\
\theta_3(u)
&=
\prod_{k=1}^\infty
\left[
(1-u^{2k})
(1+u^{2k-1})^2
\right]
,
\\
\theta_4(u)
&=
\prod_{k=1}^\infty
\left[
(1-u^{2k})
(1-u^{2k-1})^2
\right]
.
\end{split}
\end{equation}
These three relations allow us to write some particular infinite products in terms of ratios of Jacobi theta functions. 
An example of such relations is 
\begin{equation}
\label{eq:Jacobi 1}
\displaystyle \prod_{j=0}^{\infty}(1-u^{(2j+1)})=\left(\dfrac{16 u \kappa'^4}{\kappa^2} \right)^{\frac{1}{24}},
\end{equation}
where we defined
\begin{equation}
\label{eq:kappa}
\kappa(u)\equiv \dfrac{\theta_2^2(u)}{\theta_3^2(u)}, \qquad \kappa'(u)=\sqrt{1-\kappa(u)^2}=\dfrac{\theta_4^2(u)}{\theta_3^2(u)}
\end{equation}
that can be obtained properly combining the equations in  (\ref{eq:Theta2Prod}).
Other formulas that can be derived in this way are
\begin{equation}
\label{eq:ratio theta}
\frac{\theta_3(u)}{\theta_2(u)}
=
\frac{2}{u^{1/4}}
\prod_{j=0}^\infty \left(\frac{1+u^{2 j + 1}}{1+u^{2 j }}\right)^2,
\qquad
\frac{\theta_4(u)}{\theta_2(u)}
=
\frac{2}{u^{1/4}}
\prod_{j=0}^\infty \left(\frac{1-u^{2 j + 1}}{1+u^{2 j }}\right)^2,
\end{equation}
and
\begin{equation}
\label{eq:Jacobi 2}
\displaystyle \prod_{j=0}^{\infty}(1+u^{2j+1})=\left(\dfrac{16 u }{\kappa^2 \kappa'^2} \right)^{\frac{1}{24}},
\end{equation}
where $\kappa$ and $\kappa'$ are defined in Eq. (\ref{eq:kappa}).
Combining (\ref{eq:ratio theta}) with the relation \cite{book} $\theta_3^4=\theta_2^4+\theta_4^4$, we find
\begin{equation}
\label{eq:inf prod theta XXZ}
\prod_{j=0}^\infty\, (1+u^{2 j })
=
\left\lbrace
\frac{16}{u}
\left[
\prod_{j=0}^\infty\, (1+u^{2 j+1 })^8
-
\prod_{j=0}^\infty\, (1-u^{2 j+1 })^8
\right]
\right\rbrace^{1/8}.
\end{equation}
Then, using (\ref{eq:Jacobi 1}) and (\ref{eq:Jacobi 2}) we get \begin{equation}
\label{eq:inf prod theta XXZ final}
\displaystyle \prod_{j=0}^{\infty} (1+u^{2j})=\left[\frac{16^{4/3}}{( u \kappa)^{2/3}}(\kappa'^{\,-2/3}-\kappa'^{\,4/3}) \right]^{1/8}.
\end{equation}
The denominator of the Eq. (\ref{eq:ZnqXXZcountingES3}) can now be written in terms of Jacobi theta functions using (\ref{eq:Theta2Prod}), (\ref{eq:Jacobi 2}) 
and (\ref{eq:inf prod theta XXZ final}), allowing us to obtain (\ref{eq:ZnqXXZcountingES4}).

We also report the infinite product representation of $\theta_3(z|u)$ that was useful to retrieve the result of \cite{cc-04} in Eq. (\ref{eq:retrievingCC}) \cite{book}
\begin{equation}
\label{eq:Theta3Prodznonzero}
\theta_3(z|u)
=
\prod_{k=1}^\infty
\left[
(1-u^{2k})
(1+u^{2k-1}e^{2 i z})
(1+u^{2k-1} e^{-2 i z})
\right].
\end{equation}
\subsection{Some asymptotic properties of the Jacobi theta functions}
In this subsection we report some asymptotic expressions of $\theta_r(z|u)$, $r=2,3,4$ in the limit in which the variable $u \to 1$. 
These formulas are useful to derive results in the critical regime, namely for $\epsilon \to 0$.
Let us consider first the case in which the variable $z$ in the theta functions is 0.
At the leading order when $u\to 1$, we can write  \cite{asymptotic theta} 
\begin{equation}
\label{eq:asymptotictheta}
\theta_2(u)
\simeq
\,\sqrt{\frac{\pi}{\log\left(1/u\right)}},
\qquad
\theta_3(u)\simeq
\sqrt{\frac{\pi}{\log\left(1/u\right)}},
\qquad
\theta_4(u)\simeq
2 \sqrt{\frac{\pi}{\log\left(1/u\right)}} \,e^{\frac{\pi^2}{4 \log u}}.
\end{equation}
From the definition (\ref{eq:kappa}) we therefore obtain at the leading order
\begin{equation}
\label{eq:asymptotickappa}
\kappa(u)\simeq\, 1 ,
\,\,\qquad\,\,\qquad
\kappa'(u)\simeq\,
4\,e^{\frac{\pi^2}{2\, \log u}}
.
\end{equation}
Two examples in which these asymptotic formulas have been employed in the main text are, setting $u=e^{-\epsilon}$,
\begin{equation}
\left(\dfrac{16 e^{-\epsilon} \kappa'^4}{\kappa^2} \right)^{\frac{n}{24}}\simeq\,
2^{\frac{n}{2}}\,
 e^{-\frac{\pi^2 n}{12 \epsilon}},
\end{equation}
that has been exploited to obtain (\ref{eq:Fcrit}), and
\begin{equation}
\left[\frac{16^{4/3}}{( q \kappa)^{2/3}}(\kappa'^{\,-2/3}-\kappa'^{\,4/3}) \right]^{n/8}
\simeq \,
2^{\frac{n}{2}}\,
 e^{\frac{\pi^2 n}{24 \epsilon}},
\end{equation}
involved in the computation of the critical limit of (\ref{eq:ZnqXXZcountingES4}).

The asymptotic expression for $u\to 1$ of $\theta_3(z|u)$ is \cite{asymptotic theta} 
\begin{equation}
\label{eq:asymptTheta3znonzero}
\theta_3(z|u)
\simeq
\sqrt{\frac{\pi}{\log\left(1/u\right)}}\,
e^{\frac{z^2}{\log u}}\,,
\end{equation}
which reduces to the second identity of (\ref{eq:asymptotictheta}) when $z=0$. 
Plugging (\ref{eq:asymptTheta3znonzero}), setting $z=-i n \epsilon$ and $u=e^{-4 n \epsilon}$, into Eq. (\ref{eq:ChargedMomentsXXZ2}) we obtain
(\ref{eq:logZnalfaXXZcountingEScrit0}).

\section{The CTM  symmetry resolution}\label{app:C}

In the main text of the paper, we derived the symmetry resolved entropies for the most interesting case of the conserved charges $Q_A$ being the 
``electrical'' charge of the complex harmonic chain and magnetisation of the XXZ chain
(equivalently the number operator in fermion language).
Being these models integrable, there are many other conservation laws that can be used in place of these, but usually are very difficult to calculate. 
However,  a quantity we can easily deal with in the CTM approach is $Q_A=\sum_j n_j=\sum_j \beta^\dag_j\beta_j$, 
although it has {\it not} a clear physical meaning, if it has one at all. 
Indeed, since $[\rho_A,n_j]=0$ for each $j$, $Q_A$ is conserved and the symmetry resolved entanglement for the sectors with different values of this quantity may be studied. 
We will refer to $Q_A=\sum_j n_j$ as the CTM charge. 
Although these results have most likely no physical meaning at all, the details of the calculations are rather interesting and worth being presented.

\subsection{The CTM symmetry resolution in the harmonic chain}
For a single real harmonic chain, the flux resolved partition sum for the CTM charge is just $Z_n(\alpha)=e^{F_n(\alpha)}$.
Before performing the Fourier transform to get the symmetry resolved moments, it is useful to rewrite $e^{F_{n}(\alpha)}$ as
\begin{equation}
\label{eq:prods}
e^{F_n(\alpha)}=\dfrac{\displaystyle  \prod_{j=0}^{\infty}(1-e^{-  (2j+1)\epsilon})^{n}}{\displaystyle \prod_{j=0}^{\infty}(1-e^{- (2j+1)\epsilon n+i\alpha})}=\displaystyle \prod_{j=0}^{\infty}(1-e^{-  (2j+1)\epsilon})^{n} \sum_{k=0}^{\infty}\frac{e^{- k\epsilon n  +ik\alpha}}{ \prod_{j=0}^{k-1}(1-e^{-2(j+1)\epsilon n})},
\end{equation}
where in the last equality we have used the generalisation of the binomial theorem reported in Appendix \ref{app:A}.  
In addition, 
Eq. \eqref{eq:Jacobi 1} allows us to rewrite the denominator in Eq. (\ref{eq:prods}) in such a way that 
the Fourier transform $\mathcal{Z}_n(q)$ is
\begin{equation}
\label{eq:firstF}
\mathcal{Z}_n(q)=\displaystyle \int_{-\pi}^{\pi}\dfrac{d\alpha}{2\pi}e^{-iq\alpha}Z_n(\alpha)=\displaystyle \left(\dfrac{16 e^{-\epsilon} \kappa'^4}{\kappa^2} \right)^{\frac{n}{24}}\displaystyle \sum_{k=0}^{\infty}\dfrac{e^{-\epsilon  n  k} }{\prod_{j=0}^{k-1}(1-e^{-2n\epsilon(j+1)})}\displaystyle \int_{-\pi}^{\pi}\dfrac{d\alpha}{2\pi}e^{-i\alpha(q-k)}.
\end{equation}
Since $q$ and $k$ are both integer numbers,  Eq. (\ref{eq:firstF}) simplifies to 
\begin{equation}
\label{eq:secondF}
\mathcal{Z}_n(q)=\left(\dfrac{16 e^{-\epsilon} \kappa'^4}{\kappa^2} \right)^{\frac{n}{24}}\dfrac{e^{-\epsilon n q} }{\prod_{j=0}^{q-1}(1-e^{-2n\epsilon  (j+1)})}.
\end{equation}
We also provide the analytic continuation of Eq. (\ref{eq:firstF}) to real $q$
\begin{equation}
\label{eq:anal}
\mathcal{Z}_n(q)=\left(\dfrac{16 e^{-\epsilon} \kappa'^4}{\kappa^2} \right)^{\frac{n}{24}}\dfrac{e^{-\epsilon nq}}{\Gamma_{e^{-2\epsilon n}}(q+1)}(1-e^{-2\epsilon n})^{-q},
\end{equation}
where we expressed the finite product in terms of the infinite products:
\begin{equation}
\label{eq:ratioprodv}
\dfrac{1}{\prod_{j=0}^{q-1}(1-e^{-2n\epsilon  (j+1)})}=\dfrac{\prod_{j=0}^{\infty}(1-e^{-2n\epsilon  (j+q+1)})}{\prod_{j=0}^{\infty}(1-e^{-2n\epsilon  (j+1)})},
\end{equation}
and we introduced the generalised gamma function 
\begin{equation}
\label{eq:gammagen}
\Gamma_{m}(x)=\dfrac{\prod_{k=0}^{\infty}(1-m^{k+1})}{\prod_{k=0}^{\infty}(1-m^{k+x})}(1-m)^{1-x}.
\end{equation}
Eq. (\ref{eq:gammagen}) reduces to the ordinary gamma function in the limit $\epsilon\rightarrow 0$.

In the critical regime $\epsilon\to0$, as showed in Appendix \ref{app:appendixB}, Eq. (\ref{eq:secondF}) becomes
\begin{equation}
\label{eq:Fcrit}
\mathcal{Z}_n(q)\simeq\frac{2^{\frac{n}{2}}}{\Gamma(q+1)}\frac{e^{-\frac{\pi^2 n}{12 \epsilon}}}{(2 n \epsilon)^q}.
\end{equation}

The symmetry resolved R\'enyi entropies are  easily deduced from Eq. (\ref{eq:SREE1}), obtaining
\begin{equation}
\label{eq:res}
\begin{split}
S_n(q)=&\dfrac{1}{1-n}\log \left[\dfrac{\mathcal{Z}_n(q)}{\mathcal{Z}_1(q)^n} \right]=
\dfrac{1}{1-n}\log  \prod_{j=1}^{q}\dfrac{(1-e^{-2\epsilon  j})^n}{(1-e^{-2n\epsilon  j})} \\
=&\dfrac{1}{1-n}\displaystyle \sum_{j=1}^{q}\Big[n\log (1-e^{-2\epsilon  j})-\log (1-e^{-2n\epsilon  j})\Big].
\end{split}
\end{equation}
Taking the limit $n\rightarrow 1$, we get the von Neumann entropy
\begin{equation}
\label{eq:res1}
S_1(q)=-\displaystyle \sum_{j=1}^{q}\left[ \log (1-e^{-2\epsilon j})-\frac{2\epsilon j e^{-2\epsilon j}}{1-e^{-2\epsilon j}}\right],
\end{equation}
The analytic continuations of $S_n(q)$ and $S_1(q)$ to real $q$ are respectively 
\begin{eqnarray}
\label{eq:anal1}
\fl S_n(q)&=&\dfrac{1}{1-n}\log \dfrac{(1-e^{-2\epsilon})^{nq}}{(1-e^{-2n\epsilon})^q}\dfrac{\Gamma_{e^{-2\epsilon }}(q+1)^n}{\Gamma_{e^{-2n\epsilon }}(q+1)},\\
\label{eq:anal2}
\fl S_1(q)&=&-q\log(1-e^{-2\epsilon})-\log \Gamma_{e^{-2\epsilon}}(q+1)+\dfrac{2\epsilon q}{e^{2\epsilon}-1}-\dfrac{\partial_n\Gamma_{e^{-2n\epsilon}}(q+1)|_{n=1}}{ \Gamma_{e^{-2\epsilon}}(q+1)},
\end{eqnarray} 
with the leading behaviour for $\epsilon\rightarrow 0$ given by
\begin{equation}
\label{eq:snleading}
S_n(q)=-q \log 2\epsilon -\dfrac{q\log n }{1-n}+a(q)+\mathcal{O}(\epsilon), \qquad S_1(q)=-q \log 2\epsilon +b(q)+\mathcal{O}(\epsilon),
\end{equation}
where we introduced the functions $a(q)=-\log \Gamma(q+1)$ and $b(q)=a(q)+q$.

\begin{figure}
\centering
\subfigure
  {\includegraphics[width=0.45\textwidth]{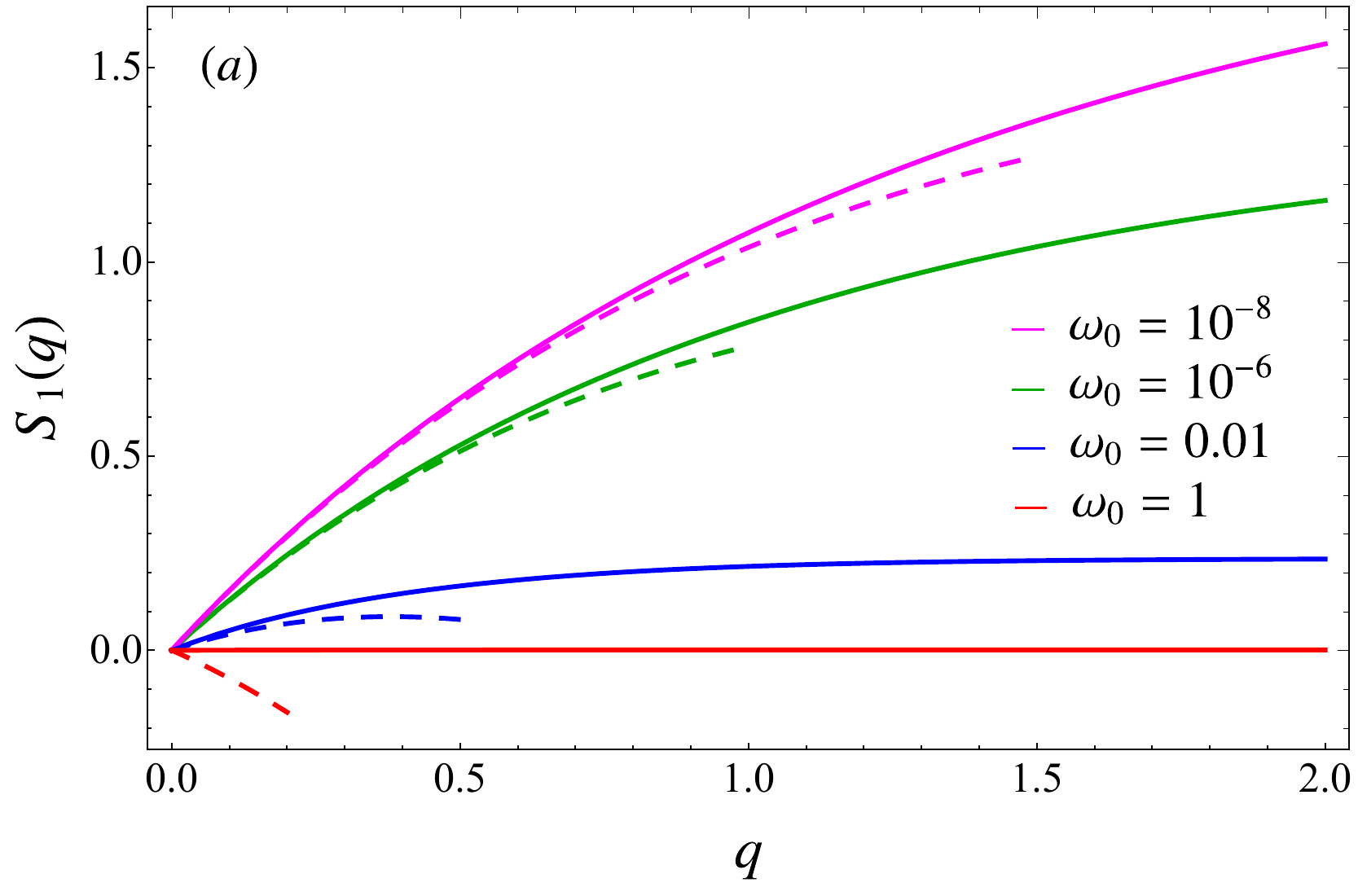}}
\subfigure
   {\includegraphics[width=0.45\textwidth]{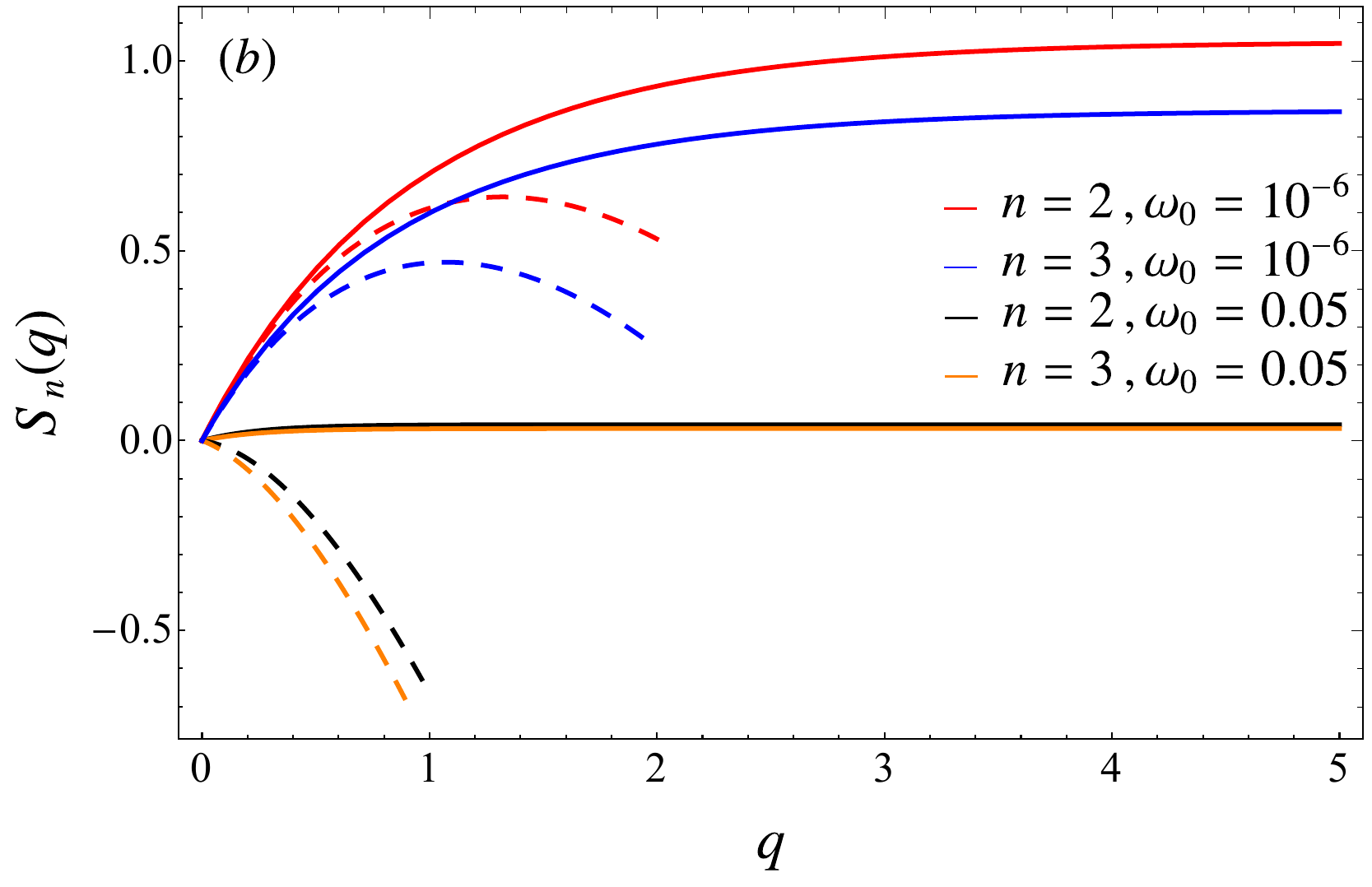}}
    \subfigure
   {\includegraphics[width=0.45\textwidth]{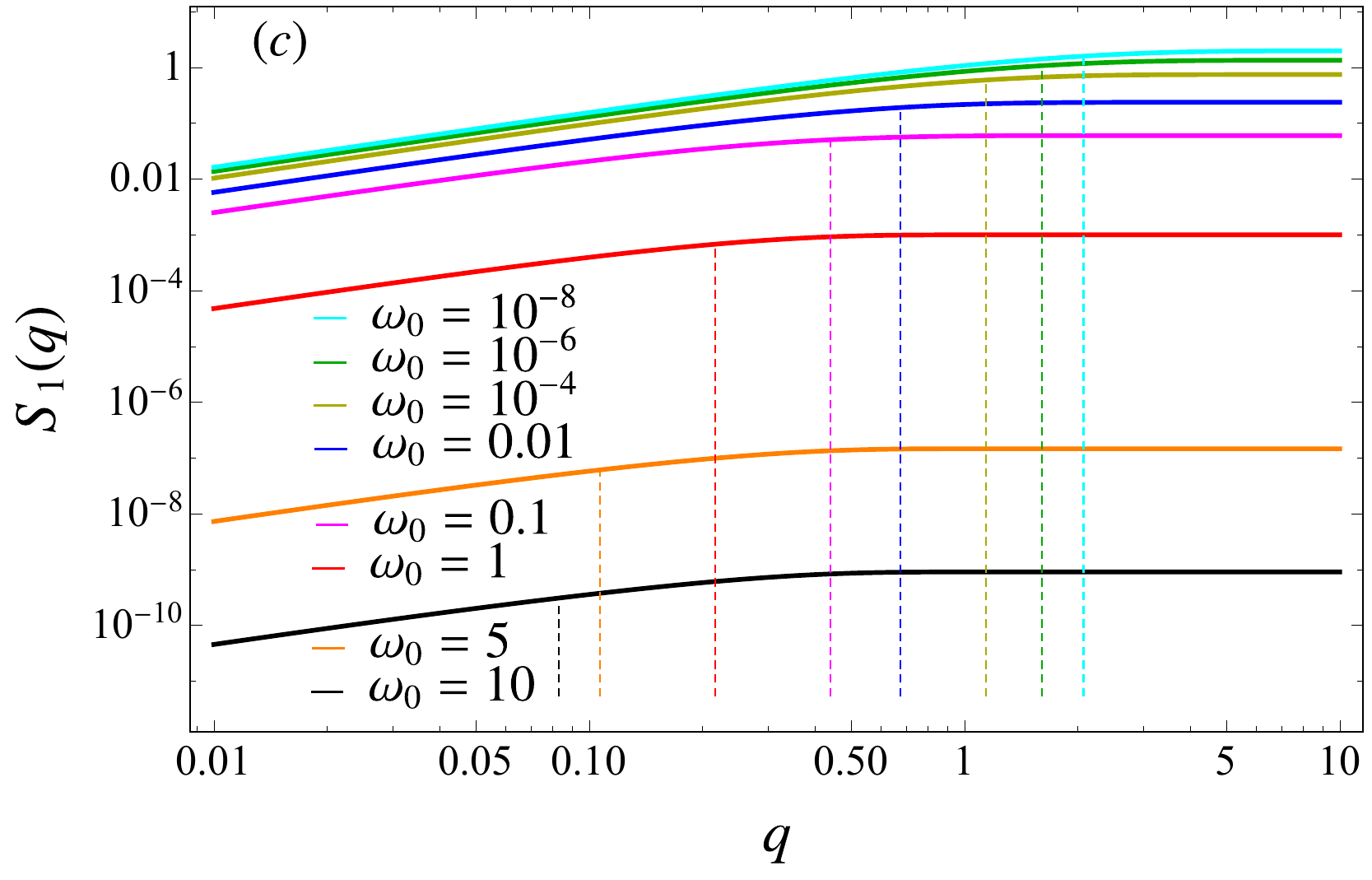}} 
   \subfigure
   {\includegraphics[width=0.45\textwidth]{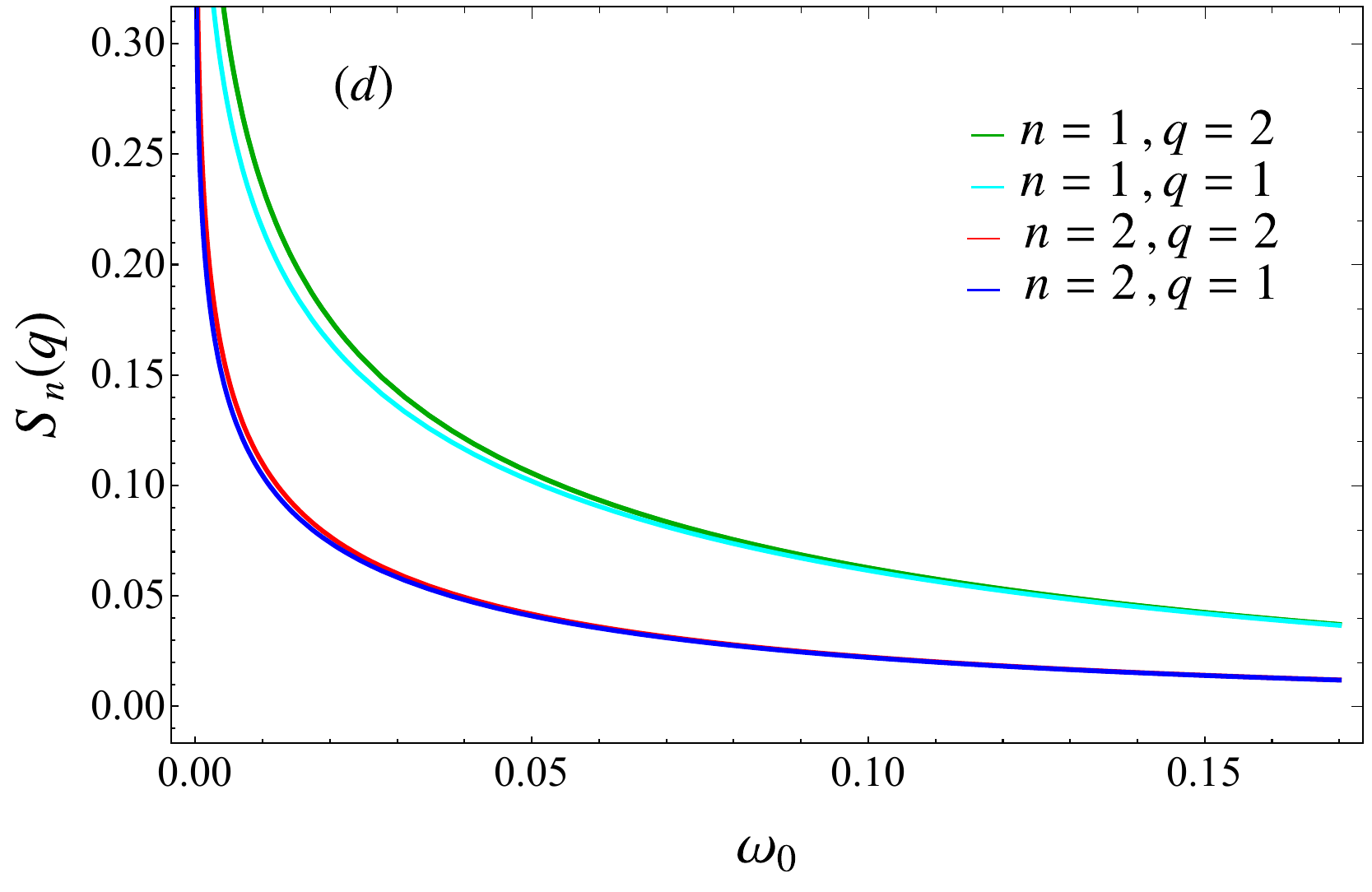}}
\caption{Symmetry resolved entanglement entropies for the CTM charge in the harmonic chain. 
The panels (a) and (b) show $S_1(q)$  and $S_n(q)$ as functions of $q$ for different values of $\omega_0$ and $n$. 
The critical limits in Eq. (\ref{eq:snleading}) are also reported as dashed lines. 
In the limit $\omega_0\to 0$, $S_n(q)$ converge non-uniformly to the results in Eq. (\ref{eq:snleading}). 
The panel (c) reports $S_1(q)$ in log-log scale to manifest the effective equipartition of entanglement for $q\gtrsim 1/\epsilon$ (these crossover values are 
reported as dashed vertical lines). 
The panel (d) shows $S_n(q)$ as function of $\omega_0$ for different values of $q$ and $n$. 
}\label{fig:pannel2}
\end{figure}

The symmetry resolved entropies do not satisfy entanglement equipartition, like the one for the true charge 
of the complex chain. 
However, the breaking of equipartition is rather different: 
in this case the leading term for $\epsilon\to0$ which grows linearly in $q$ and is proportional to $\log \epsilon$, 
while for the complex chain the first term breaking equipartition is subleading and goes like $\epsilon^2$ 
(the sums for the entanglement entropies are finite because the probabilities decay fast with $q$, cf. Eq. \eqref{eq:Fcrit} for $n=1$).
Anyhow, from the expressions as sums over $q$ in Eqs. \eqref{eq:res} and \eqref{eq:res1}, it is clear that all the terms with $2\epsilon j \gg 1$
are exponentially suppressed.  
Practically, the total sum is more or less the same for all $q$ such that $\epsilon q \gtrsim 1$ (from Eq. \eqref{eq:unscaling} this is equivalent to $ q \pi^2 \gtrsim  \log \xi$
in the critical region).
Hence, there is an {\it effective equipartition} among all $q\gtrsim 1/\epsilon$. 
Actually, since the only physical values of $q$ are the integers, this fact implies that there is an almost exact equipartition (with the trivial exception of $S_n(0)=0$)
of the entropy if $\epsilon \gtrsim 1$, which corresponds to $\omega_0\gtrsim 10^{-4}$. 

Some results for the symmetry resolved moments and entropies are reported in Figure \ref{fig:pannel2} as continuous functions of real $q$, 
although only the integer values are physical.
It is evident from the figure that, as $q$ becomes large enough, the entropies $S_n(q)$ do not depend on $q$ anymore, as from the previous argument about 
effective equipartition. 
In panel (c) we explicitly report the (approximate) crossover values for $q\sim 1/\epsilon$ (as function of $\omega_0$ is given by Eq. \eqref{eq:CTMsf}), 
showing that it correctly captures the change of behaviour. 
In panels (a) and (b) we report $S_1(q)$ and $S_n(q)$ respectively, together with the critical limit (\ref{eq:snleading}). 
As expected, the approach to the critical behaviour is highly non-uniform in $q$: 
as $q$ becomes larger we need smaller values of $\epsilon$.

As a final non-trivial consistency check of our results we compute the total von Neumann entanglement entropy starting from the symmetry resolved ones 
using Eq. \eqref{eq:SvN}. 
The probability $p(q)$ is given by Eq. (\ref{eq:secondF}) with $n=1$ while the symmetry resolved entropies are in Eq. (\ref{eq:res1}). Plugging these two results  
into Eq. \eqref{eq:SvN} leads to
\begin{multline}
\label{eq:checkent_s1}
S_1
=
-\sum_{j=0}^{\infty} 
\log\left(
1-e^{-\epsilon(2j+1)}
\right)
+\\ +
\prod_{j=0}^\infty
\left(
1-e^{-\epsilon(2j+1)}
\right)
\sum_{q=0}^{\infty}
\left[
\frac{\epsilon e^{-\epsilon q}}{\prod_{k=1}^q \left(1-e^{-2\epsilon k} \right)}
\left( 
q
+
\sum_{k=1}^q
\frac{2k}{e^{2\epsilon k}-1}
\right)
\right].
\end{multline}
The sum over $q$ in (\ref{eq:checkent_s1}) can be written as the following derivative
\begin{equation}
\label{eq:checkent_s2}
\frac{ e^{-\epsilon q}}{\prod_{k=1}^q (1-e^{-2\epsilon k} )}
\Big( 
q +
\sum_{k=1}^q
\frac{2k}{e^{2\epsilon k}-1}
\Big)
=
-\frac{d}{d\epsilon}\Big[ \frac{ e^{-\epsilon q}}{\prod_{k=1}^q \left(1-e^{-2\epsilon k} \right)}\Big].
\end{equation}
Using (\ref{eq:checkent_s2}) in (\ref{eq:checkent_s1}) and exchanging the derivative with respect to $\epsilon$ with the sum over $q$, we can exploit that
\begin{equation}
\label{eq:checkent_s3}
\sum_{q=0}^{\infty}
\left[ \frac{ e^{-\epsilon q}}{\prod_{k=1}^q \left(1-e^{-2\epsilon k} \right)}\right]
=\frac{1}{
\prod_{j=0}^\infty
\left(
1-e^{-\epsilon(2j+1)}
\right)},
\end{equation}
reflecting that $\mathcal{Z}_1(q)$ is normalised to 1.
Taking now the derivative with respect to $\epsilon$, we finally obtain 
\begin{equation}
S_1
=
\sum_{j=0}^{\infty} 
\left[
\frac{ \epsilon (2j+1)}{e^{\epsilon (2j+1)}-1}
-\log\left(
1-e^{-\epsilon(2j+1)}
\right)
\right],
\end{equation}
which is the known result from the CTM calculation in Ref. \cite{cc-04}, i.e. Eq. \eqref{eq:first} for $\alpha=0$.

\subsection{The CTM charge for the XXZ spin chain}
\subsubsection{Charged CTM moments.} 

We now consider the CTM charge in the XXZ spin chain.  As a difference compared to the main text, in this appendix we focus on the state that does 
not break the symmetry, i.e. with entanglement Hamiltonian given by Eq. \eqref{eq:diagonalform} with the sum over $j$ starting from $0$.
As usual, we first compute the charged moments $Z_n(\alpha)$: 
\begin{equation}
\label{eq:stepsXXZ}
Z_n(\alpha)=\dfrac{\mathrm{Tr}e^{-\sum_{j=0}^{\infty}(\epsilon_j n  -i\alpha)n_j}}{\left( \mathrm{Tr} e^{-\sum_{j=0}^{\infty}\epsilon_j  n_j}\right)^n}=\dfrac{\displaystyle \prod_{j=0}^{\infty} \sum_{k=0,1}e^{-(2\epsilon j n  -i\alpha)k}}{\left( \displaystyle \prod_{j=0}^{\infty} \sum_{k=0,1} e^{-2 \epsilon j k}\right)^n}=
\dfrac{\displaystyle \prod_{j=0}^{\infty}(1+e^{-2\epsilon j n+i\alpha})}{\displaystyle \prod_{j=0}^{\infty}(1+e^{-2\epsilon  j})^{n}},
\end{equation}
where we used that for this model the $n_j$'s are fermionic number operators. 
Taking the logarithm of Eq. (\ref{eq:stepsXXZ}) we have
\begin{equation}
\label{eq:firstXXZ}
\log Z_n(\alpha)=\sum_{j=0}^{\infty}\log [1+e^{-2 j n\epsilon+i\alpha}]-\sum_{j=0}^{\infty} n\log [1+e^{-2j \epsilon}].
\end{equation}

The asymptotic expansion for small $\epsilon$  is obtained applying the Poisson resummation formula (\ref{eq:poisson}). 
Defining $f_{n,\alpha}(x)$ as
\begin{equation}
\label{eq:fnXXZ}
f_{n,\alpha}(x)
=
\log(1+e^{-2nx+i\alpha}),
\end{equation}
 we can write (\ref{eq:firstXXZ}) as
 \begin{equation}
 \label{eq:secondXXZ}
 \begin{split}
 \log Z_n(\alpha)&=\sum_{j=0}^{\infty}\left[\,f_{n,\alpha}(\epsilon j)- n f_{1,0}(\epsilon j)\,\right]
 \\
&=\frac{1}{2}
 \sum_{j=-\infty}^{\infty}\left[\,f_{n,\alpha}(|\epsilon j|)- n f_{1,0}(|\epsilon j|)\,\right]
 +
 \frac{\log\left(e^{i\alpha}+1\right)-n\log  2}{2},
 \end{split}
 \end{equation}
 where we used $f_{n,\alpha}(0)={\log(e^{i\alpha}+1)}/{2}$.
 The cosine-Fourier transform (\ref{eq:cosineFourier}) of (\ref{eq:fnXXZ}) is
 \begin{equation}
 \hat{f}_{n,\alpha}(y)=\dfrac{ie^{i\alpha}}{2y}\left[\Phi(-e^{i\alpha},1,1+\frac{iy}{2n})-\Phi(-e^{i\alpha},1,1-\frac{iy}{2n})\right],
\end{equation}
with the function $\Phi$ defined in \eqref{eq:lerch}. For $\alpha=0$ and $n=1$, it reduces to
\begin{equation}
\hat{f}_{1,0}(y)=\dfrac{1}{y^2}-\dfrac{\pi}{2y} \mathrm{csch} \left( \dfrac{\pi y}{2}\right).
\end{equation}  
We now apply to (\ref{eq:secondXXZ}) the Poisson resummation formula (\ref{eq:poisson}) with $b=1$ and $a=0$ and we isolate the term $k=0$,  finding
\begin{multline}
\label{eq:leadinghcXXZexact}
\log Z_{n}(\alpha)=-\dfrac{\mathrm{Li}_2(-e^{i\alpha})}{2\epsilon n}-\dfrac{n\pi^2}{24 \epsilon}
+
\frac{\log\left(e^{i\alpha}+1\right)-n\log  2}{2}\,
+\\
\displaystyle \sum_{k=1}^{\infty}   
\left[
\dfrac{n}{2k} \mathrm{csc h} \frac{\pi^2 k}{\epsilon}
-\dfrac{n \epsilon}{2\pi^2 k^2}   
+ \right.
\left.\dfrac{ie^{i\alpha}}{2\pi k}
\left(\Phi(-e^{i\alpha},1,1+\frac{i\pi k}{\epsilon n})-\Phi(-e^{i\alpha},1,1-\frac{i \pi k}{\epsilon n})\right)
 \right].
\end{multline}
For $\epsilon\to 0$, the leftover sum over $k$ is vanishing. In particular, the last part behaves as 
\begin{equation}\label{eq:divergentXXZ}
\dfrac{ie^{i\alpha}}{2\pi}\displaystyle \sum_{k=1}^{\infty} \dfrac{(-1)^k}{k}\left[\Phi(-e^{i\alpha},1,1-\frac{i\pi k}{\epsilon n})-\Phi(-e^{i\alpha},1,1+\frac{i \pi k}{\epsilon n}) \right] \rightarrow \dfrac{n\epsilon}{6}\dfrac{e^{i\alpha}}{1+e^{i\alpha}}.
\end{equation} 
Thus, we get
\begin{equation}\label{eq:leadinghcXXZcritico}
\log Z_{n}(\alpha)=-\dfrac{\mathrm{Li}_2(-e^{i\alpha})}{2\epsilon n}-\dfrac{n\pi^2}{24 \epsilon}
+
\frac{\log\left(e^{i\alpha}+1\right)-n\log  2}{2}
+
O(\epsilon).
\end{equation}
It is worth to observe that, for this model, the limit $\alpha\to 0$ can be taken after the expansion close to $\epsilon= 0$ retrieving the result found in \cite{cal2010}
\begin{equation}
\log Z_{n}=\left(\frac{1}{n}-n\right)\dfrac{\pi^2}{24 \epsilon}
+
(1-n)\,\frac{\log  2}{2}
+
O(\epsilon).
\end{equation}

\begin{figure}
\centering
\subfigure
  {\includegraphics[width=0.325\textwidth]{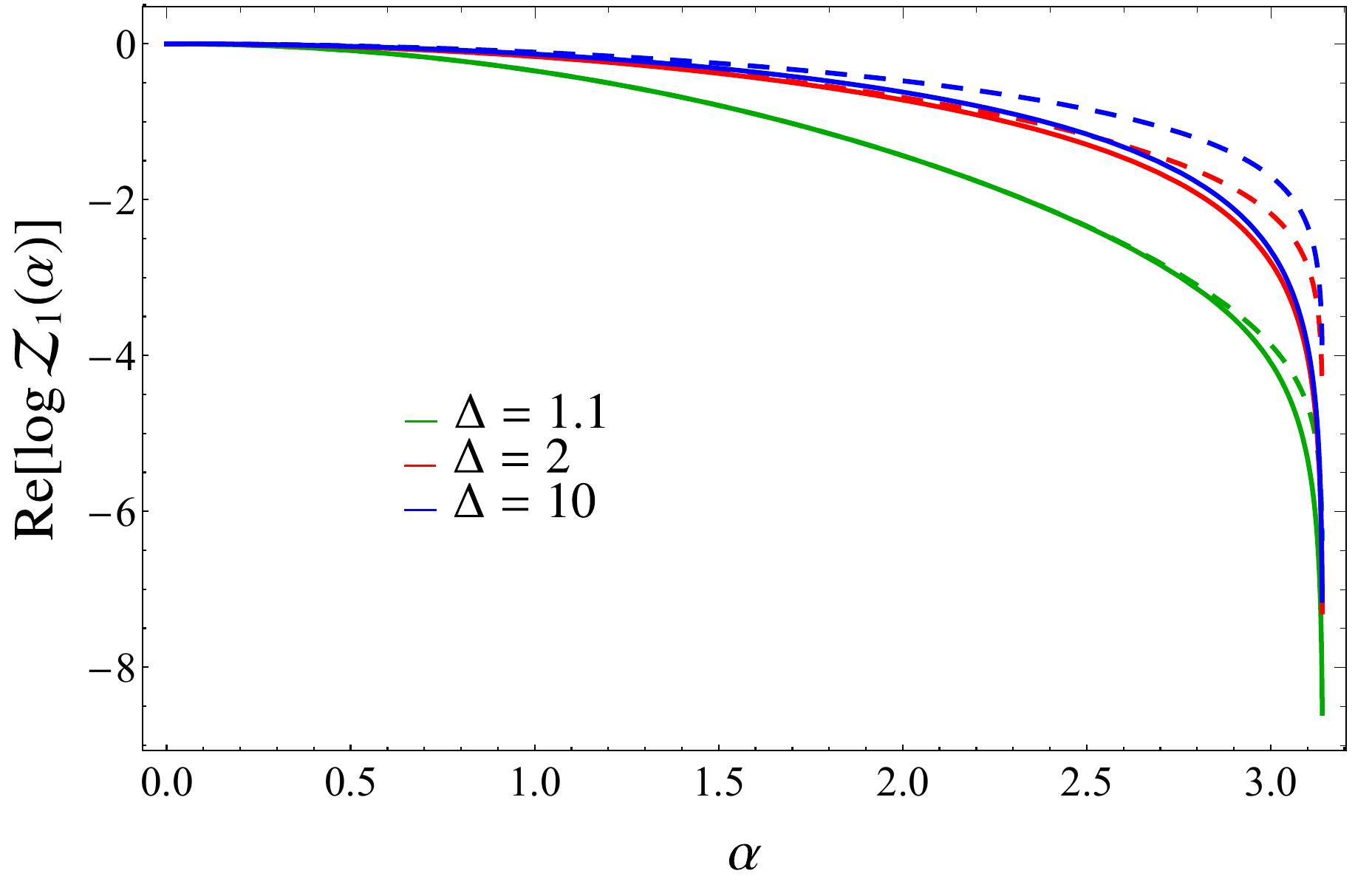}}
\subfigure
   {\includegraphics[width=0.325\textwidth]{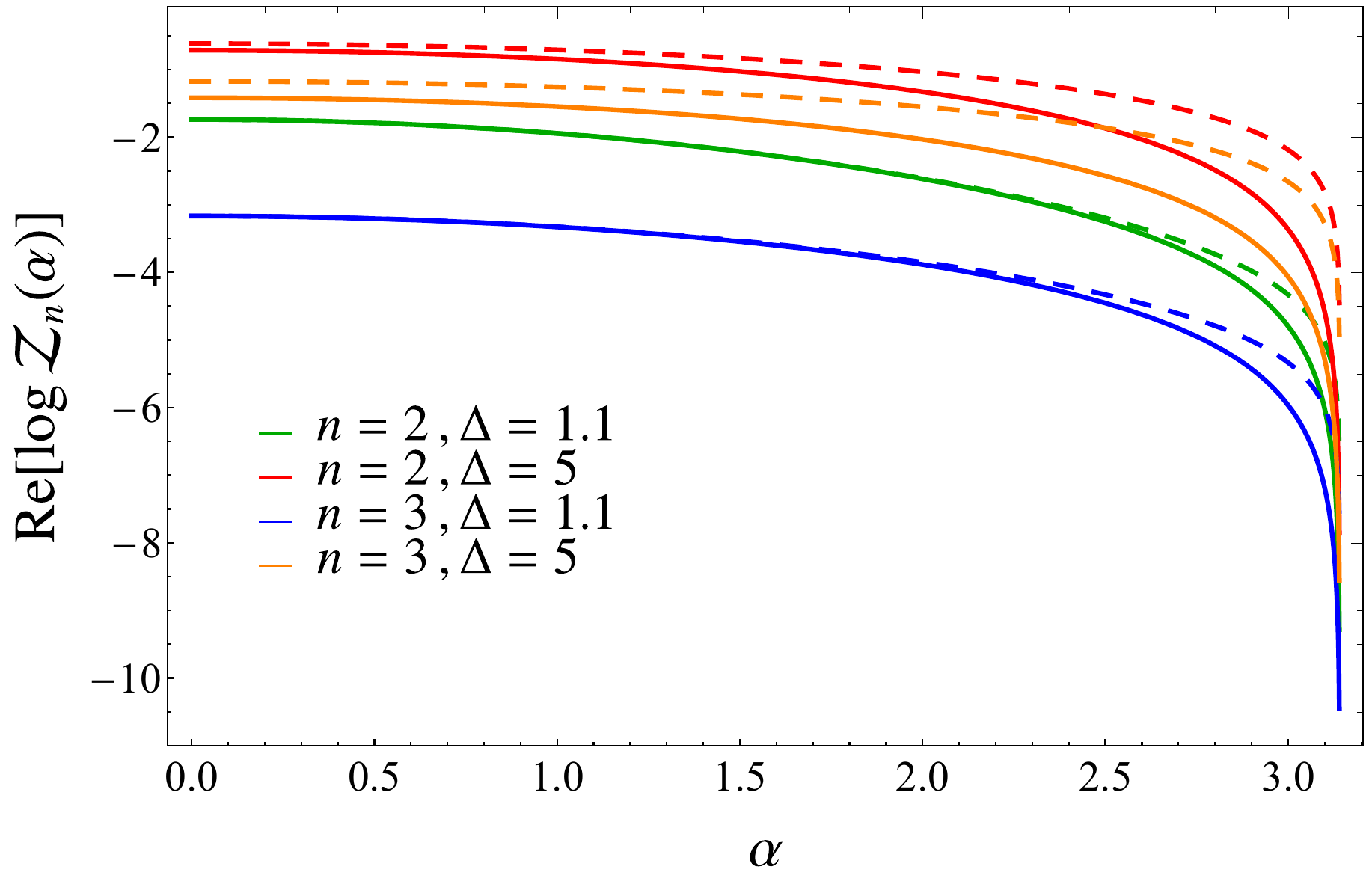}}
   \subfigure
  {\includegraphics[width=0.325\textwidth]{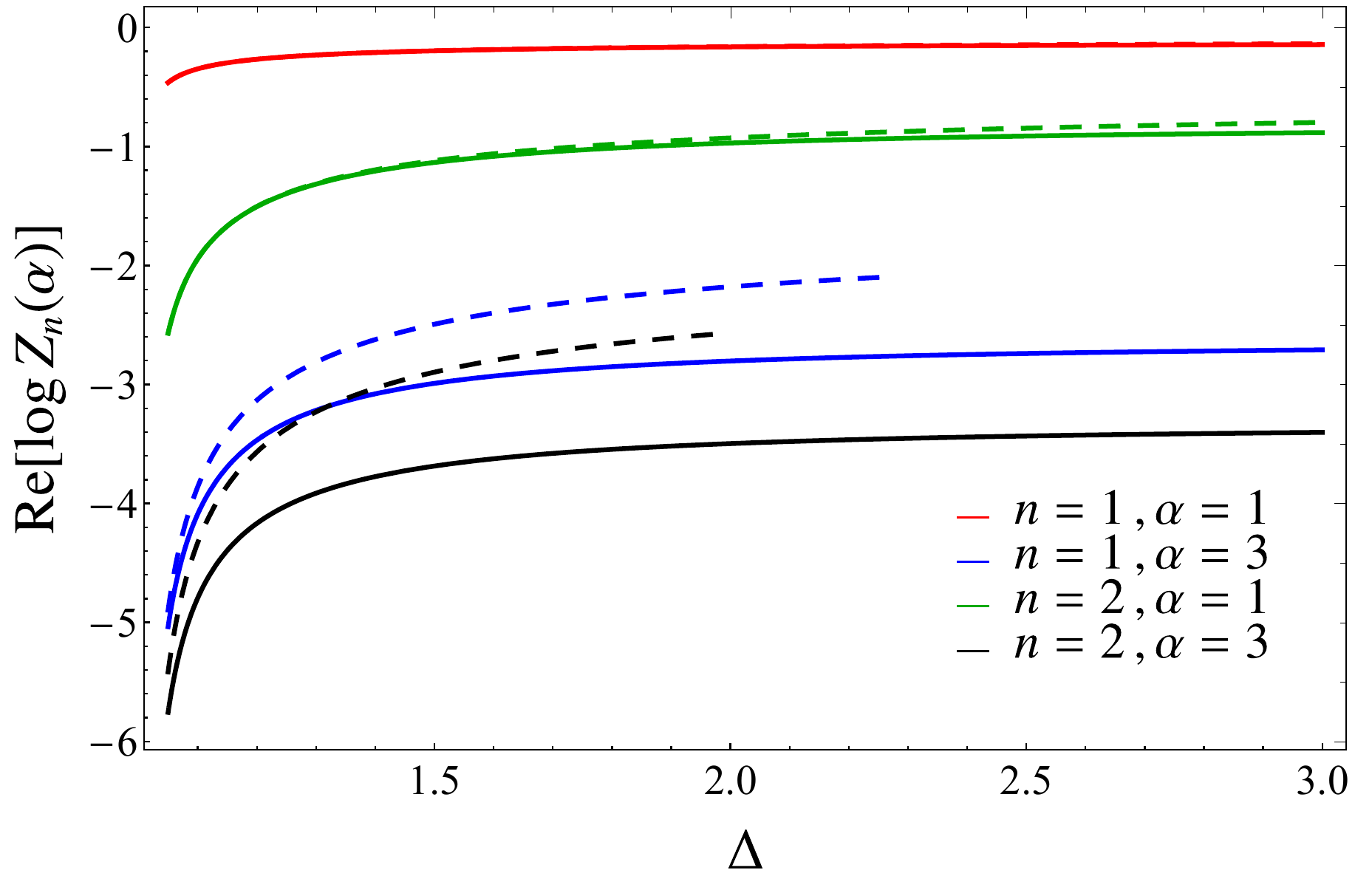}}
  \subfigure
  {\includegraphics[width=0.325\textwidth]{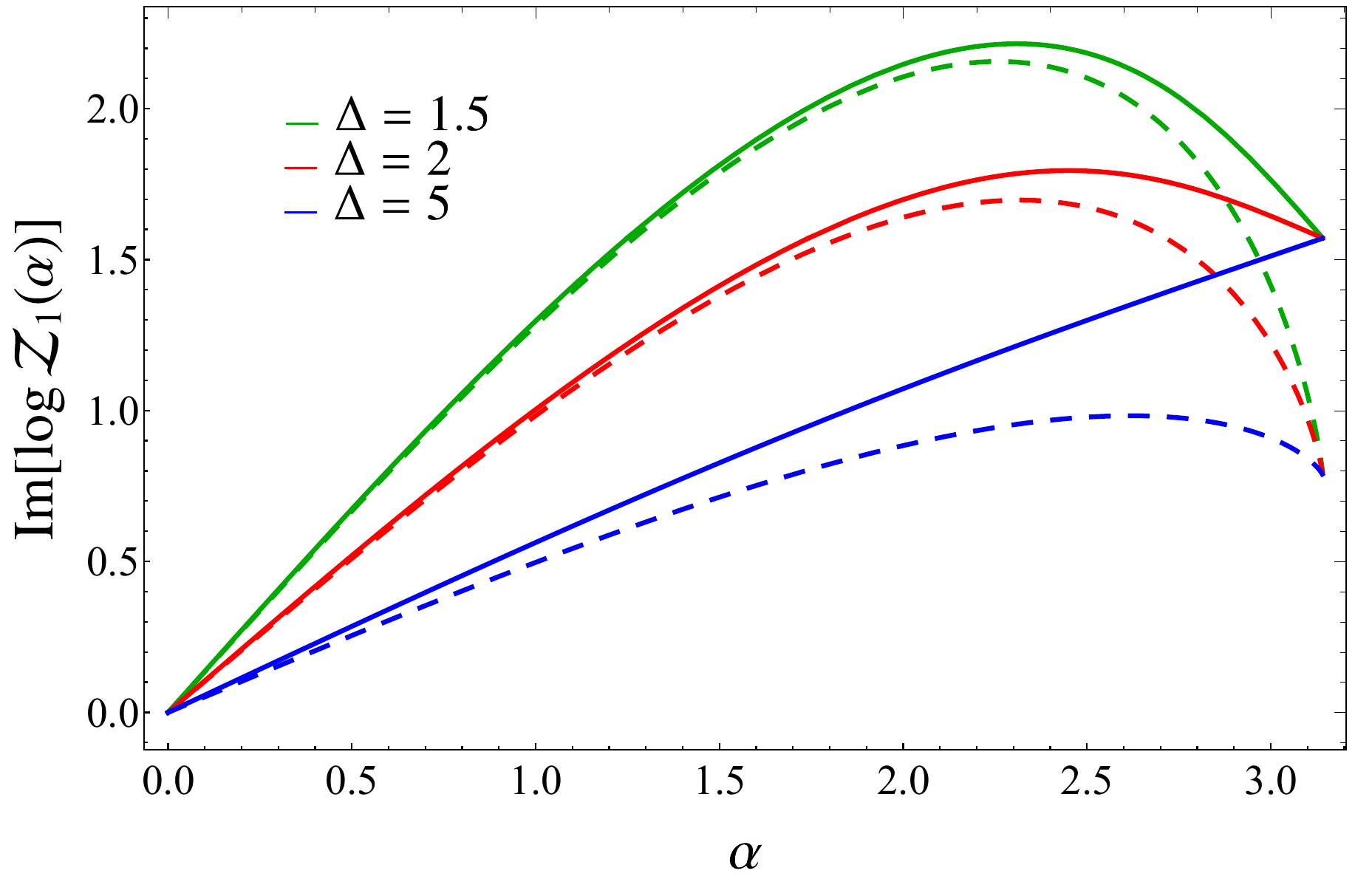}}
\subfigure
  {\includegraphics[width=0.325\textwidth]{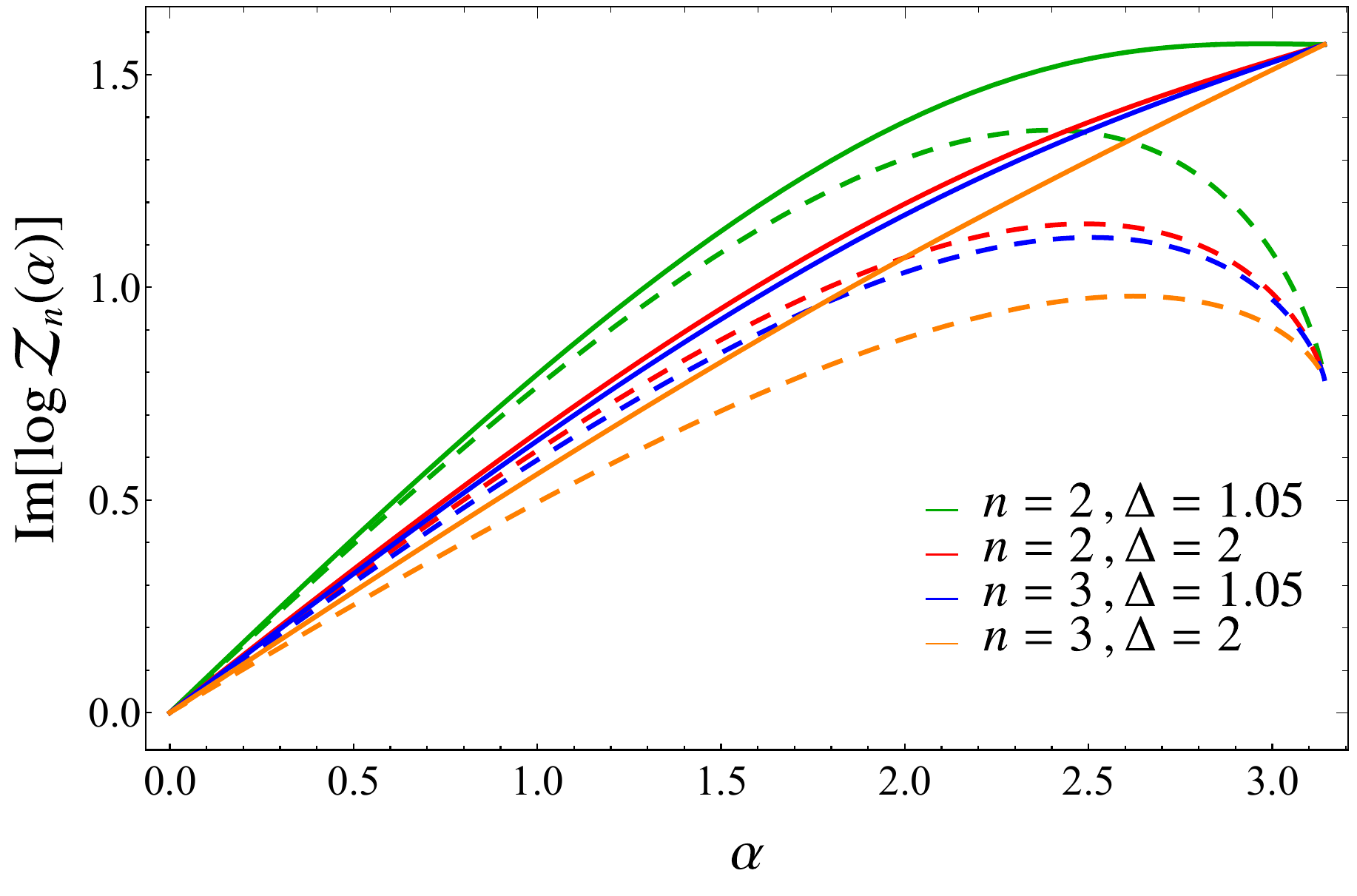}}
  \subfigure
  {\includegraphics[width=0.325\textwidth]{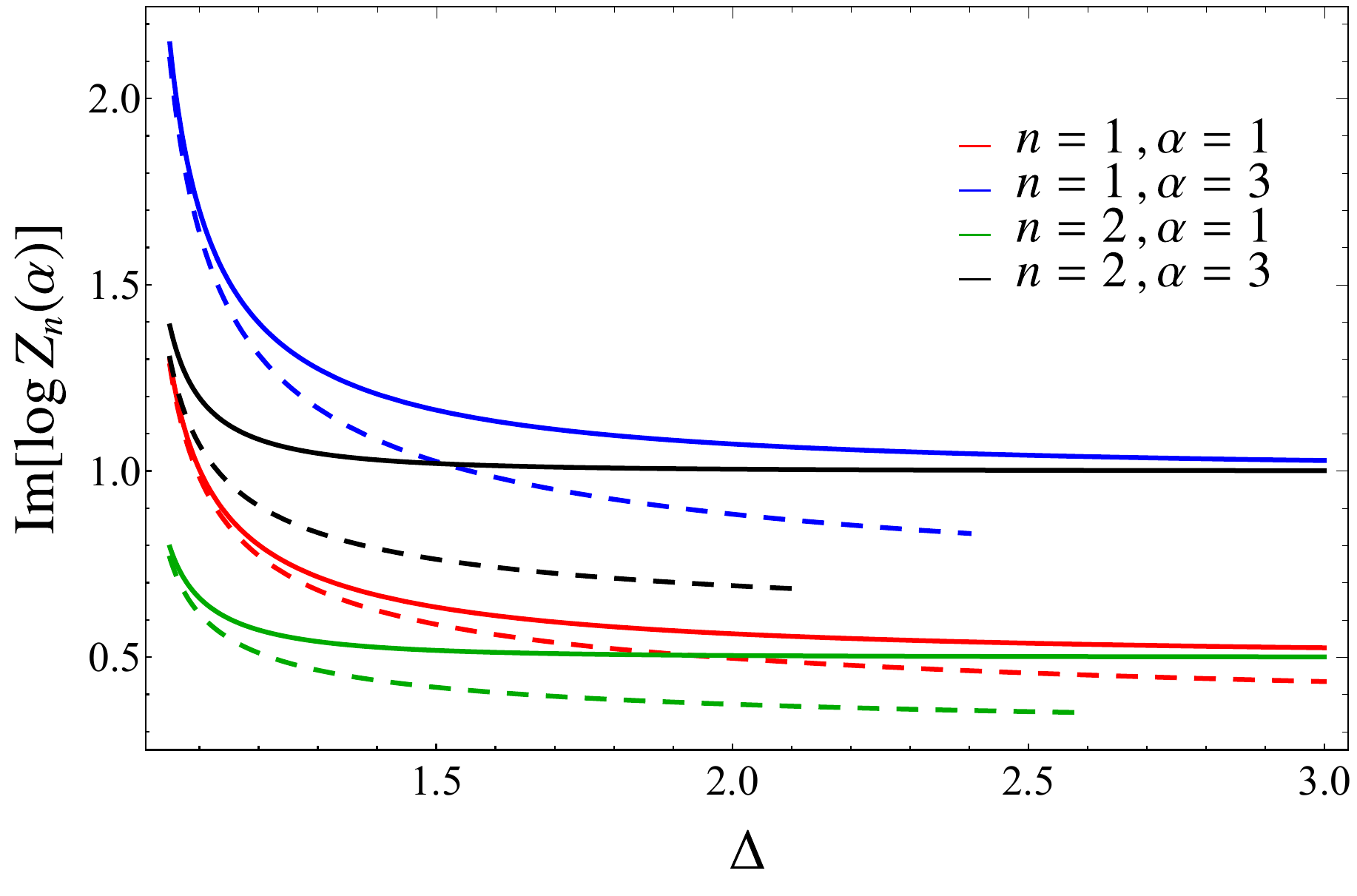}}
\caption{Charged moments for the CTM number in the XXZ spin chain. 
The real and the imaginary part of $\log Z_n(\alpha)$ as functions of $\alpha$ for different values of $\Delta$ and $n$.
As $\Delta$ approaches its critical value, i.e. $\Delta \rightarrow 1$, the exact result (\ref{eq:leadinghcXXZexact}) is well described by the asymptotic
expansion (\ref{eq:leadinghcXXZcritico}), the dashed lines.
The rightmost panels show the same results as functions of $\Delta$ for different $\alpha$ and $n$.}\label{fig:pannelXXZ}
\end{figure}

In Figure \ref{fig:pannelXXZ}, we report the $\alpha$ dependence of the charged moments for different values of $\Delta$ and $n$. 
We also shows the comparison between the exact result (\ref{eq:leadinghcXXZexact}) and its critical limit (\ref{eq:leadinghcXXZcritico}). 
As expected, the latter gets very close to the former as $\Delta$, therefore $\epsilon$, is close to its critical value. 

\subsubsection{Resolved moments via Fourier trasform.}\label{sec:XXZ1}
The Fourier transform of $Z_n(\alpha)$ is obtained by first rewriting (\ref{eq:stepsXXZ}) exploiting Eqs. (\ref{eq:qanalog}) and (\ref{eq:inf prod theta XXZ final})
\begin{equation}\label{eq:prods1}
Z_n(\alpha)
=
\left[\frac{16^{\,4/3}}{( q \kappa)^{2/3}}(\kappa'^{-2/3}-\kappa'^{4/3}) \right]^{-n/8}
\sum_{k=0}^{\infty}\frac{ e^{-\epsilon n k (k-1)+i\alpha k}}{ \prod_{j=0}^{k-1}\left(1-e^{-2\epsilon n  (j+1)}\right)}\,.
\end{equation}
The Fourier transform (\ref{eq:defF}) then reads
\begin{equation}
\label{eq:ZnqXXZ}
\begin{split}
\mathcal{Z}_n(q)
&=
\left[\frac{16^{4/3}}{( q \kappa)^{2/3}}(\kappa'^{-2/3}-\kappa'^{4/3})\right]^{-n/8}
\sum_{k=0}^{\infty}\frac{ e^{-\epsilon n k (k-1)}}{\prod_{j=0}^{k-1}\left(1-e^{-2\epsilon n  (j+1)}\right)}\int_{-\pi}^\pi  \frac{d\alpha}{2\pi} \, e^{-i\alpha(q-k)}\\
&=
\left[\frac{16^{4/3}}{( q \kappa)^{2/3}}(\kappa'^{-2/3}-\kappa'^{4/3})\right]^{-n/8}
\frac{ e^{-\epsilon n q (q-1)}}{ \prod_{j=0}^{q-1}\left(1-e^{-2\epsilon n  (j+1)}\right)}.
\end{split}
\end{equation}
The analytic continuation of $\mathcal{Z}_n(q)$ to real $q$ is achieved by using Eq. (\ref{eq:gammagen}):
\begin{equation}\label{eq:XXZplot1}
\mathcal{Z}_n(q)
=
\left[\frac{16^{4/3}}{( q \kappa)^{2/3}}(\kappa'^{-2/3}-\kappa'^{4/3})\right]^{-n/8}\,
\dfrac{e^{-\epsilon nq(q-1)}}{\Gamma_{e^{-2\epsilon n}}(q+1)}(1-e^{-2n\epsilon})^{-q}.
\end{equation}
In the critical regime $\epsilon\to 0$ we get (see Appendix \ref{app:appendixB})
\begin{equation}\label{eq:ZncriticaXXZ}
\mathcal{Z}_n(q)
\simeq \dfrac{2^{-\frac{n}{2}}\,e^{-\frac{\pi^2n}{24 \epsilon}}}{\Gamma(q+1)\, (2 \epsilon n)^q}.
\end{equation}

We can check Eqs. \eqref{eq:XXZplot1} and \eqref{eq:ZncriticaXXZ} computing $\mathcal{Z}_n(q)$ directly from the entanglement spectrum, as 
done in Sec. \ref{sec:RMXXZ} for the case of $Q_A$ being the magnetisation. 
In the symmetry sector with charge $q$, the degeneracy of the level $2\epsilon s$ is $\mathcal{P}_{q}(s)$, i.e. the number of partitions of an integer $s$ in exactly $q$ parts, 
not exceeding $s$. The partition function $\mathcal{Z}_n(q)$ then is
\begin{equation}
\label{eq:ZnqXXZ combinatorial}
\mathcal{Z}_n(q)
=
\frac{\sum_{s=0}^\infty \mathcal{P}_{q}(s) e^{-2\epsilon s n}}{\left(\sum_{s=0}^\infty  2\mathcal{Q}(s) e^{-2\epsilon s }\right)^n} ,
\end{equation}
which is equivalent to (\ref{eq:ZnqXXZ}): $\prod_{j=0}^{\infty}(1+e^{-2\epsilon  j})^{-n} 
$, as already said, is linked to the partitions of integers into distinct parts, while  $\frac{ e^{-\epsilon n q (q-1)}}{ \prod_{j=0}^{q-1}\left(1-e^{-2\epsilon n  (j+1)}\right)} $ is the generating function for the number of partitions of $s$ into $q$ positive integers  (see Appendix \ref{app:A}). Therefore
\begin{equation}
\dfrac{ e^{-\epsilon n q (q-1)}}{ \prod_{j=0}^{q-1}\left(1-e^{-2\epsilon n  (j+1)}\right)}=\sum_{s} \mathcal{P}_{q}(s)\, e^{-2\epsilon n s}.
\end{equation} 
For $n=1$ Eq. (\ref{eq:ZnqXXZ combinatorial}) is  normalised since $\sum_{q=0}^\infty \mathcal{P}_q(s)=2 \mathcal{Q}(s)$, 
as it should since $\mathcal{Z}_1(q)$ is a probability.

\begin{figure}
\centering
\subfigure
  {\includegraphics[width=0.45\textwidth]{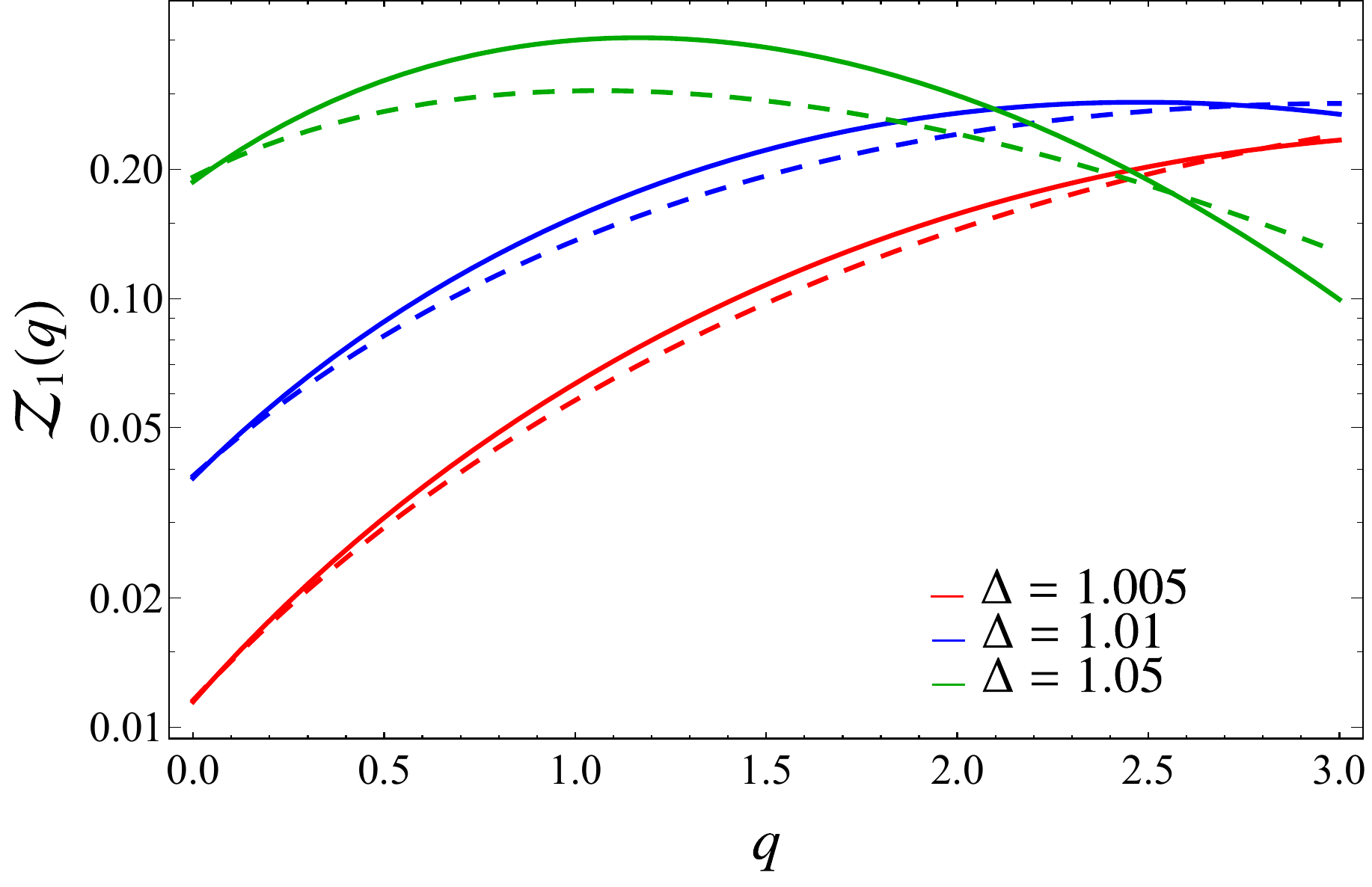}}
\subfigure
  {\includegraphics[width=0.45\textwidth]{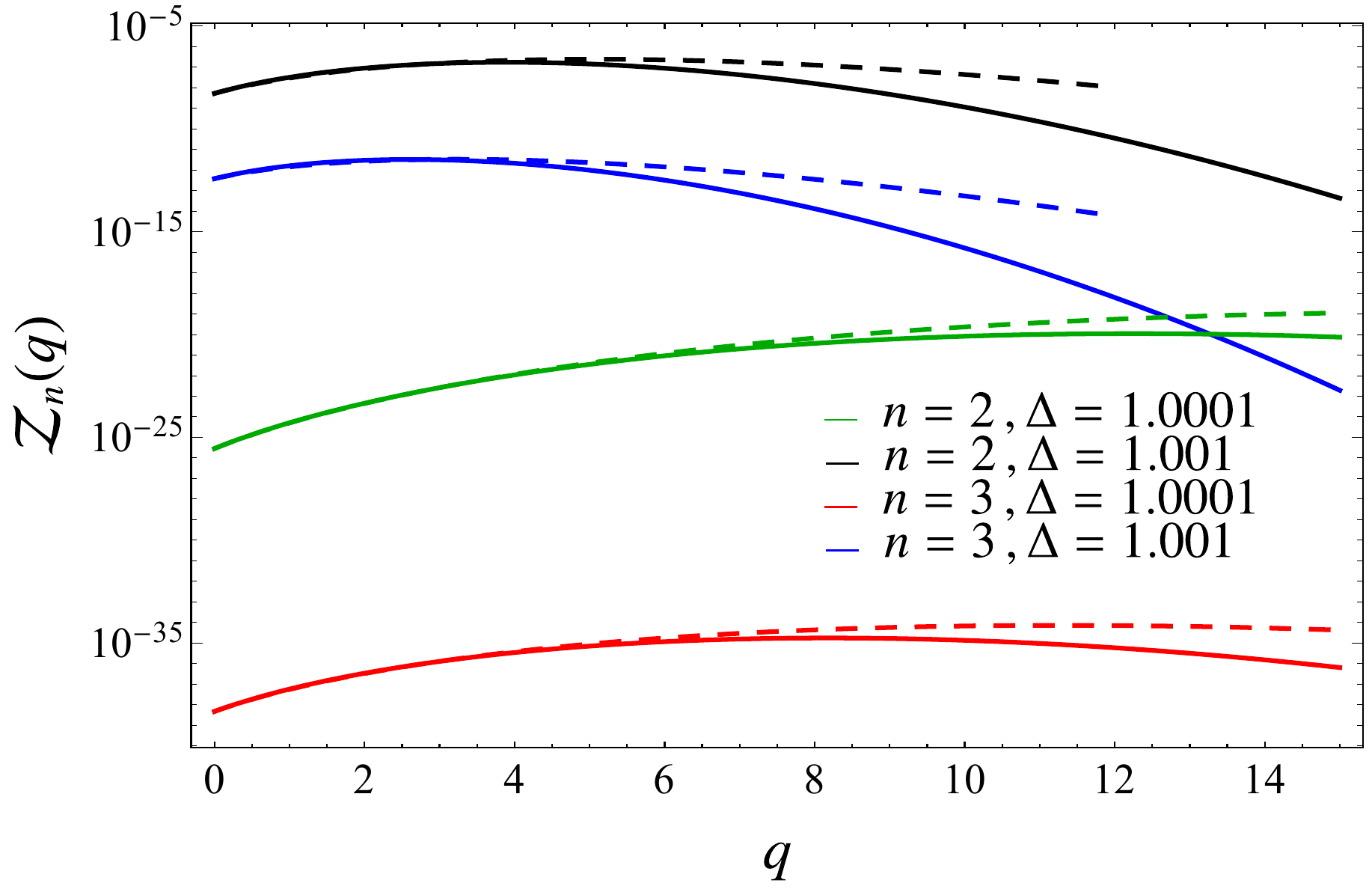}}
\subfigure
   {\includegraphics[width=0.45\textwidth]{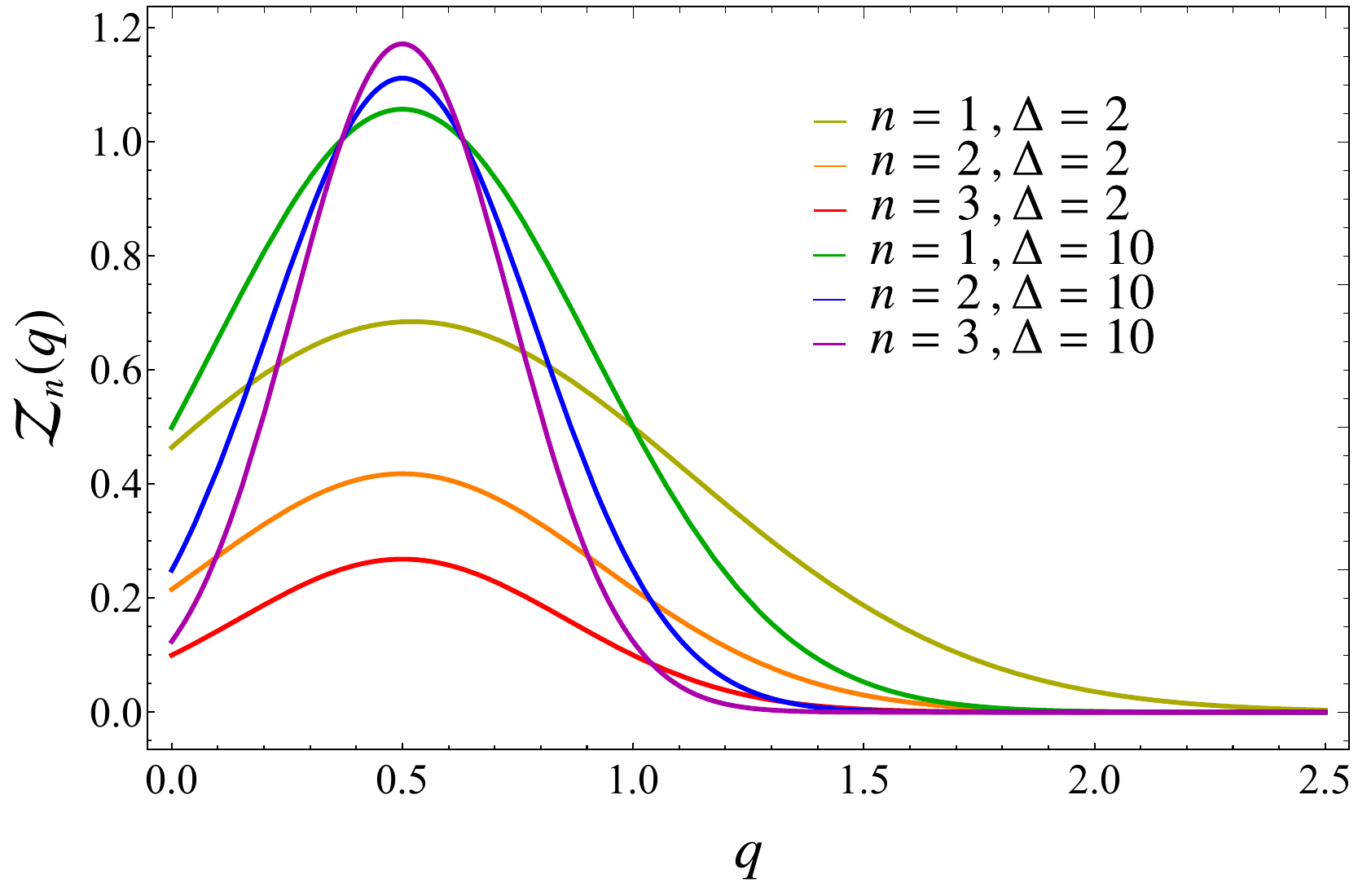}}
  \subfigure
   {\includegraphics[width=0.45\textwidth]{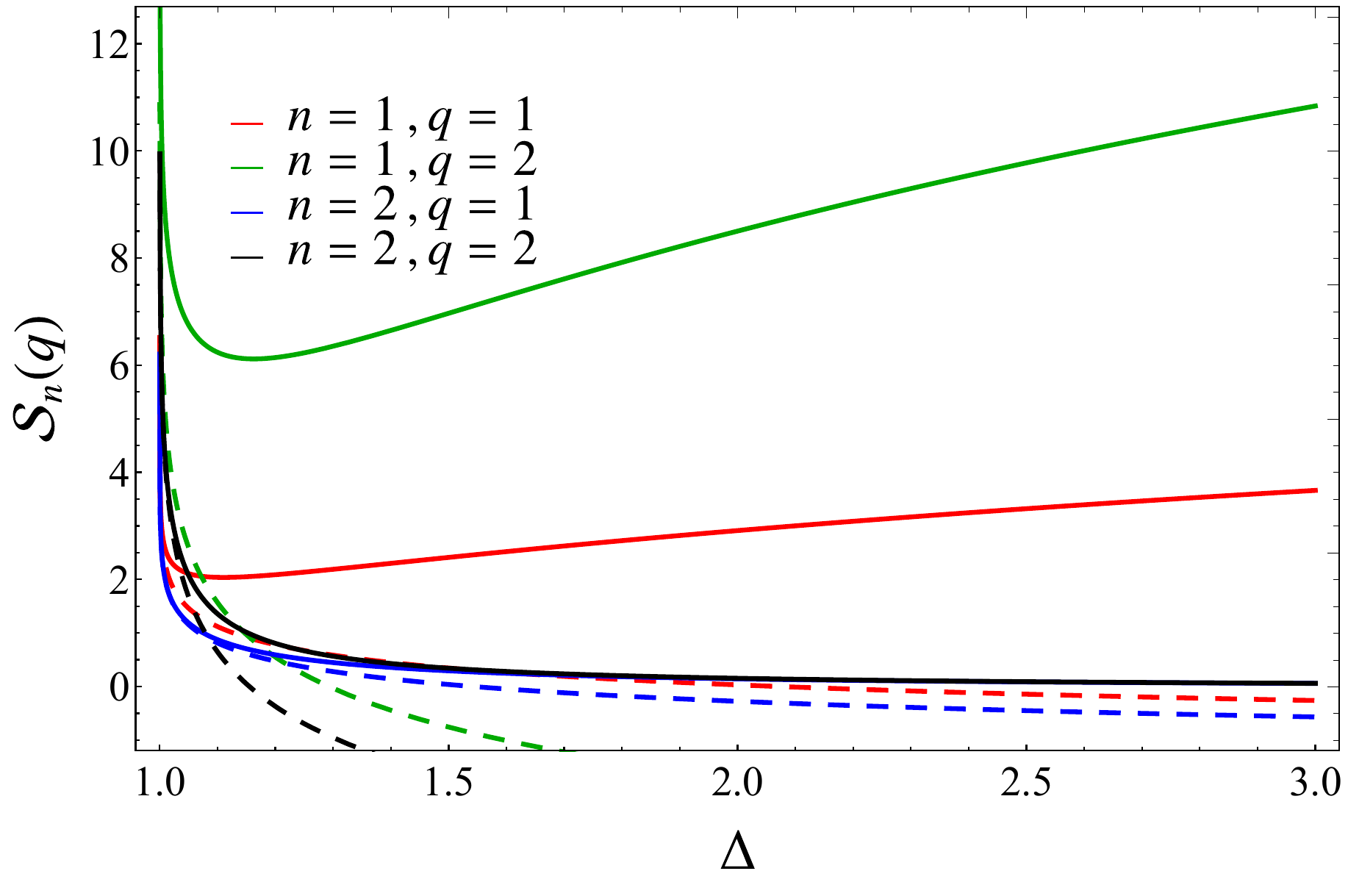}}
\caption{Symmetry resolved moments and entropies for the CTM number in the XXZ spin chain. 
The top two panels report the exact results for $\mathcal{Z}_n(q)$ (\ref{eq:XXZplot1})--full lines-- for $\Delta$ close to $1$ and the comparison with the critical limit, Eq. (\ref{eq:ZncriticaXXZ})--dashed lines--, as a function of $q$, for different values of $n = 1, 2, 3$. 
At fixed $\Delta$, the approach is not uniform and the smaller values of $q$ converges faster. 
In the third panel, we report $\mathcal{Z}_n(q)$ for $\Delta$ far from the critical point, where a peak at $q=1/2$ is developed. 
In the last panel, the exact and critical limit of $S_n(q)$, respectively Eqs. (\ref{eq:anal1}) and (\ref{eq:snleading}), are shown against $\Delta$ for different $q$.
}\label{fig:plotXXZ}
\end{figure}

From $\mathcal{Z}_n(q)$, we  compute the symmetry resolved entropies (\ref{eq:SREE1})
\begin{equation}
\label{eq:resXXZ}
\begin{split}
S_n(q)=&\dfrac{1}{1-n}\log  \prod_{j=1}^{q}\dfrac{(1-e^{-2\epsilon j})^n}{(1-e^{-2n\epsilon  j})} 
=\dfrac{1}{1-n}\displaystyle \sum_{j=1}^{q}[n\log (1-e^{-2\epsilon j})-\log (1-e^{-2n\epsilon  j})].
\end{split}
\end{equation}
These symmetry resolved entropies have the same form as the ones  for the harmonic chain in Eq. (\ref{eq:res}) except for the explicit expression of $\epsilon$. 
Thus, the analytic continuation to real $q$, the von Neumann limit, and the behaviour in the critical regime are the same as those obtained in the previous section 
and we do not report here. 
Notice that these symmetry resolved entropy do not satisfy entanglement equipartition. However, as for the harmonic chain, equipartition is effectively recovered as 
$q\gtrsim 1/\epsilon$. 

Finally, notice the similarity between these symmetry resolved entropies and the ones for the magnetisation in Eq. \eqref{eq:SREEXXZn}. 
Apart from a reparametrisation and an additive term, the main difference is that in the case of the CTM charge the sum is up to $q$ and in the magnetisation case it is 
up to $\infty$ (and that is why the former does not satisfy  equipartition while the latter does). When the upper limits in the former do not matter, the two become
practically equivalent.

 In Figure \ref{fig:plotXXZ} we plot $\mathcal{Z}_n(q)$, showing also a comparison between the exact result (\ref{eq:XXZplot1}) and its critical limit, Eq. (\ref{eq:ZncriticaXXZ}).  It is interesting to observe that the maxima of $\mathcal{Z}_n(q)$ are increasing or decreasing with $n$ depending on the considered values of $\Delta$.
In the last panel we report the exact expression  of $S_n(q)$ and its critical limit, respectively Eq. (\ref{eq:anal1}) and Eq. (\ref{eq:snleading}), as function of $\Delta$ 
for different $q$. The agreement improves for $\Delta$ close to $1$, as it should.

\end{appendices}

\end{document}